\documentclass{aa}  
\usepackage{scrextend}
\usepackage[dvipsnames]{xcolor}
\usepackage{rotating}
\usepackage{multirow}
\usepackage{ulem}
\usepackage{natbib,twoopt}
\usepackage{anyfontsize}
\usepackage{ccaption}
\usepackage[breaklinks=true]{hyperref} %% to avoid \citeads line fills
\bibpunct{(}{)}{;}{a}{}{,}             %% natbib format for A&A and ApJ
\makeatletter
  \newcommandtwoopt{\citeads}[3][][]{\href{http://adsabs.harvard.edu/abs/#3}%
    {\def\hyper@linkstart##1##2{}%
     \let\hyper@linkend\@empty\citealp[#1][#2]{#3}}}
  \newcommandtwoopt{\citepads}[3][][]{\href{http://adsabs.harvard.edu/abs/#3}%
    {\def\hyper@linkstart##1##2{}%
     \let\hyper@linkend\@empty\citep[#1][#2]{#3}}}
  \newcommandtwoopt{\citetads}[3][][]{\href{http://adsabs.harvard.edu/abs/#3}%
    {\def\hyper@linkstart##1##2{}%
     \let\hyper@linkend\@empty\citet[#1][#2]{#3}}}
  \newcommandtwoopt{\citeyearads}[3][][]%
    {\href{http://adsabs.harvard.edu/abs/#3}
    {\def\hyper@linkstart##1##2{}%
     \let\hyper@linkend\@empty\citeyear[#1][#2]{#3}}}
\makeatother

\usepackage{xcolor}
\usepackage{pdflscape}
\usepackage{multirow}
\usepackage{hyperref}
\usepackage{amsmath}
\usepackage{txfonts}

\usepackage{xintcore}
\usepackage{xintfrac}

\usepackage{cleveref}
\crefdefaultlabelformat{\textsubscript{\textbf{(#2#1#3.)}}}
\usepackage[inline]{enumitem}

\usepackage{tikz}
\usetikzlibrary{shapes, arrows.meta, positioning}
\usepackage{longtable}

\newcommand{\phabs}{\texttt{phabs}}
\newcommand{\diskbb}{\texttt{diskbb}}
\newcommand{\powerlaw}{\texttt{powerlaw}}
\newcommand{\gaussian}{\texttt{gaussian}}

\newcommand{\FeKav}{\ion{Fe}{xxv} K$\alpha$}
\newcommand{\FeKavi}{\ion{Fe}{xxvi} K$\alpha$}
\newcommand{\FeKan}{\ion{Fe} K$\alpha$}

\newcommand{\FeKbv}{\ion{Fe}{xxv} K$\beta$}
\newcommand{\FeKbvi}{\ion{Fe}{xxvi} K$\beta$}
\newcommand{\FeKbn}{\ion{Fe} K$\beta$}

\newcommand{\NiKavii}{\ion{Ni}{xxvii} K$\alpha$}

\newcommand{\FeKgv}{\ion{Fe}{xxv} K$\gamma$}
\newcommand{\FeKgvi}{\ion{Fe}{xxv} K$\gamma$}

\newcommand{\msun}{$M_\odot$}

\newcommand{\nustar}{\textit{NuSTAR}}
\newcommand{\chandra}{\textit{Chandra}}
\newcommand{\xmm}{XMM-\textit{Newton}}

\newcommand\T{\rule{0pt}{2.6ex}}       % Top strut
\newcommand\B{\rule[-1.2ex]{0pt}{0pt}} % Bottom strut

\begin{document} 

   \title{The current state of disk wind observations in BHLMXBs\\ through X-ray absorption lines in the iron band}
\titlerunning{The current state of disk wind observations in BHLMXBs through X-ray absorption lines in the iron band}

   \author{M. Parra\inst{1,2}, P.-O. Petrucci \inst{1}, S. Bianchi \inst{2}, V. E. Gianolli\inst{1,2}, F. Ursini \inst{2}, G. Ponti \inst{3,4}
          }
          \authorrunning{M. Parra et al.} 

   \institute{
   Univ. Grenoble Alpes, CNRS, IPAG, 38000 Grenoble, France\\
   e-mail:\href{mailto:maxime.parra@univ-grenoble-alpes.fr}{maxime.parra@univ-grenoble-alpes.fr}
   \and
   Dipartimento di Matematica e Fisica, Università degli Studi Roma Tre, via della Vasca Navale 84, 00146 Roma, Italy
   \and
   INAF – Osservatorio Astronomico di Brera, Via Bianchi 46, I-23807
Merate (LC), Italy
   \and
    Max Planck Institute fur Extraterrestriche Physik, D-85748 Garching, Germany
}

   \date{}

% \abstract{}{}{}{}{} 
% 5 {} token are mandatory
 
  \abstract
  % context heading (optional)
   {The presence of blueshifted absorption lines in the X-ray spectra of Black Hole Low Mass X-ray Binaries is the telltale of massive outflows called winds. These signatures are found almost exclusively in soft states of high-inclined systems, hinting at equatorial ejections originating from the accretion disk and deeply intertwined with the evolution of the outburst patterns displayed by these systems. In the wake of the launch of the new generation of X-ray spectrometers, studies of wind signatures remain mostly restricted to single sources and outbursts, with some of the recent detections departing from the commonly expected behaviors. We thus give an update to the current state of iron band absorption lines detections, through the analysis of all publicly available \xmm{}-PN and \chandra{}-HETG exposures of known Black Hole Low-Mass X-ray Binary candidates. Our results agree with previous studies, with wind detections exclusively found in dipping, high-inclined sources, and almost exclusively in bright ($L_{X}>0.01L_{Edd}$) soft ($HR<0.8$) states, with blueshift values generally restricted to few 100 km s$^{-1}$. The line parameters indicate similar properties between objects and outbursts of single sources, and despite more than 20 years of data, very few sources have the HID sampling necessary to properly study the evolution of the wind during single outbursts. We provide an online tool with details of the wind signatures and outburst evolution data for all sources in the sample.}

   \maketitle

\section{Introduction} \label{sec:intro}

In X-ray Binaries, the accretion of matter from a main sequence star onto a compact object (either a Neutron Star [NS] or a Black Hole [BH]) produces spectral signatures peaking in the X-ray band. For the sub-population of Low Mass X-ray Binaries (LMXBs), this accretion is sustained via Roche Lobe overflow of the K-M spectral type donor star, and forms an accretion disk around the compact object. The vast majority of BHLMXBs (the focus of this study) are transients \citep[][]{King1996_LMXB_BHNS_proportion,Corral-Santana2016_blackcat}, alternating between long-term phases of quiescence and brief periods of outburst, lasting from a few months to a few years. These events are characterized by a raise of several orders of magnitude in luminosity accross all wavelengths \citep[][]{Fender2004_BHXRB_jet,Remillard2006_BHXRB_properties}, interpreted as the consequence of instabilities due to ionisation of hydrogen in the disk \citep[see e.g.][for a review]{Done2007_BHXRB_accretion}. These outbursts show many remarkable spectral and timing properties, especially in the X-ray and radio bands, the most obvious being a common hysteresis pattern between two distinct spectral states \citep[see e.g.][for a review]{Dunn2010_BHXRB_spectral}.

The beginning of an outburst is marked by a rise of the X-ray luminosity of several orders of magnitude. In this phase, the X-ray spectrum is dominated by a hard ($\Gamma\sim1.5$) power law with an exponential cutoff around $100$kev\citep[][]{Remillard2006_BHXRB_properties,Done2007_BHXRB_accretion}. This so-called \textit{hard} spectral state is associated to non-thermal processes in an extremely hot and optically thin plasma close to the BH (the `corona'). At low energies, the Spectral Energy Distribution (SED) is dominated by a component associated with jets, which extends from the radio to the infrared. These hard states also exhibit strong variability rms, with values of several 10\% in the X-ray band, often accompanied by  type C Quasi-Perdiodic Oscillations (QPOs, see e.g. \citealt{Ingram2019_QPO_review} for a review). When the source reaches high luminosities (up to several percent of $L_{Edd}$\footnote{the Eddington luminosity $L_{Edd}$ is defined as the maximum isotropic emission above which the radiation pressure evens out the gravitational force of the source, stopping the accretion.}), a state transition occurs in the span of a few days, coinciding with the appearance of type A and B QPOs. The X-ray powerlaw index raises to $\Gamma\geq2.5$, and the spectrum transitions to being largely dominated by a bump appearing at around $\sim1-2$kev. This is commonly modeled as a multi-temperature blackbody and interpreted as the thermal emission of an optically thick and geometrically thin accretion disk, extending close to the innermost stable circular orbit (ISCO) of the BH. Concurrently, the radio emission becomes strongly suppressed  (see e.g. \citealt{Fender1999,Corbel2001,Gallo2003,Fender2004, Coriat2009_GX339-4_IRX_correl}), pointing towards a partial or complete quenching of the jet component, and the variability rms is greatly reduced to values of a few percents. After a period of time in this so-called \textit{soft} spectral state, the luminosity of the source decreases by 1-2 orders of magnitude, after which the inverse transition happens, bringing the source back to the hard state, before a final descent to quiescence.

The physical mechanisms behind this outburst cycle are hardly understood. The transitions can be linked to a shift in geometry of the disk, from a truncated accretion flow, a hot corona and a jet in the hard state, to a disk extending to the ISCO and no jets in the soft state\citep[][]{Gallo2003}. However, the geometry of the hard state is hard to distinguish with spectral information alone and thus remains heavily debated \citep[although see ][for recent X-ray polarisation constraints]{Krawczynski2022_CygX-1_pola}, as the expected accretion configurations (corona, accretion disk) have difficulties to reproduce the entire cycle. For instance, the hot plasma required to reproduce the hard state struggles to reach the high luminosities for which the hard to soft transition occurs before collapsing \citep[][]{Yuan2014_BH_accretion_review,Dexter2021_GRMHD_hardstate_sim}. On the other hand, the `cold' accretion disk present in the soft state, such as the standard solution of \citet{Shakura1973_classico}, remains stable well below the accretion rates at which the soft to hard transition should occur.

Coincidentally, the description of the jet in itself is far from complete.
It is now well admitted that a poloidal magnetic field is needed to produce large scale jets (e.g. \citealt{Beckwith2008}), and that they can be powered by two mechanisms. These two processes, namely Blandford \& Znajek \citep[][]{Blandford1977} and Blandford \& Payne \citep[][]{Blandford1982_Blandford-Payne}, extract rotational energy from  the black hole or its accretion disk respectively. However, the relative importance of each in the formation of the global accretion-ejection structure remains a very debated question. Numerical simulations of such structures threaded by a large scale magnetic field around black holes have now become quite standard, with computations on a large number of dynamical time scales (e.g., \citealt{Narayan2003,McKinney2009,Ohsuga2009_MHD_disk_jet_simu,Tchekhovskoy2011,Liska2018,Liska2022}), but including realistic radiative processes remains a difficult task (see e.g. \citealt{Liska2022} for recent results). Direct comparison of these numerical simulations to observational data is thus far from being achieved. On the other hand, stationary, self-similar accretion-ejection solution threaded by large scale magnetic field have been developed since more than 20 years now (e.g. \citealt{Ferreira1993_magnetized_accretion_ejection_1,Ferreira1997_MHW_wind,zanni2007}). While less general than the previously mentioned GRMHD numerical simulations, they have the advantage of being more easily comparable to observations (see, e.g., \citealt{Petrucci2010_JED_XRB,Marcel2018_JED-SAD_III,Marcel2019_JED-SAD_IV,Marcel2020_JED-SAD_V}). Unfortunately, no matter the numerical approach, and even if a few scenarii have been proposed (e.g. \citealt{Meyer2000,Petrucci2008,Begelman2014,Kylafis2015,Cao2016}), none of the current simulations are able to reproduce the hard-to-soft
and soft-to-hard transitions observed during the outbursts and/or how it is related to the jet appearance/disappearance.

Nevertheless, jets are hardly the last piece of the puzzle. Starting  25 years ago \citep[][]{Ueda1998_GROJ1655-40_wind_ASCA,Kotani2000_GRS1915+105_winds_ASCA}, X-ray absorption lines have been detected in a number of LMXBs, mostly with \FeKav{} and \FeKavi{} transitions. These are the signature of a new class of outflows, far from the relativistic speeds of jets but much more massive: \textit{winds} \citep[see][for reviews]{DiazTrigo2016_winds_XRB_review,Ponti2016_winds_XRB_review}. Winds are deeply intertwined with other accretion and ejection processes, their detection being generally mutually exclusive with jet signatures \citep[][]{Neilsen2009_GRS1915+105_wind_jet_connection}. They are also generally observed in the soft states of high inclined BHLMXB, the latter pointing to significant detections mainly along equatorial line of sight \citep[][]{Ponti2012_ubhw}. This, combined with an ability to eject matter at rates potentially comparable -if not higher- to the accretion rate \citep[][]{Ponti2012_ubhw}, makes their understanding essential to fully grasp the accretion-ejection processes in LMXBs.

However, the picture depicted by the observations is becoming increasingly complex. In the last ten years, X-ray absorption lines have been reported also in some hard state observations of BHLMXBs \citep[][]{Shidatsu2013_MAXIJ1305-704_wind_soft_hard_Swift_Suzaku,King2015_GS2023+338_wind_x,Xu2018_SwiftJ1658.2_i_winds,Reynolds2018_GRS1758-258_winds_hard_COSPAR,Wang2020_EXO1846-031_winds_hard_i}, sometimes for potentially low inclined sources \citep[][]{Chiang2012_XTEJ1652-453_i_winds_2sigma,Wang2018_IGRJ17091_winds_hard_i_low,Chakraborty2021_MAXIJ1348-630_i_winds_soft_hard}. Meanwhile, a wealth of P-cygni line profiles are being detected, also in the hard state, but this time in the visible band  \citep[][]{Rahoui2014_GX339-4_wind_visible_soft_hard,Munoz-Darias2016_GS2023+338_wind_hard_visible,Munoz-Darias2018_SAXJ1819.3-2525_winds_hard_optical,Munoz-Darias2019_MAXIJ1820+070_winds_hard_visible,Jimenez-Ibarra2019_SwiftJ1357.2-0933_winds_visible_1,Cuneo2020_GRS1716-249_winds_hard_optical}. These findings imply that the wind is also present in the hard state, but is  preferentially seen at high energies (in the X-rays) in the soft state and at larger wavelengths (in the optical) in the hard state. This new depiction of a state-independent outflow agrees with recent observations of absorption lines in optical and infrared observations in the soft state \citep[][]{Panizo-Espinar2022_MAXIJ1348-630-winds_optical} \cite[see also][]{Sanchez-Sierras2020_MAXIJ1820+070_wind_emission_infrared_soft_hard}, and of simultaneous X-ray and optical absorption lines in the hard state with compatible origin \citep[][]{Munoz-Darias2022_GS2023+338_wind_xray-optical_2015_details}.

In parallel, theoretical and modeling efforts are starting to catching up. Recent studies have showed that the thermal stability of the ionized material in the wind is heavily dependant on the spectrum of the source \citep[see e.g.][]{Chakravorty2013_thermal_stability,Chakravorty2016_JED-SAD_warm_wind_thermal_stability,Bianchi2017_stability_NS,Dyda2017_wind_thermal_stability,Higginbottom2020_thermal_wind_hard_modeling}. This supports the idea that thermal instabilities may play a role in the disappearance of the X-ray absorption lines in the  hard state, independently of the physical state of the wind \citep[][]{Petrucci2021_outburst_wind_stability}. The picture is much less clear during state transitions however and thermal instabilities may not be the only process at work \citep[e.g.][]{Gatuzz2019_4U1630-47_wind_2012-13Chandra}. Furthermore, for a thermal stability analysis to be applied, photoionization equilibrium must be achieved throughout the wind, a constraint that must be carefully checked (see e.g. \citealt{Dyda2017_wind_thermal_stability} and other caveats discussed in \citealt{Petrucci2021_outburst_wind_stability}).

Meanwhile, the physical process powering the wind remains widely debated. The disks of BHLMXBs radiate more in X-rays, unlike  Active Galactic Nuclei (AGNs) in which the thermal emission peaks in the UV \citep[][]{Proga2002_wind_line_driving_issue_XRB}. This rules out line-driving as a driving mechanism in XRBs, since the wind is expected to be strongly ionised by the illuminating X-ray continuum. The two remaining mechanisms, thermal driving, where the central SED heats up the surface of the disk until the material exceeds its escape velocity (e.g., \citealt{Begelman1983_wind_thermal_init_1,Woods1996_wind_thermal_init_2,Done2018_thermal_winds_modeling_H1743_GROJ1655,Tomaru2022_GROJ1655-40_wind_thermal}), and magnetic driving, where the material is lifted by large-scale magnetic fields threading the disk (e.g., \citealt{Konigl1994_MHD_win_AGNs,Fukumura2010_MHD_wind_AGNs,Fukumura2017GROJ1655-40_wind_magnetic,Chakravorty2016_JED-SAD_warm_wind_thermal_stability,Chakravorty2023_MHD_wind_prefictions_Athena_XRISM}), are both viable for X-ray Binaries, and can affect one another (e.g., \citealt{Proga2003_wind_simulation_mhd_thermal,Waters2018_wind_magnetic_thermal_modeling}). However, these two driving mechanisms predict very different absorption line properties. Thermal driving is effective much further away from the BH, and thus results in lower outflow velocities, densities, and variability on longer timescales. On the other hand MHD winds can be produced everywhere on the disk where the magnetisation is large enough (e.g. \citealt{Jacquemin-Ide2019_wind_weak_magnetic_JEDSAD_modeling}). Thus, strong wind signatures with high blueshifts, density and high variability have been traditionally associated to magnetic winds (see e.g. \citealt{Miller2006_GROJ1655-40_winds,Miller2015_HETG_analysis_multi,Trueba2019_4U1630-47_wind_2012-13Chandra}).
 
Nevertheless, numerical simulations of thermal winds  \citep[][]{Higginbottom2015_wind_thermal_XRB_new}, and, more recently, of hybrid thermal-radiative winds \citep[e.g.,][]{Done2018_thermal_winds_modeling_H1743_GROJ1655,Higginbottom2018_GROJ1655-40_wind_thermal,Higginbottom2020_thermal_wind_hard_modeling}, are now able to reproduce the observed absorption features  with a high degree of fidelity \citep[e.g.,][]{Tomaru2020_H1743-322_wind_model_thermalradiative_2,Tomaru2022_GROJ1655-40_wind_thermal}. In parallel, spectrum predictions for XRBs from MHD models, which have only been achieved recently \citep[e.g.,][]{Chakravorty2016_JED-SAD_warm_wind_thermal_stability,Fukumura2017GROJ1655-40_wind_magnetic,Chakravorty2023_MHD_wind_prefictions_Athena_XRISM}, can successfully recreate absorption line features in standard observations \citep[][]{Fukumura2021_magneticwind_BHLMXB}. Although these comparisons are only beginning, it is becoming apparent that the quality of current datasets might not allow to distinguish directly between these new solutions of MHD and thermal-radiative winds, with their wide range of possible signatures, as even for the highest quality observations, both processes now only differ on very fine degrees \citep[][]{Tomaru2022_GROJ1655-40_wind_thermal}. Thankfully, the new generation of X-ray telescopes should soon definitely settle the debate \citep[][]{Chakravorty2023_MHD_wind_prefictions_Athena_XRISM,Gandhi2022}.

However, many answers can still be found in the existing observations and more constrains can be put through with comparisons to much larger datasets. Indeed, both observational studies and modeling efforts often focus on either single observations or select samples, with very precise analysis or modeling of the existing features, but mostly in observations with the most prominent lines. Moreover, no detailed study of large samples of sources, with exhaustive, multi-instrument data coverage have been performed since the seminal work of \citep[][]{Ponti2012_ubhw}, despite a huge increase in number of observation, sources, and of our understanding of the winds. 

In this work, we analyse all  \xmm{} and \chandra{} X-ray observations of current BHLMXBs candidates public as of October 2022, in order to have a global view of the wind signatures in a large sample of objects and observations. After explaining our sample selection and data reduction in   Sect.~\ref{sec:data_reduction}, we detail the process of line detection in   Sect.~\ref{sec:linedet}. Following this, we present our results in   Sect.~\ref{sec:results_glob}, and discuss some physical implications in   Sect.~\ref{sec:discussion}, before concluding. More information regarding individual sources will be presented in a separate work. Finally, beyond providing the main detection and non-detection of lines in appendix \ref{sec:table_exposures}, we also provide an online tool \footnote{\href{https://visual-line.streamlit.app/}{https://visual-line.streamlit.app/}} for both interactive visualisation of our results, and to provide easy access to all spectral and line parameters obtained in the study  (see appendix \ref{sec:visualisation}).

\section{Observations \& Data Reduction}\label{sec:data_reduction}

\subsection{Sample and Data Selection}\label{sub:sample}

In order to maximize the number of BHLMXBs candidates, we draw our sample from both the BlackCAT \citep[][]{Corral-Santana2016_blackcat} and WATCHDOG \citep[][]{Tetarenko2016_watchdog} BH catalogs. The BlackCAT catalog has been continuously updated since its release but is voluntarily restricted to transient sources, which is why some archetypal binaries are missing from it and present in WATCHDOG. On the other side, the WATCHDOG catalog also includes high-mass X-ray binaries (HMXBs), and after its publication in 2016, some of its sources have been identified as NSs. Thus, our parent sample is composed of 79 sources: 67 from BlackCAT (in which we only exclude Cen X-2 due to a weak position determination and possible mismatch with GS 1354-64 according to \citealt{Kitamoto1990_CenX-2_mismatch}), and 12 from WATCHDOG (as 11 of the 23 sources not overlapping with BlackCAT are either HMXBs or NSs). 

In this work, we further restrict the analysis to sources with observations from the two X-ray instruments with the highest sensitivity and energy resolution in the iron K band, namely \xmm{}'s EPIC-PN and \chandra{} HETG. After selecting spectra with sufficiently high statistics to apply the line detection process (see Sect.~\ref{sec:linedet}), we are left with a final sample of 42 sources. Details about their physical properties, previous detections of iron K wind signatures in the literature, and number of exposures in our sample, are given in Tab.~\ref{table:sources}.

We draw the source physical properties, namely mass, distance and inclination, primarily from the more up-to-date references of BlackCAT, and use WATCHDOG otherwise, except for the most recent updates we found in the literature. We assume a distance of 8 kpc if unknown otherwise. As for the masses, we only use estimates resulting from dynamical measurements, and otherwise consider a fixed BH mass of 8 \msun{}. We stress than only 11 sources in our final sample have been confirmed as BHs through dynamical measurements (and are noted as such in Tab.~\ref{table:sources}). We refer to the two referred catalogs for the arguments in favor or against BHs in the other binaries. Among these, we highlight that IGR J17451-3022, whose origin remains very debated and exhibits absorption lines \citep[][]{Bozzo2016_IGRJ17451-3022_XMM_wind_2014}, is still included in BlackCAT, so we keep it in our sample. 

\begin{table*}[p]
\scriptsize{
\caption[blabla]{Sources included in our final sample, relevant physical parameters, and number of spectra of sufficient quality for our analysis.\label{table:sources}}
\fontsize{8}{8}
\begin{center}
\begin{tabular}{c || c c c || c || c c }
\hline
\hline
       \multirow{2}{*}{Name}
     & \multirow{2}{*}{mass (\msun)}
     & \multirow{2}{*}{distance (kpc)}
     & \multirow{2}{*}{inclination (°)}
     & absorption lines 
     & \multicolumn{2}{c}{exposures in the sample}

     \T \B \\
     
     &
     & 
     &
     & reported in the iron band
     & EPIC PN
     & HETG
     
     \T \B \\
\hline
\hline

1E 1740.7-2942 
& 8
& 8
& $>50^R$\labelcref{ref_source:1E1740.7_i} 
& X
& 6 
& 1 \T \B \\

4U 1543-475$^D$\labelcref{ref_source:4U1543-475_D} 
&$8.4\pm1$\labelcref{ref_source:4U1543-475_i_mass}
& $7.5\pm0.5$ \labelcref{ref_source:4U1543-47_d}
& \textbf{dips}\labelcref{ref_source:4U1543-47_dips},    $20.7\pm1.5^D$\labelcref{ref_source:4U1543-475_i_mass} 
& X
& 0 
& 1 \T \B \\

4U 1630-47 
& 8
& $8.1\pm3.4$\labelcref{ref_source:4U1630-47_d}
& \textbf{dips}\labelcref{ref_source:4U1630-47_i}, $[60-75]^D$\labelcref{ref_source:4U1630-47_i} 
& \checkmark
& 8 
& 12 \T \B \\

4U 1957+115 
& 8
& 8
& $\sim 13^D$\labelcref{ref_source:4U1957+115_i_2}/$77.6(+1.5-2.2)^R$\labelcref{ref_source:4U1957+115_i_1} 
& X 
& 2 
& 4 \T \B \\

AT 2019wey 
& 8
& 8
& $<30^R$\labelcref{ref_source:AT2019wey_i} 
& X
& 0 
& 1 \T \B \\

EXO 1846-031 
& 8
& $\sim7$\labelcref{ref_source:EXO1846-031_d}
& / $\sim73^R$\labelcref{ref_source:EXO1846-031_i_1}/$\sim40^R$\labelcref{ref_source:EXO1846-031_winds_hard_i} 
& \checkmark
& 2 
& 6 \T \B \\

 GRO J1655-40$^D$\labelcref{ref_source:GROJ1655-40_D} 
& $5.4\pm0.3$\labelcref{ref_source:GROJ1655-40_i_mass}
& $3.2\pm0.2$\labelcref{ref_source:GROJ1655-40_d}
& \textbf{dips}\labelcref{ref_source:GROJ1655-40_dips} $69\pm2^D$\labelcref{ref_source:GROJ1655-40_i_mass} 
& \checkmark
& 6 
& 2 \T \B \\

GRS 1716-249 
& 8
& $2.4\pm0.4$\labelcref{ref_source:GRS1716-249_d}
& $\sim 40-60^R$\labelcref{ref_source:GRS1716-249_i} 
& X
& 0 
& 1 \T \B \\

GRS 1739-278 
& 8
& $7.3\pm1.3$\labelcref{ref_source:GRS1739-278_d}
& $\sim33^R$\labelcref{ref_source:GRS1739-278_i}
& X
& 0 
& 1 \T \B \\

GRS 1758-258 
& 8
& 8
& / 
& \checkmark
& 3 
& 1 \T \B \\

GRS 1915+105$^D$\labelcref{ref_source:GRS1915+105_D_i_mass_d} 
&$12.4_{-1.8}^{+2}$\labelcref{ref_source:GRS1915+105_D_i_mass_d} 
&$8.6_{-1.6}^{+2}$\labelcref{ref_source:GRS1915+105_D_i_mass_d} 
& \textbf{dips}\labelcref{ref_source:GRS1915+105_IGRJ17091_dips}$60\pm5^J$\labelcref{ref_source:GRS1915+105_D_i_mass_d} 
& \checkmark
& 17 
& 22 \T \B \\

GS 1354-64$^D$\labelcref{ref_source:GS1354-64_D_d} 
& 8
& $\sim25$\labelcref{ref_source:GS1354-64_D_d}
& $<79^D$\labelcref{ref_source:GS1354-64_D_d}/ $\sim 70^R$\labelcref{ref_source:GS1354-64_i} 
& X
& 2 
& 0 \T \B \\

GX 339-4$^D$\labelcref{ref_source:GX339-4_D_mass} 
& $5.9\pm3.6$\labelcref{ref_source:GX339-4_D_mass} 
& 8
& $[37-78]^{D}$\labelcref{ref_source:GX339-4_D_mass} 
& X
& 21 
& 4 \T \B \\

H 1743-322 
& 8
& $8.5\pm0.8$\labelcref{ref_source:H1743-322_i_d}
& \textbf{dips}\labelcref{ref_source:H1743-322_winds_dips}, $75\pm3^J$\labelcref{ref_source:H1743-322_i_d} 
& \checkmark 
& 8 
& 9 \T \B \\

IGR J17091-3624 
& 8
& 8
& \textbf{dips}\labelcref{ref_source:GRS1915+105_IGRJ17091_dips}$\sim70^H$\labelcref{ref_source:IGRJ17091_i_high_1}\labelcref{ref_source:IGRJ17091_i_high_2}/$\sim30-40^R$\labelcref{ref_source:IGRJ17091_i_low}\labelcref{ref_source:IGRJ17091_winds_hard_i_low} 
& \checkmark 
& 6 
& 9 \T \B \\

IGR J17098-3628 
& 8
& $\sim10.5$\labelcref{ref_source:IGRJ17098-3628_d}
& / 
& X
& 2 
& 0 \T \B \\

IGR J17285-2922 
& 8
& 8
& / 
& X
& 1 
& 0 \T \B \\

IGR J17451-3022 
& 8
& 8
& \textbf{dips}\labelcref{ref_source:IGRJ17451-3022_wind_dips_Suzaku_2014}$>70^D$\labelcref{ref_source:IGRJ17451-3022_wind_dips_Suzaku_2014} 
& \checkmark
& 1 
& 0 \T \B \\

IGR J17497-2821 
& 8
& 8
& / 
& X
& 1 
& 1 \T \B \\

MAXI J0637-430 
& 8
& 8
& $64\pm6^{R}$\labelcref{ref_source:MAXIJ0637-430_i} 
& X
& 1
& 0 \T \B \\

MAXI J1305-704$^D$\labelcref{ref_source:MAXIJ1305-704_D_i_mass_d} 
&$8.9_{-1.}^{+1.6}$\labelcref{ref_source:MAXIJ1305-704_D_i_mass_d} 
&$7.5_{-1.4}^{+1.8}$\labelcref{ref_source:MAXIJ1305-704_D_i_mass_d}
& \textbf{dips}\labelcref{ref_source:MAXIJ1305-704_wind_soft_hard_soft+softX_dip},${72^{+5}_{-8}}^D$\labelcref{ref_source:MAXIJ1305-704_D_i_mass_d} 
& \checkmark
& 0 
& 1 \T \B \\

MAXI J1348-630 
& 8
& $3.4_{-0.4}^{+0.4}$ \labelcref{ref_source:MAXIJ1348-630_d}
& $28\pm3^J$\labelcref{ref_source:MAXIJ1348-630_i_jet}/$65\pm7$ \labelcref{ref_source:MAXIJ1348-630_mass_i_scaling}/ $[30-45]^R$\labelcref{ref_source:MAXIJ1348-630_i_winds_soft_hard}
& \checkmark
& 2 & 3 \T \B \\

MAXI J1535-571 
& 8
& $4.1_{-0.5}^{+0.6}$\labelcref{ref_source:MAXIJ1535-571_d}
& $\leq 45^J$\labelcref{ref_source:MAXIJ1535-571_i_jet}/ $70-74^R$\labelcref{ref_source:MAXIJ1535-571_i_refl} 
& X
& 6 
& 5 \T \B \\

MAXI J1659-152 
& 8
& $8.6\pm3.7$\labelcref{ref_source:MAXIJ1659-152_i_d}
& \textbf{dips}\labelcref{ref_source:MAXIJ1659-152_i_d},$70\pm10^D$\labelcref{ref_source:MAXIJ1659-152_i_d} 
& X
& 2 
& 0 \T \B \\

MAXI J1803-298 
& 8
& 8
& \textbf{dips}\labelcref{ref_source:MAXIJ1803-298_i},$>70^D$\labelcref{ref_source:MAXIJ1803-298_i} 
& \checkmark
& 0 
& 4 \T \B \\

MAXI J1820+070 $^D$\labelcref{ref_source:MAXIJ1820+070_D} 
& $6.9\pm1.2$\labelcref{ref_source:MAXIJ1820+070_i_mass}
& $2.96\pm0.33$\labelcref{ref_source:MAXIJ1820+070_d}
& \textbf{dips}\labelcref{ref_source:MAXIJ1820+070_dips},$[67-81]^D$\labelcref{ref_source:MAXIJ1820+070_i_mass} 
& X
& 14
& 0 \T \B \\

SAX J1711.6-3808 
& 8
& 8
& / 
& X 
& 1 
& 0 \T \B \\

Swift J1357.2-0933
& 8
& 8
& \textbf{dips}\labelcref{ref_source:Swiftj1357.2-0933_dips},$\geq 80^D$\labelcref{ref_source:SwiftJ1357.2-0933_i} 
& X
& 1 
& 0 \T \B \\

Swift J1658.2-4242 
& 8
& 8
& \textbf{dips}\labelcref{ref_source:SwiftJ1658.2_4242_dips},$64(+2-3)^R$\labelcref{ref_source:SwiftJ1658.2_i_winds} 
& \checkmark
& 8 
& 1 \T \B \\

Swift J174510.8-262411 
& 8
& 8
& / 
& X
& 1 
& 0 \T \B \\

Swift J1753.5-0127 
& 8
& $6\pm2$\labelcref{ref_source:SwiftJ1753.5-0127_d}
& $55(+2-7)^R$\labelcref{ref_source:SwiftJ1753.5-0127_i} 
& X
& 6 
& 1 \T \B \\

Swift J1910.2-0546 
& 8
& 8
& / 
& X
& 1  
& 1 \T \B \\

V404 Cyg$^D$\labelcref{ref_source:GS2023+338_D_i_mass} 
&$9_{-0.6}^{+0.2}$\labelcref{ref_source:GS2023+338_D_i_mass} 
& $2.4\pm0.2$\labelcref{ref_source:GS2023+338_d}
& $67(+3-1)^D$\labelcref{ref_source:GS2023+338_D_i_mass}
& \checkmark 
& 0 
& 2 \T \B \\

V4641 Sgr$^D$\labelcref{ref_source:SAXJ1819.3_2525_D} 
&$6.4\pm 0.6$\labelcref{ref_source:SAXJ1819.3-2525_i_mass_d}
&$6.2\pm0.7$\labelcref{ref_source:SAXJ1819.3-2525_i_mass_d}
& $72\pm4^D$\labelcref{ref_source:SAXJ1819.3-2525_i_mass_d} 
& X
& 0 
& 2 \T \B \\

XTE J1550-564$^D$\labelcref{ref_source:XTEJ1550-564_D_i_m_d} 
&$11.7\pm3.9$\labelcref{ref_source:XTEJ1550-564_D_i_m_d} 
&$4.4_{-0.4}^{+0.6}$\labelcref{ref_source:XTEJ1550-564_D_i_m_d}
& $75\pm4^D$\labelcref{ref_source:XTEJ1550-564_D_i_m_d}/$\sim40^R$\labelcref{ref_source:XTEJ1550-564_i_2} 
& X
& 0 
& 2 \T \B \\

XTE J1650-500$^D$\labelcref{ref_source:XTEJ1650-500_D} 
& 8
& $2.6\pm0.7$\labelcref{ref_source:XTEJ1650-500_d}
& $\geq47^D$\labelcref{ref_source:XTEJ1650-500_D} 
& X
& 1 
& 2 \T \B \\

XTE J1652-453 
& 8
& 8
& $\leq32^R$\labelcref{ref_source:XTEJ1652-453_i_winds_sigma} 
& \checkmark 
& 1 
& 0 \T \B \\

XTE J1720-318 
& 8
& $6.5\pm3.5$\labelcref{ref_source:XTEJ1720-318_d}
& /
& X
& 1 
& 0 \T \B \\

XTE J1752-223 
& 8
& $6\pm2$\labelcref{ref_source:XTEJ1752-223_d}
& $<49^J$\labelcref{ref_source:XTEJ1752-223_i}/$35\pm4^R$\labelcref{ref_source:XTEJ1752-223_i_2} 
& X
& 2 
& 2 \T \B \\

XTE J1817-330 
& 8
& $5.5\pm4.5$\labelcref{ref_source:XTEJ1817-330_d}
& \textbf{dips}\labelcref{ref_source:XTEJ1817-330_dips} 
& X
& 1
& 4 \T \B \\

XTE J1856+053 
& 8
& 8
& / 
& X
& 1 
& 0 \T \B \\

XTE J1901+014 
& 8
& 8
& / 
& X
& 1 
& 0 \T \B \\

\end{tabular}
\end{center}
\footnotesize
Notes: The letter $D$ in the object name column identifies dynamically confirmed BHs. A fiducial mass of 8 \msun{} and distance of 8 kpc are used when not reliably known, including when dynamical constraints are only lower limits, according to the properties of the bulk of the Galactic BHLMXB population (see e.g. \citealt{Corral-Santana2016_blackcat}). For inclination measurements, we highlight dippers, and letters $D$, $J$, $H$, $R$ refer respectively to dynamical inclination measurements (dips/eclipses/modulations), jets, heartbeats, and reflection fits.
Details and references for line detection reports are provided in Tab.~\ref{table:sources_det_states}.References:\\
\scriptsize
\begin{enumerate*}[label=\arabic{enumi}]
    
    %1E 1740.7-2942
    \item \citep[][]{Stecchini2020_1E1740.7_i}\label{ref_source:1E1740.7_i}
    
    %4U 1543-475
    \item \citep[][]{Orosz1998_4U1543-475_D}\label{ref_source:4U1543-475_D}
    \item \citep[][]{Orosz2003_4U1543-475_i_mass}\label{ref_source:4U1543-475_i_mass}
    \item \citep[][]{Jonker2004_4U1543-47_d}\label{ref_source:4U1543-47_d}
    \item \citep[][]{Park2004_4U1543-47_dips}\label{ref_source:4U1543-47_dips}
    
    %4U 1630-47
    \item \citep[][]{Kalemci2018_4U1630-47_d}\label{ref_source:4U1630-47_d}
    \item \citep[][]{Tomsick1998_4U1630-47_i}\label{ref_source:4U1630-47_i}
    
    %4U 1957+115
    \item \citep[][]{Gomez2015_4U1957+115_i_2}\label{ref_source:4U1957+115_i_2}
    \item \citep[][]{Maitra2013_4U1957+115_i_1}\label{ref_source:4U1957+115_i_1}
    
    %AT 2019wey
    \item \citep[][]{Yao2021_AT2019wey_i}\label{ref_source:AT2019wey_i}
    
    %EXO 1846-031
    \item \citep[][]{Parmar1993_EXO1846-031_d}\label{ref_source:EXO1846-031_d}
    \item \citep[][]{Draghis2020_EXO1846-031_i_1}\label{ref_source:EXO1846-031_i_1}
    \item \citep[][]{Wang2020_EXO1846-031_winds_hard_i}\label{ref_source:EXO1846-031_winds_hard_i}
    %GRO J1655-40
    \item \citep[][]{VanDerHooft1998_GROJ1655-40_D}\label{ref_source:GROJ1655-40_D}
    \item \citep[][]{Beer2002_GROJ1655-40_i_mass}\label{ref_source:GROJ1655-40_i_mass}
    \item \citep[][]{Hjellming1995_GROJ1655-40_d}\label{ref_source:GROJ1655-40_d}
    \item \citep[][]{Kuulkers1998_GROJ1655-40_dips}\label{ref_source:GROJ1655-40_dips}
    
    %GRS 1716-249
    \item \citep[][]{DellaValle1994_GRS1716-249_d}\label{ref_source:GRS1716-249_d}
    \item \citep[][]{Bharali2019_GRS1716-249_i}\label{ref_source:GRS1716-249_i}
   
    %GRS 1739-278
    \item \citep[][]{Greiner1996_GRS1739-278_d}\label{ref_source:GRS1739-278_d}
    \item \citep[][]{Miller2015_GRS1739-278_i}\label{ref_source:GRS1739-278_i}
    
    %GRS 1758-258

    %GRS 1915+105
    \item \citep[][]{Reid2014_GRS1915+105_D_i_mass_d}\label{ref_source:GRS1915+105_D_i_mass_d}
    %dips
    \item \citep[][]{Pahari2013_GRS1915+105_IGRJ17091-3624_dips_study}\label{ref_source:GRS1915+105_IGRJ17091_dips}

    %GS 1354-64
    \item \citep[][]{Casares2009_GS1354-64_D_d}\label{ref_source:GS1354-64_D_d}
    \item \citep[][]{Pahari2017_GS1354-64_i}\label{ref_source:GS1354-64_i}
    
    %GX339-4
    \item \citep[][]{Heida2017_GX339-4_D_mass}\label{ref_source:GX339-4_D_mass}
    
    %H1743-322
    \item \citep[][]{Steiner2012_H1743-322_i_d}\label{ref_source:H1743-322_i_d}
    \item \citep[][]{Miller2006_H1743-322_winds}\label{ref_source:H1743-322_winds_dips}

    %IGR J17091-3624
    %high inclination estimates due to heartbeats
    \item \citep[][]{Capitanio2012_IGRJ17091_i_high_1}\label{ref_source:IGRJ17091_i_high_1}
    \item \citep[][]{Rao2012_IGRJ17091_i_high_2}\label{ref_source:IGRJ17091_i_high_2}
    %low inclination estimate with reflection
    \item \citep[][]{Xu2017_IGRJ17091_i_low}\label{ref_source:IGRJ17091_i_low}
    \item \citep[][]{Wang2018_IGRJ17091_winds_hard_i_low}\label{ref_source:IGRJ17091_winds_hard_i_low}
    
    %IGR J17098-3628
    \item \citep[][]{Grebenev2006_IGRJ17098-3628_d}\label{ref_source:IGRJ17098-3628_d}

    %IGR J17285-2922
    
    %IGR J17451-3022
    \item \citep[][]{Jaisawal2015_IGRJ17451-3022_wind_dips_Suzaku_2014}\label{ref_source:IGRJ17451-3022_wind_dips_Suzaku_2014}

    %IGR J17497-2821

    %MAXI J0637-430
    \item \citep[][]{Lazar2021_MAXIJ0637_430_i}\label{ref_source:MAXIJ0637-430_i}

    %MAXI J1305-704
    \item \citep[][]{Sanchez2021_MAXIJ1305-704_D_i_mass_d}\label{ref_source:MAXIJ1305-704_D_i_mass_d}
    \item \citep[][]{Shidatsu2013_MAXIJ1305-704_wind_soft_hard_Swift_Suzaku}\label{ref_source:MAXIJ1305-704_wind_soft_hard_soft+softX_dip}
    
    %MAXIJ1348-630
    \item \citep[][]{Lamer2021_MAXIJ1348-630_distance}\label{ref_source:MAXIJ1348-630_d}
    \item \citep[][]{Carotenuto2022_MAXIJ1348-630_i_jet}\label{ref_source:MAXIJ1348-630_i_jet}
    \item \citep[][]{Titarchuk2022_MAXIJ1348-630_mass_i_scaling}\label{ref_source:MAXIJ1348-630_mass_i_scaling}
    \item \citep[][]{Chakraborty2021_MAXIJ1348-630_i_winds_soft_hard}\label{ref_source:MAXIJ1348-630_i_winds_soft_hard}
    
    %MAXI J1535-571
    \item \citep[][]{Chauhan2019_MAXIJ1535-571_d}\label{ref_source:MAXIJ1535-571_d}
    \item \citep[][]{Russell2019_MAXIJ1535-571_i_jet}\label{ref_source:MAXIJ1535-571_i_jet}
    \item \citep[][]{Dong2022_MAXIJ1535-571_i_refl}\label{ref_source:MAXIJ1535-571_i_refl}

    %MAXIJ1659-152
    \item \citep[][]{Kuulkers2013_MAXIJ1659-152_i_d}\label{ref_source:MAXIJ1659-152_i_d}
    
    %MAXI J1803-298
    \item \citep[][]{Homan2021_MAXIJ1803-298_dips_NICER_hintwind_2021}\label{ref_source:MAXIJ1803-298_i}
    
    %MAXIJ1820+070
    \item \citep[][]{Torres2019_MAXIJ1820+070_D}\label{ref_source:MAXIJ1820+070_D}
    \item \citep[][]{Torres2020_MAXIJ1820+070_i_mass}\label{ref_source:MAXIJ1820+070_i_mass}
    \item \citep[][]{Atri2020_MAXIJ1820+070_d}\label{ref_source:MAXIJ1820+070_d}
    \item \citep[][]{Homan2018_MAXIJ1820-070_dips}\label{ref_source:MAXIJ1820+070_dips}

    %SAX J1711.6-3808
    
    %Swift J1357.2-0933
    \item \citep[][]{Corral-Santana2013_SwiftJ1357.2-0933_dips}\label{ref_source:Swiftj1357.2-0933_dips}
    \item \citep[][]{Sanchez2015_SwiftJ1357.2-0933_i}\label{ref_source:SwiftJ1357.2-0933_i}

    %Swift J1658.2-4242
    \item \citep[][]{Xu2018_SwiftJ1658.2-4242_dips}\label{ref_source:SwiftJ1658.2_4242_dips}
    \item \citep[][]{Xu2018_SwiftJ1658.2_i_winds}\label{ref_source:SwiftJ1658.2_i_winds}

    %Swift J174510.8-262411
    
    %Swift J1753.5-0127
    \item \citep[][]{CadolleBel2007_SwiftJ1753.5-0127_d}\label{ref_source:SwiftJ1753.5-0127_d}
    \item \citep[][]{Reis2009_SwiftJ1753.5-0127_i}\label{ref_source:SwiftJ1753.5-0127_i}

    %Swift J1910.2-0546

    %V404 Cyg (GS 2023+338)
    \item \citep[][]{Khargharia2010_GS2023+338_D_i_mass}\label{ref_source:GS2023+338_D_i_mass}
    \item \citep[][]{Miller-Jones2009_GS2023+338_d}\label{ref_source:GS2023+338_d}
    %V 4641 Sgr (SAX J1819.3-2525)
    \item \citep[][]{Orosz2001_SAXJ1819.3_2525_D}\label{ref_source:SAXJ1819.3_2525_D}
    \item \citep[][]{Macdonald2014_SAXJ1819.3-2525_i_mass_d}\label{ref_source:SAXJ1819.3-2525_i_mass_d}

    %XTE J1550-564
    \item \citep[][]{Orosz2011_XTEJ1550-564_D_i_m_d}\label{ref_source:XTEJ1550-564_D_i_m_d}
    \item \citep[][]{Connors2019_XTEJ1550-564_i_2}\label{ref_source:XTEJ1550-564_i_2}

    %XTE J1650-500
    \item \citep[][]{Orosz2004_XTEJ1650-500_D}\label{ref_source:XTEJ1650-500_D}
    \item \citep[][]{Homan2006_XTEJ1650-500_d}\label{ref_source:XTEJ1650-500_d}
    
    %XTE J1652-453
    \item \citep[][]{Chiang2012_XTEJ1652-453_i_winds_2sigma}\label{ref_source:XTEJ1652-453_i_winds_sigma}

    %XTE J1720-318
    \item \citep[][]{Chaty2006_XTEJ1720-318_d}\label{ref_source:XTEJ1720-318_d}
    
    %XTE J1752-223
    \item \citep[][]{Ratti2012_XTEJ1752-223_d}\label{ref_source:XTEJ1752-223_d}
    \item \citep[][]{Miller-Jones2011_XTEJ1752-223_i}\label{ref_source:XTEJ1752-223_i}
    \item \citep[][]{Garcia2018_XTEJ1752-223_i_2}\label{ref_source:XTEJ1752-223_i_2}

    %XTE J1817-330
    \item \citep[][]{Sala2007_XTEJ1817-330_d}\label{ref_source:XTEJ1817-330_d}
    \item \citep[][]{Sriram2012_XTEJ1817-330_dips}\label{ref_source:XTEJ1817-330_dips}
    
    %XTE J1856+053
    
    %XTE J1901+014

\end{enumerate*}}

\end{table*}
\subsection{\xmm{}}\label{sub:xmm}

Data reduction for \xmm{} observations was performed with the Science Analysis System (SAS\footnote{\href{https://www.cosmos.esa.int/web/XMM-Newton/sas}{https://www.cosmos.esa.int/web/XMM-Newton/sas}}) version 19.1.0, following the standard analysis threads\footnote{see \href{https://www.cosmos.esa.int/web/XMM-Newton/sas-threads}{https://www.cosmos.esa.int/web/XMM-Newton/sas-threads}}. Observation Data Files (ODFs) were reduced with the $epproc$ task.

In order to optimize the absorption line detection, we maximize the signal to noise ratio (SNR) of the final spectra through an automated procedure, whose main steps are detailed below:

\begin{enumerate}
\item Extract an image centered on the sky coordinates of the source from the event files in the 4-10kev band
\item Compute initial source and background regions:\\
    $\bullet$ In imaging, we start with the source sky coordinates and fit with a point spread function (PSF) to optimize the source localisation. The background region is computed from the largest circular region not intersecting with the brightest 2$\sigma$ of the source PSF in the source/neighboring CCD, with an area between 1 and 2 times the source region's. For exposures with background rates above 100 times the value of standard blank fields \footnote{obtained from \href{https://www.cosmos.esa.int/web/XMM-Newton/bs-countrate}{https://www.cosmos.esa.int/web/XMM-Newton/bs-countrate}}, the background contribution is disregarded. Although recent work has shown that dust scattering halos can significantly alter the broad band SEDs of XRBs (see e.g. \citealt{Jin2017_DSC,Jin2019_DSC}), this effect is smaller at high energies, and is not expected to affect the detection of narrow absorption lines. We thus do not apply such corrections in this work for simplicity. \\
    $\bullet$ In Timing/Burst mode, we center the source at the brightest column, and we do not use the background to avoid source contamination
\item Compute the radius of the circular region for the source and the filtering of high background periods in a self-consistent way to reach the highest SNR. For this, we select increasingly large circular (rectangular in timing) regions, each of which is independently filtered for good time intervals (GTIs) in order to maximize their individual SNRs against the background computed previously (following the method of \citealt{Piconcelli2004_SNR_opti}).
    
\item If necessary, excise increasingly larger circular regions from the center of the source region until the pile-up value (estimated with $epatplot$) falls below 5\%\footnote{ Up to 7\% is accepted for 4 exposures, highlighted in Table. \ref{tab:obs_details}}, a level at which no significant effect on the line detection process is expected. We note that the majority of the spectra remain actually below 1\%. The first 2 steps are then repeated, this time starting with the filtered GTIs and excised image, in order to refine the region and filtering of the events.

\item Extract the source and background spectra from the final region files and GTIs, and generate response matrices and ancillary response files with the standard SAS tasks $arfgen$ and $rmfgen$

\item Group the source spectra with the \cite{Kaastra2016_binning_opt} optimized binning.
\end{enumerate}

\subsection{\chandra{}}

The reduced, science-ready spectra of the first order of all grating observations are publicly available on the \chandra{} Transmission Grating Data Archive and Catalog (TGCat,\citealt{Huenemoerder2011_TGCat}), and observations of BHLMXBs have been recently updated according to recent improvements in data reduction. We only consider the first order spectra, and regroup the products according to the \cite{Kaastra2016_binning_opt} optimized binning. Background spectra are not computed as they are often contaminated by the PSF wings \footnote{see \href{https://cxc.cfa.harvard.edu/ciao/threads/xspec_phabackground/}{https://cxc.cfa.harvard.edu/ciao/threads/xspec\_phabackground/}}.

\section{Spectral Analysis}\label{sec:linedet}

In order to filter out spectra with unsufficient SNR necessary to detect absorption lines in the iron band, we apply a predefined count threshold of 5000 counts in the $4-10$kev band to both XMM and \chandra{} exposures. For XMM, simulations of template spectra from soft state SEDs ofGRO J1655-40 in the soft state show than observations fainter than the chosen threshold cannot detect \FeKavi{} upper limits below 75 eVs, which coincides with the high-end tail of the EW distribution in our sample and reports in the literature (see  Sect.~\ref{sub:param_distrib}). While such simulations are less straightforward for \chandra{} HETG, manual inspection of excluded spectra confirms than their SNR is always insufficient to detect lines with EWs below 100 eV at 7kev. 242 exposures remain after this final cut: 137 EPIC-pn spectra and 105 HETG spectra.\\

The line detection process can be split in four main steps, which are detailed in the subsections below:
\begin{enumerate}
\item Fit the continuum with a broad band model (Sect.~\ref{sub:broadband}) 
\item Perform a blind search for line features in the high energy (6-10kev) band (Sect.~\ref{sub:blind_search})
\item Fit the line features in this energy range with the strongest absorption and emission lines expected in this band (Sect.~\ref{sub:line_fit})
\item Check the absence of remaining line features with a second blind search from the best fit model including the lines
\item Assess the true significance of the absorption lines via Monte-Carlo (MC) simulations (Sect.~\ref{sub:sign_fakes})
\end{enumerate}

In the following, we use Xspec version 12.12.0 \citep[][]{Arnaud1996_xspec}, via Pyxspec version 2.0.5, along with wilm abundances \citep[][]{Wilms2000_xspec_abundances} and the Cash statistic \citep[][]{Cash1979_Cstat}. Uncertainties for all the reported parameters are estimated drawing a Monte-Carlo chain from the final fit, using the internal Xspec \texttt{Chain} commands. Due to the great number of spectra to be analyzed, and the use of multiple runs during the line detection process, for a given number of free parameters $n_{free}$, we only use $2\cdot n_{free}$ parameters, for $4000\cdot n_{free}$ steps, and discarding the first $2000\cdot n_{free}$ steps of each chain. 
Unless specified otherwise, all uncertainties are quoted at a 90\% confidence level.

\subsection{Broad-band modeling}\label{sub:broadband}

We use a simple fitting procedure, in which a list of components is added recursively to converge to the best fit. Adding or choosing a component over its peers is deemed statistically significant through F-tests, with a threshold fixed at 99\% confidence level. For the broad band modelization of the continuum, three components can be combined: a \powerlaw{}, a \diskbb{}, and an absorption component \phabs{}, which is applied to all of the additive components together. As the goal is to get a precise (although phenomenological) estimate of the continuum, we initially limit the contamination due to iron band features by ignoring the 6.7-7.1 and 7.8-8.3kev bands in this step only.

In order to limit the effect of low energy spurious features, we restrict the broad band fit to 2-10kev for XMM-pn and 1.5-10kev for HETG\footnote{In HETG exposures in timed mode, there can be issues with event resolution at high energy due to an overlap between the default HEG and MEG spatial masks. Thus, whenever necessary, we restrict the upper limit of all energy bands to 7.5keV, so as to minimize the effect on the continuum while keeping the ability to at least analyse lines of the K$\alpha$ complex.}. While with this choice of energy band the N$_H$ value may not be estimated perfectly, notably for sources with low absorption, it still allows for a good measure of the intrinsic unabsorbed 3-10 keV luminosity and 6-10/3-6 keV Hardness Ratio (HR). Following this, we then fix the neutral absorption column density, in order to perform the blind search in the 4-10kev range as a second step, as described in the next Section.

\subsection{Blind search}\label{sub:blind_search}

\begin{figure*}[p!]
\includegraphics[width=1.\textwidth]{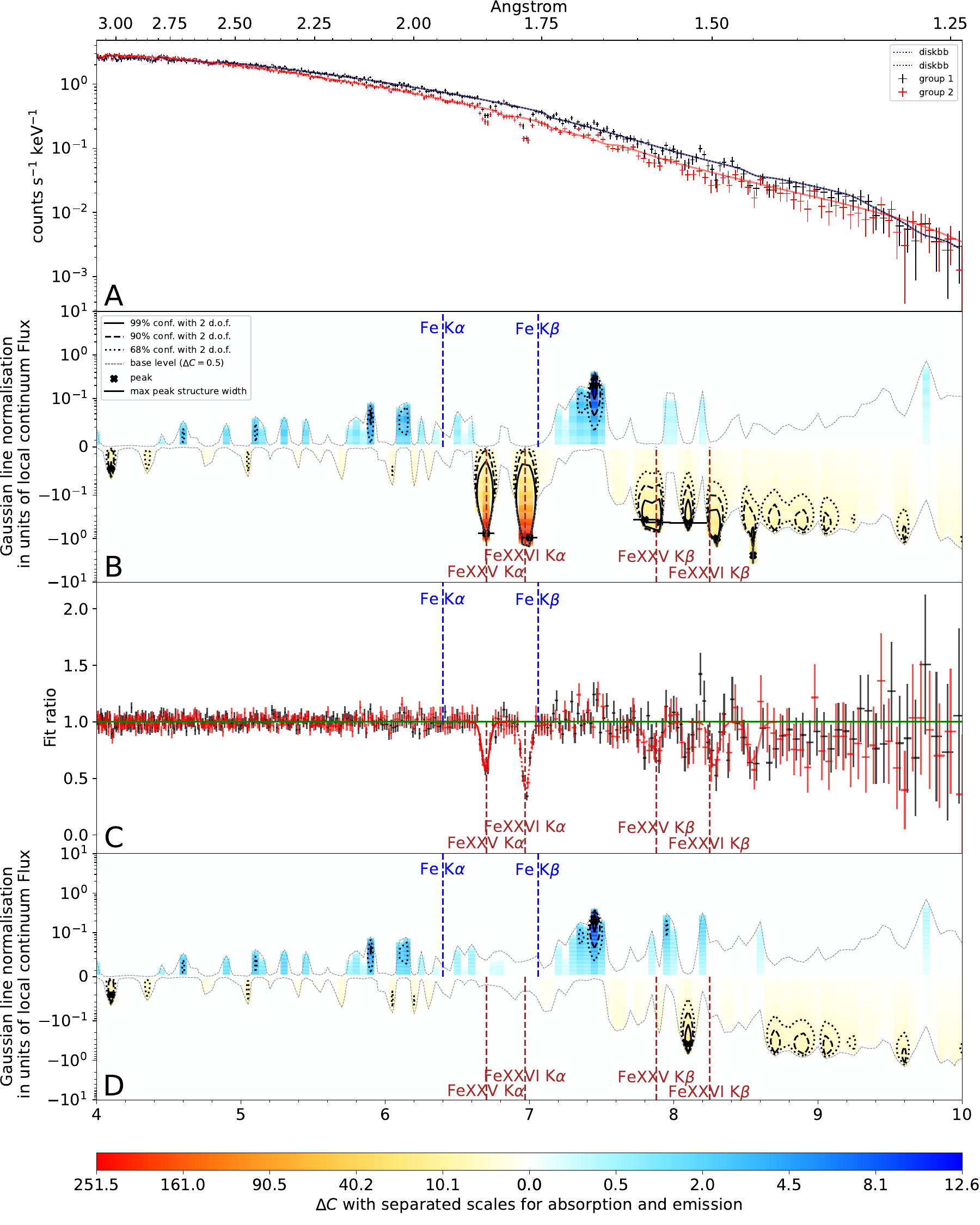}
\vspace{-2em}
\caption{Steps of the line detection procedure for a standard 4U130-47 \chandra{}-HETG spectrum.\\
\textbf{(A)} [4-10]kev spectrum after the first continuum fit in thiskev band. \textbf{(B)} $\Delta C$ map of the line blind search, restricted to positive (i.e. improvements) regions. Standard confidence intervals are highlighted with different line styles, and the colormap with the $\Delta C$ improvements of emission and absorption lines. \textbf{(C)} Ratio plot of the best fit model once absorption lines are added. \textbf{(D)} Remaining residuals seen through a second blind search.}
\label{fig:autofit_example}
\end{figure*}

Once the continuum is fixed, we carry out a standard blind search of narrow emission/absorption features in the 4-10kev band: we measure the change in $\Delta C$ of adding a narrow (width fixed at 0) \gaussian\ line with varying normalisation and energy on the fit and map out the resulting 2D $\Delta C$ surface in the line normalisation-line energy plan. Regions of strong and relatively narrow ($<$1keV) $\Delta C$ excess indicate the possible presence of lines. On the other hand, broader regions ($>$1keV) of $\Delta C$ excess could rather reflect the limit of our simple continuum fit process.

The Gaussian energy varies between 4 and 10kev with linear energy steps of 50 eV for \xmm{}, which is around a third of the EPIC pn spectral resolution at those energies \footnote{see \href{https://xmm-tools.cosmos.esa.int/external/xmm_user_support/documentation/uhb/basics.html}{https://xmm-tools.cosmos.esa.int/external/xmm\_user\_support/\\documentation/uhb/basics.html}} and 20 eV for \chandra{} HETG, which is slightly below HETG's energy resolution at 4keV and a half at 6keV \footnote{see \href{https://cxc.cfa.harvard.edu/proposer/POG/html/chap8.html}{https://cxc.cfa.harvard.edu/proposer/POG/html/chap8.html}}. 
Meanwhile, we dynamically scale the line normalisation in an interval of [$10^{-2}$,$10^1$] times the best fit continuum flux in each energy step, split in 500 logarithmic steps for both positive and negative normalisation.

We show in Fig.\ref{fig:autofit_example} an example of the result of the procedure for 4U 1630-47, a source well known for its absorption lines. Panels (A) and (B) show the spectrum and model after the first continuum fit in the [4-10] band and the $\Delta C$ map obtained with our blind search procedure. The contours over plotted in black highlight $\Delta C$ levels of 68\%, 90\% and 99\% confidence intervals with 2 parameters. The position of the ``maxima'' in $\Delta C$ improvement are highlighted for visualisation. In this example, the blind search clearly identifies two very significant (more than 99\%) absorption features at $\sim$ 6.7 and $\sim$7kev, compatible with the \FeKav{} and \FeKavi{} absorption lines, as well as fainter absorption features at higher energies, compatible with the K$\beta$ complex. The significant emission residual identified at 7.5kev does not seem to affect the absorption regions.

\subsection{Line fitting procedure}\label{sub:line_fit}

\begin{figure*}[t!]
\includegraphics[width=0.5\textwidth]{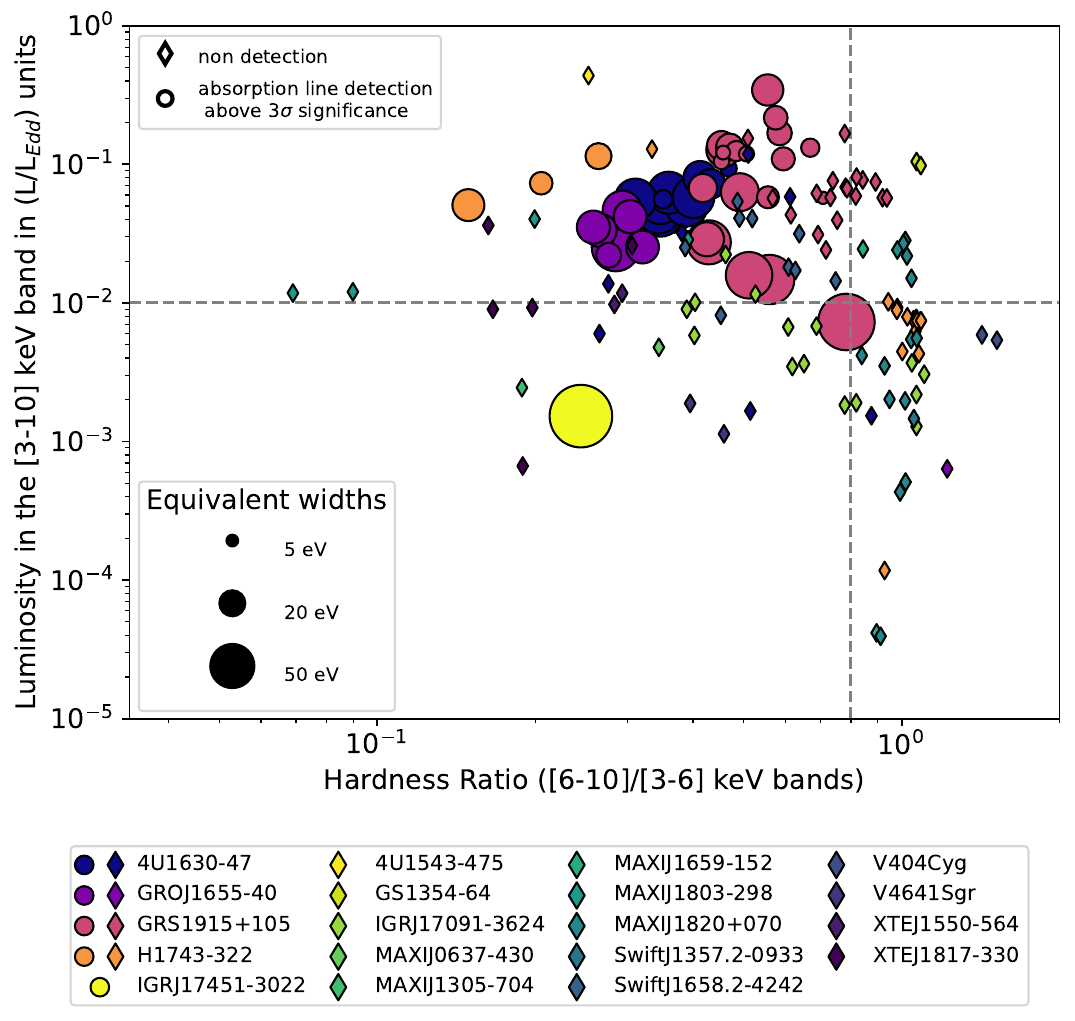}
\includegraphics[width=0.5\textwidth]{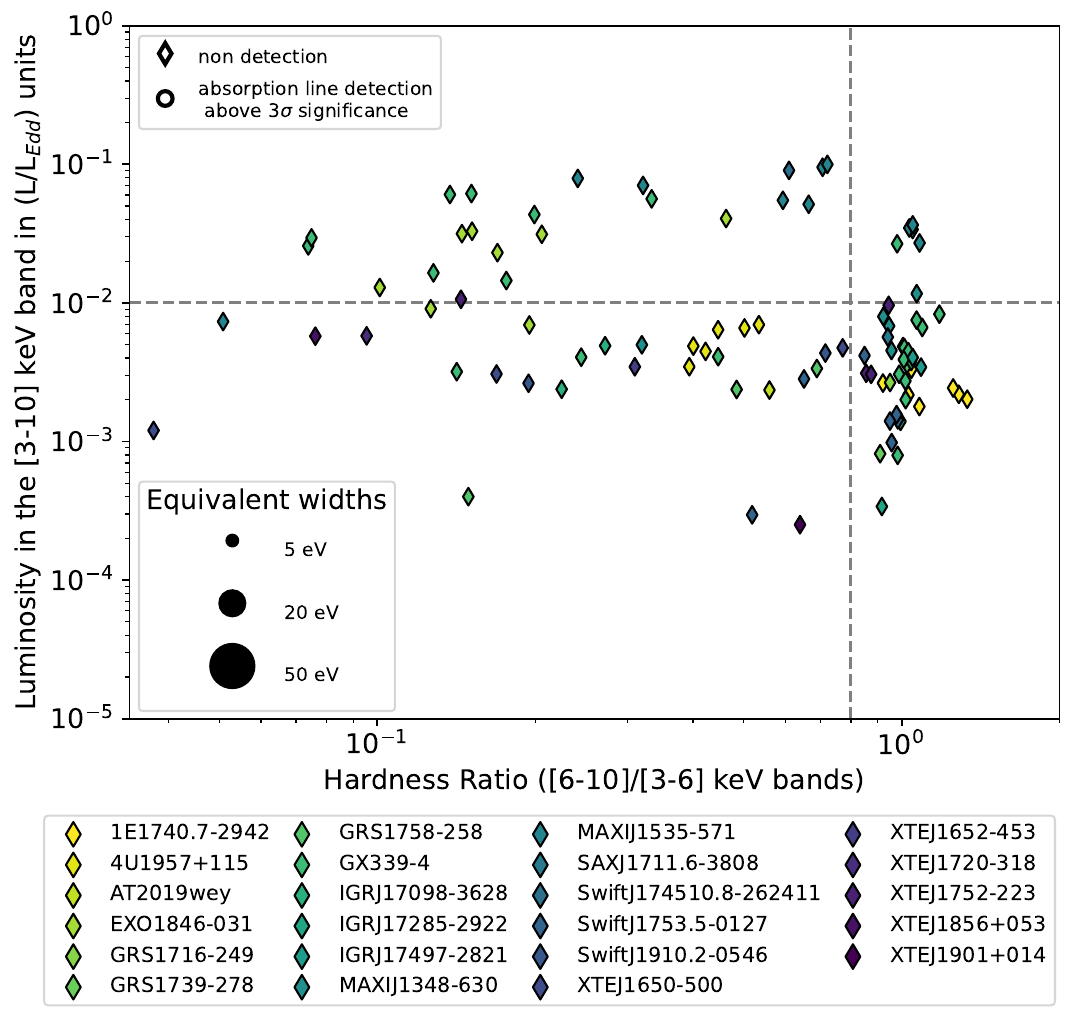}
\caption{HID diagram with the position of all line detections in the sample. The sample is split according to the viewing angle: the left panel is restricted to dippers or sources with $i>55\degr$, while the right panel shows all other sources. The vertical and horizontal line highlight the luminosity and HR thresholds proposed in   Sect.~\ref{sub:glob_results_dichotomy}.}\label{fig:glob_hid_sources}
\end{figure*}

While the blind search described above simply gives a semi-quantitive visualization of the possible presence of line-like features in the spectra, the goal of the next step is to identify the main individual absorption lines, and to derive their physical parameters.
Thus, we start from the continuum fit, and add up to 7 potential line features, using the same F-test threshold as used for continuum components. Among these, 5 are the strongest absorption lines in the iron complex, namely \FeKav{} (6.70kev\footnote{The energy of the \FeKav{} line is set equal to the  resonant transition because the intercombination line is signicantly weaker. XMM-epic is unable to resolve the two lines, neither is HETG unless with extremely high statistics.}), \FeKavi{} (6.97kev), \FeKbv{} (7.88kev), \FeKbvi{} (8.25kev) and \FeKgvi{} (8.70kev).  The remaining 2 are fluorescent emission lines from neutral iron, \FeKan{} (6.40kev) and \FeKbn{} (7.06kev). We do not consider the \NiKavii{} and \FeKgv{} absorption lines, as they can be blended with the stronger \FeKbv{} and \FeKbvi{} respectively at our resolutions.

All lines are modelled with a simple \texttt{gaussian} component, convolved with \texttt{vashift} in order to allow for a shift of the lines, limited to $[-10000,5000]$ km s$^{-1}$. Indeed, we do not expect significantly redshifted absorption lines, nor speeds beyond ~0.03c, as the vast majority of wind observations up until now have shown wind speeds compatible with 0 or of at most a few thousands of km s$^{-1}$ (see references in Table. \ref{table:sources_det_states}). Moreover, allowing for higher blueshifts would produce degeneracy between neighboring lines (\FeKav{} reaches \FeKavi{}{}'s energy at $v\sim$12000 km s$^{-1}$, and \FeKbv{} reaches \FeKbvi{} at v$\sim$ 14000 km s$^{-1}$). We assume that all lines of a single ion are produced in the same region of the wind, and consequently have the same velocity shift. All absorption lines are considered narrow, allowing their width to vary only up to $\sigma<=50$ eV. A line is considered resolved only if its width is larger than 0 with a $3\sigma$ level of confidence.

While we are not interested in characterizing emission lines in detail, a good portion of observations show significant broad emission features in the iron region, which we model using up to two simple phenomenological neutral \FeKan{} and \FeKbn{} broad \gaussian{} components, restricting their blueshift to the same interval taken for absorption lines and limiting their widths to $[0.2,0.7]$kev. The lower limit prevents an overlapping between narrower emission and absorption features, while the upper limit prevents the broad emission features from modeling large parts of the continuum.

In very few XMM observations of GRS 1915+105 andGRO J 1655-40 however, such as the exposures analyzed in \citet{Trigo2007_GROJ1655-40_wind_XMM_2005}, the presence of extreme emission features requires a more complex modeling. In these spectra, we follow the same approach as \citet{Trigo2007_GROJ1655-40_wind_XMM_2005}, using a \texttt{laor} component with energy free in the range [6.4,7.06]kev, inclination in the range $[50,90]$ degrees (consistent with the sources being highly inclined), and $R_{in}$ and $R_{out}$ fixed at their default values. 

We show in panel (C) of Fig.\ref{fig:autofit_example} an example of the result of the procedure for a standard observation. In this case, all 5 Fe absorption components are sufficiently significant to be added in the model and reproduce very well the absorption features. Nevertheless, once the line fit is complete, we perform a second blind search to check the presence of remaining line features in the residuals, following the procedure described in the previous section. We show in panel (D) of Fig.\ref{fig:autofit_example} the result of this step for our example spectra. While all 5 main absorption features are indeed perfectly reproduced, a significant narrow feature at $\sim8.1$kev remains, which can be identified with the K$\alpha$ transition from \ion{Ni}{xxviii}. Similar residual features are only found in the highest SNR \chandra{} spectra, suggesting the presence of other weaker transitions not included in our 5 main components. However, these further absorption features are present only in combination with the much stronger lines considered in our analysis, and their detailed characterization is beyond the scopes of this paper.

For all the observations with no detected absorption lines, we compute the $3\sigma$ (99.7\%) upper limit of each line's EW, using the highest value in the line's range of velocity shift. All EW measurements and upper limits are reported in Tab. \ref{tab:obs_details}.

\subsection{Line significance assessment }\label{sub:sign_fakes}

Goodness and F-test methods have long been known to overestimate the detection significance of lines \citep[][]{Protassov2002_stat_problems}.
Reliable estimates can only be obtained through Monte-Carlo simulations \citep[][]{Porquet2004_stat_MC}, which have been adopted as the standard in the last decade \citep[][]{Tombesi2010_wind_AGN_XMM,Gofford2013_wind_AGN_Suzaku,Parker2020_wind_AGN_rms,Chartas2021_wind_AGN_XMM_highz}. We follow a similar procedure, adopting the same methodology as for the real data by putting similar constraints in energy and width as described in   Sect.~\ref{sub:line_fit}.

We thus generate 1000 distributions of parameters within the uncertainties of the final model from 1000 runs of the \texttt{simpars} xspec command. We then delete all absorption line components from the models, before repeating the following steps for 1000 iterations:
\begin{enumerate}
    \item Load a set of model parameters from the simulated distribution
    
    \item Simulate a spectrum from the current model using the \texttt{fakeit} xspec command, retaining all of the observational parameters (exposure, response files, background) of the initial spectrum
    
    \item Fit the continuum + emission lines model to the simulated spectrum, to obtain a baseline C-stat
    
    \item Compute the maximal possible $\Delta C$ gained from the addition of an absorption line in each line's allowed blueshift bands (exactly as done for the real data, and described in Sect.~\ref{sub:line_fit}).
    
\end{enumerate}

The $\Delta C$ of the line detected in the real data can then be compared to the distribution of the 1000 maximal $\Delta C_{sim}$ of the simulated spectra, and the statistical significance of the line is defined by:
$P=1-N/1000$, with N the number of $\Delta C_{sim}$ larger than the real value. Only lines with a significance larger than 3 $\sigma$ (99.7\%) in their blueshift range, as derived from this procedure, are considered detections, and will be considered as such in the following sections, as well as reported in Tab. \ref{tab:obs_details}. 

\begin{figure*}[p!]

\includegraphics[width=0.5\textwidth]{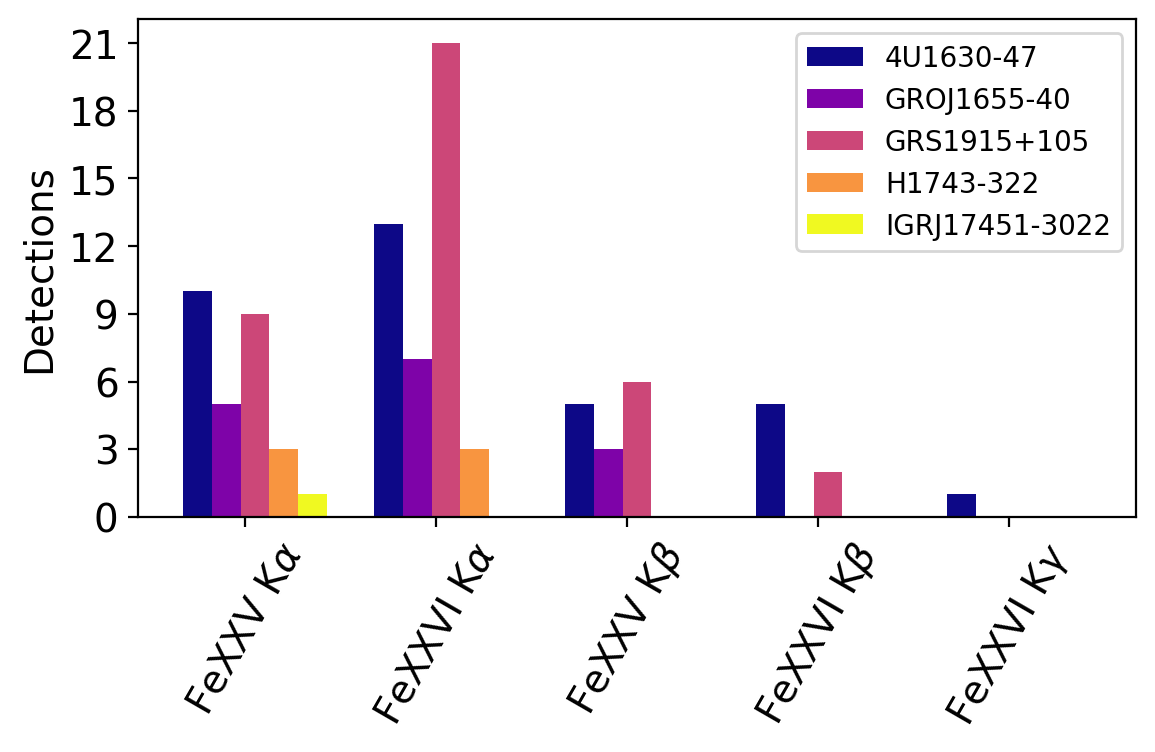}
\includegraphics[width=0.5\textwidth]{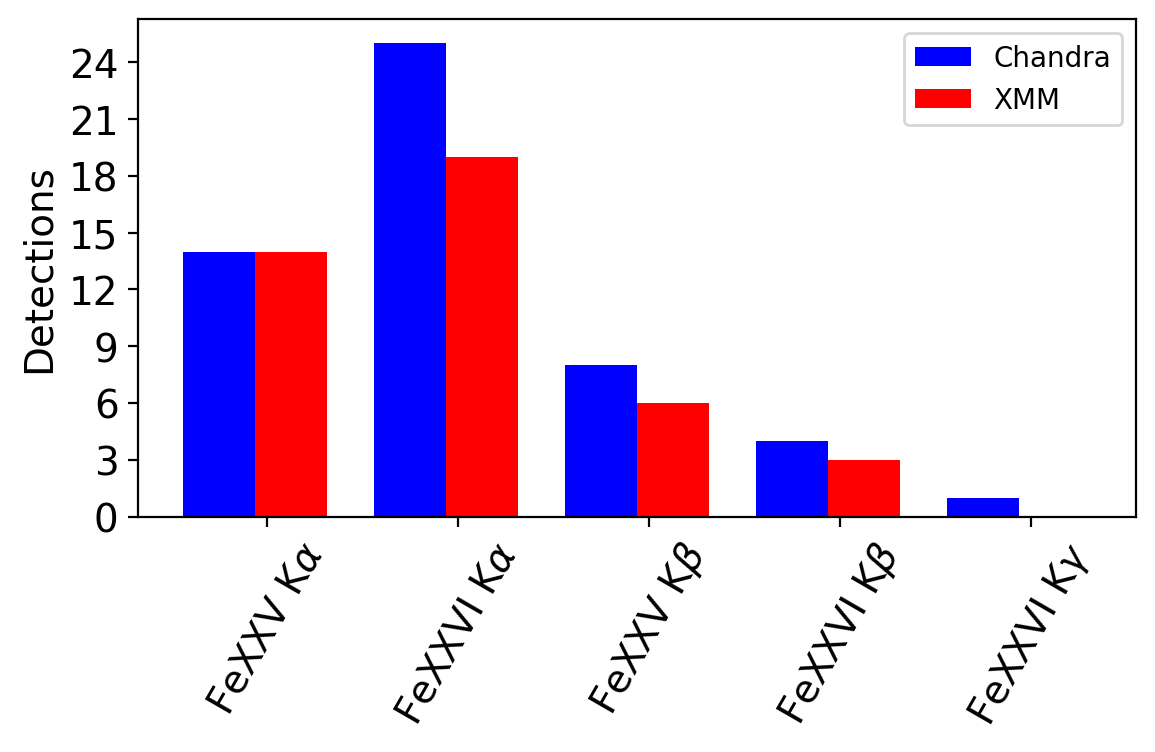}

\includegraphics[width=0.5\textwidth]{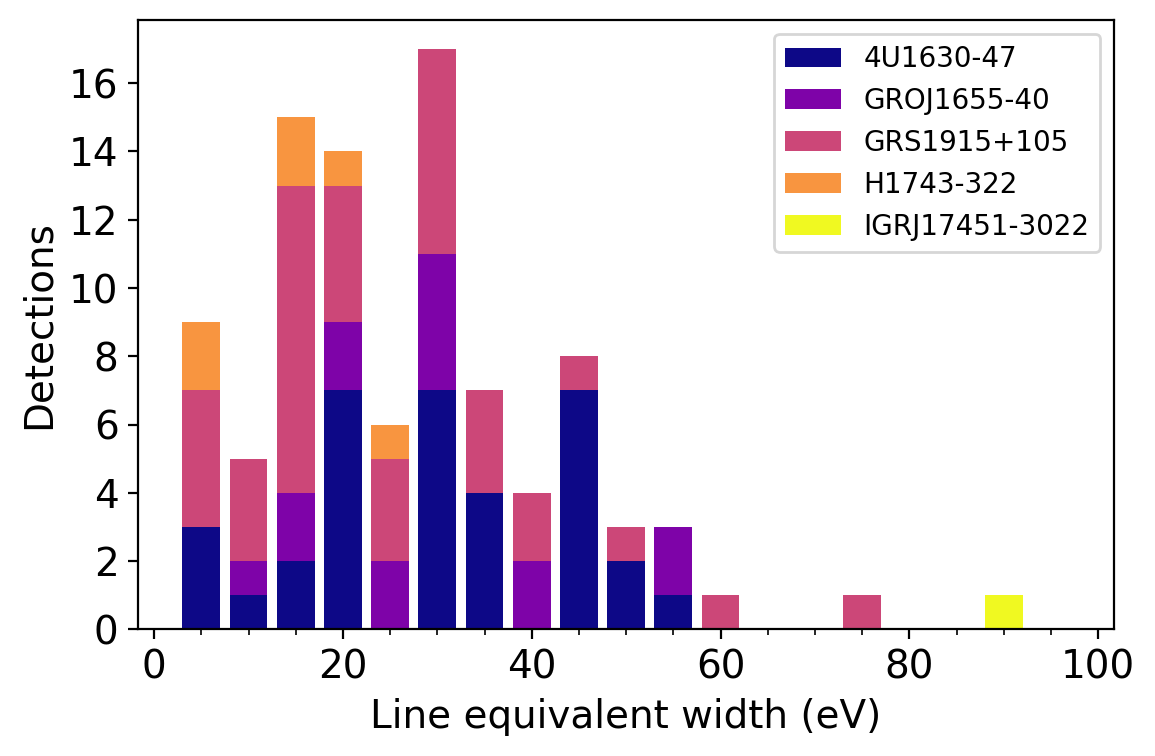}
\includegraphics[width=0.5\textwidth]{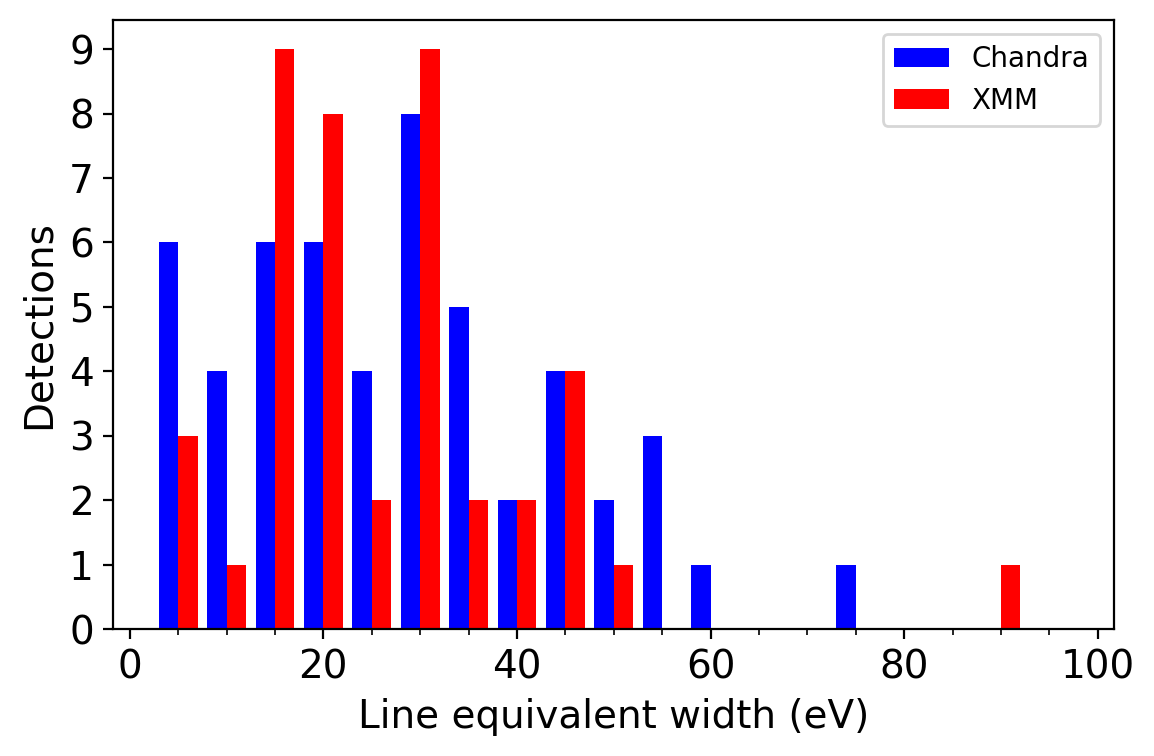}

\includegraphics[width=0.5\textwidth]{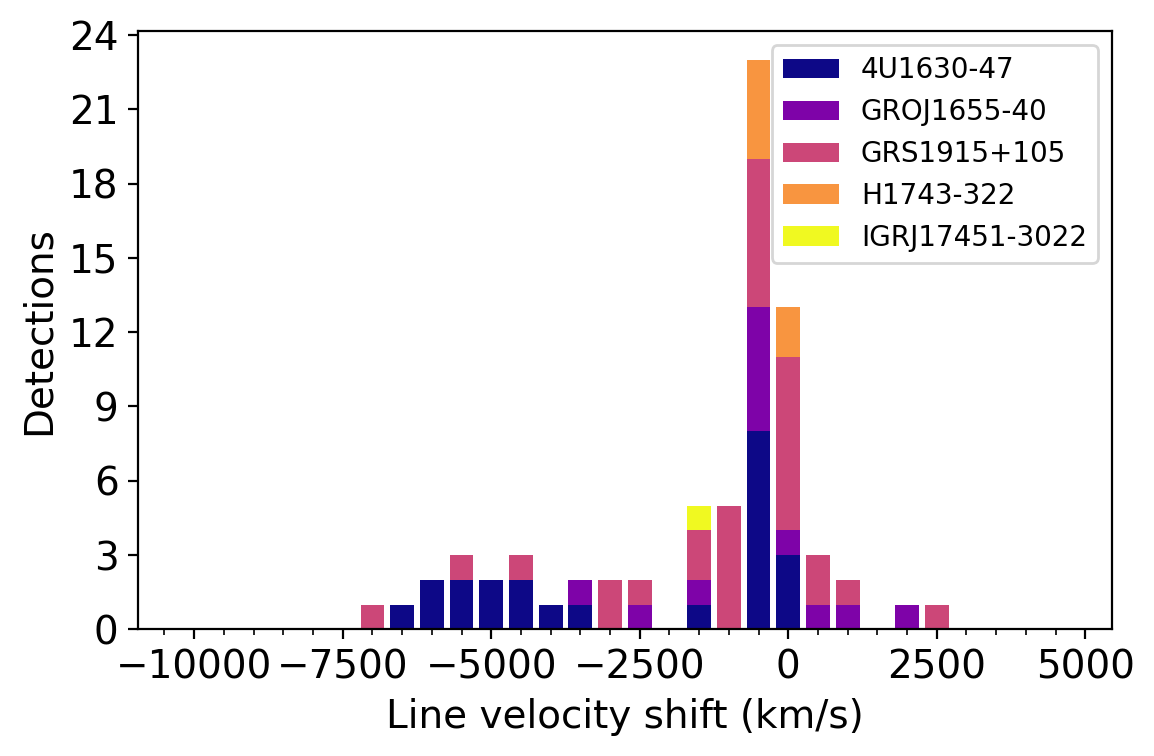}
\includegraphics[width=0.5\textwidth]{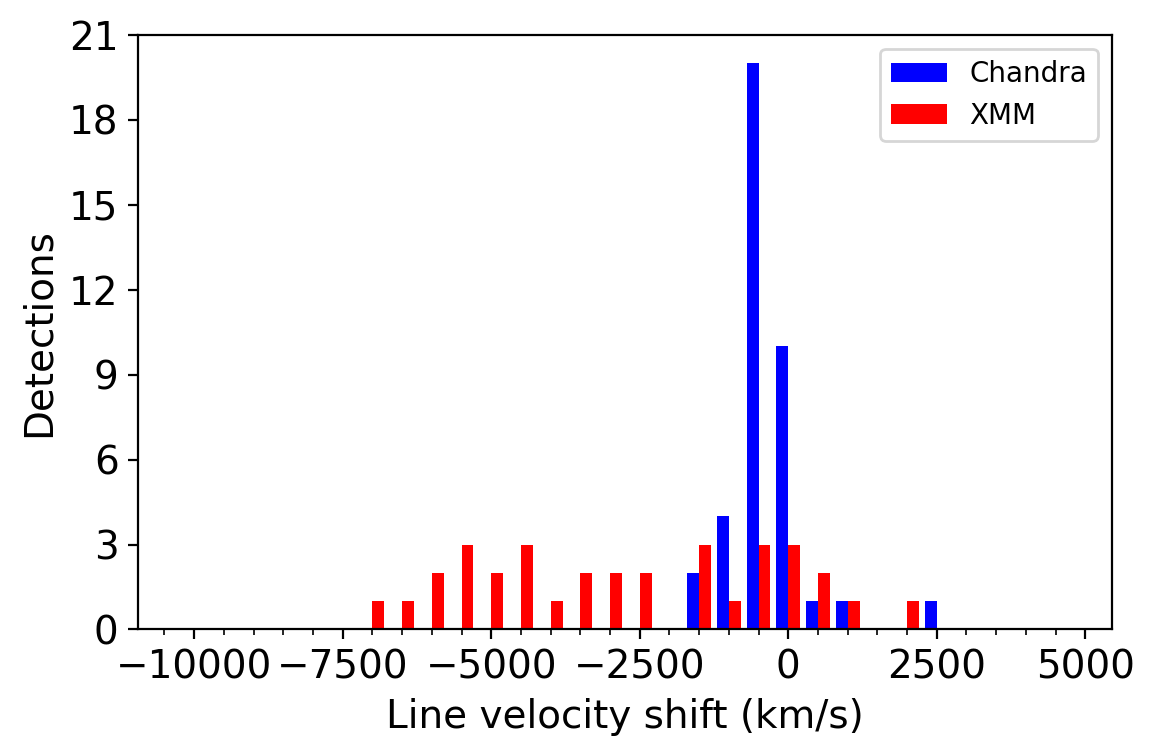}

\includegraphics[width=0.5\textwidth]{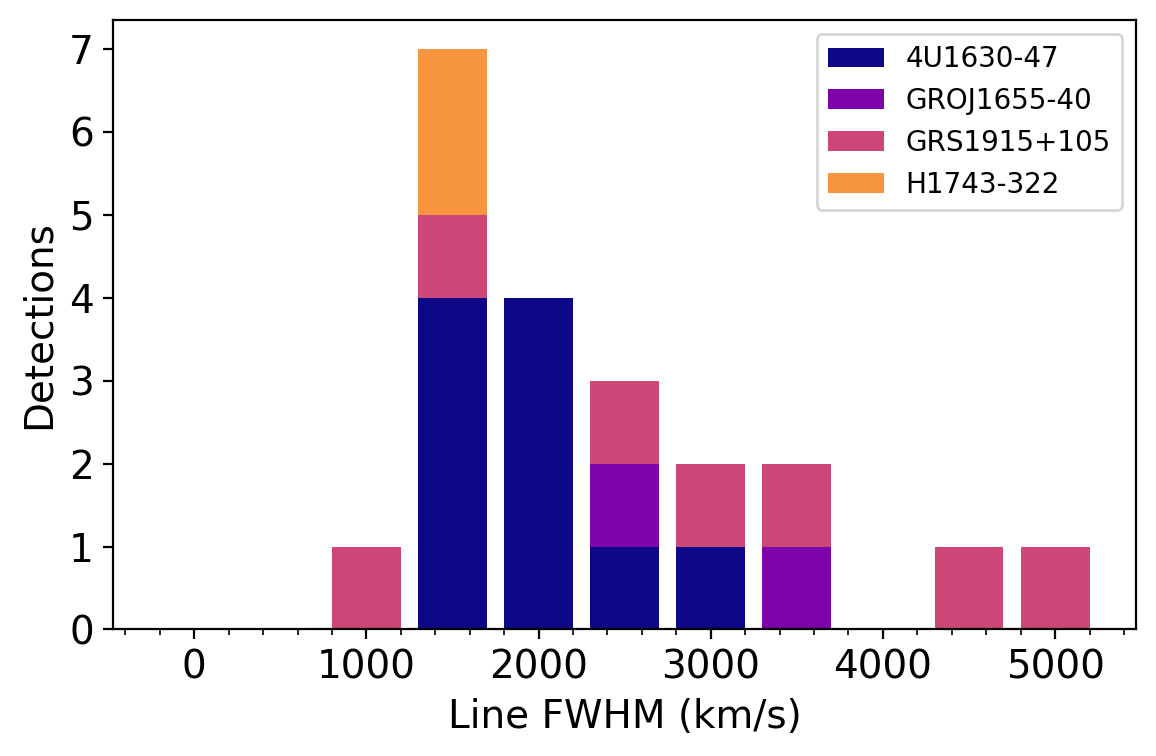}
\includegraphics[width=0.5\textwidth]{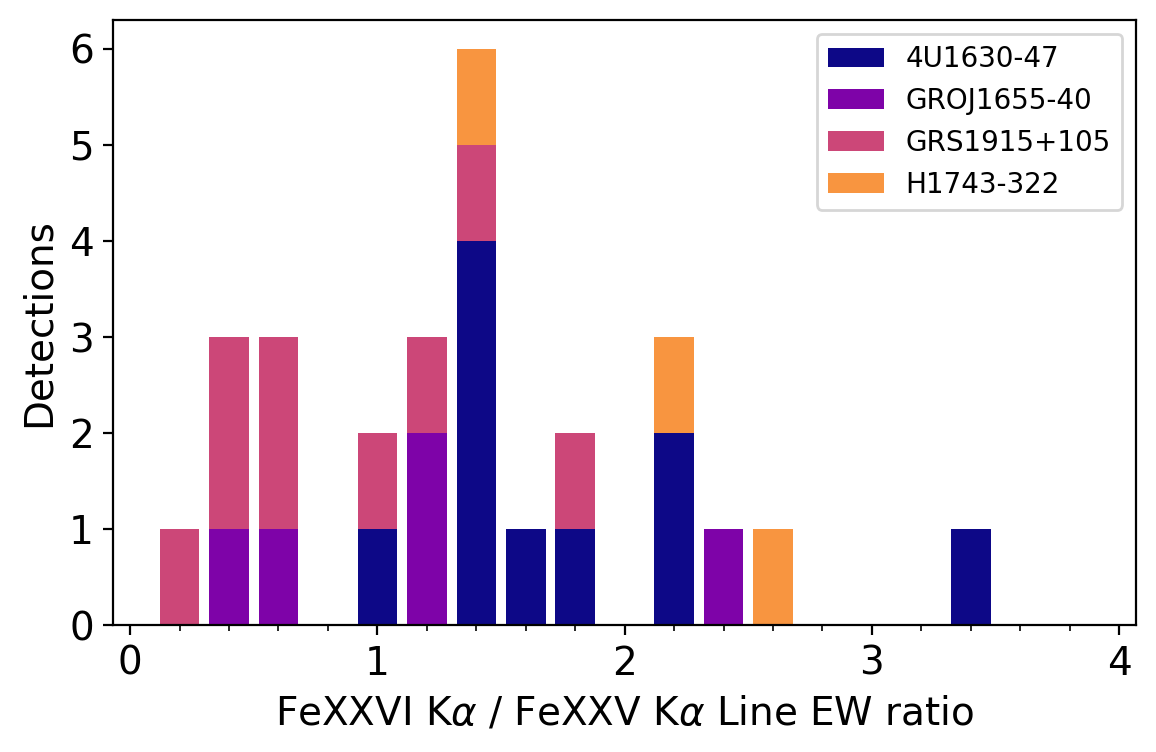}

\caption{Distribution of intrinsic line parameters (detections of each line, EW, blueshift, widths and K$\alpha$ EW ratio) for the  entire sample, split by source, and instrument whenever relevant. The blueshift distributions are restricted to the K$\alpha$ complex.}\label{fig:glob_distrib}

\end{figure*}
\section{Global Results}\label{sec:results_glob}

The Hardness Intensity Diagram (HID) of the full sample is shown in Fig.~\ref{fig:glob_hid_sources}. Despite a sample of 42 sources, absorption line detections remain restricted only to a very small subset of objects, i.e. the highly-inclined 4U 1630-472, GRO J1655-40, GRS 1915+105, H 17432-322, and IGR J17451-3022 (IGRJ17451 hereafter). The detections follow the same trend as previously reported in  \cite{Ponti2012_ubhw}, without any detection in ``pure'' hard states  (corresponding to HR$\sim$1, see \ref{sub:glob_results_dichotomy} for details). In the case of GRS 1915+105, which does not follow the standard outburst evolution, absorption lines  are generally detected when the jet is quenched, with one single exception for ObsID 660 (\citealt{Lee2002_GRS1915+105_winds_first} and see \citealt{Neilsen2009_GRS1915+105_wind_jet_connection,Neilsen2012_GRS1915+105_wind_variability} for details).

\subsection{Parameter distribution and correlation}

In order to study in more details the behavior of the absorption lines and their interplay with the continuum SED, we analyzed the distribution of their main parameters, and identified statistically significant correlations. To identify the correlations between individual parameters, we compute the Spearman coefficients, which trace general monotonic relations between two parameters. For that purpose, and in order to take into account the uncertainties of each parameter, we apply MC simulations to estimate the distribution of the correlation coefficients and associated p-values, following the perturbation method of \cite{Curran2014_correl_MC}. This is implemented through the python library \textit{pymccorrelation} \citep[][]{Privon2020_pymmccorrelation}. In the following, we focus on all correlations with $p<0.001$ found in our sample.

%DISTRIBUTIONS

\begin{figure*}[]

\includegraphics[width=0.49\textwidth]{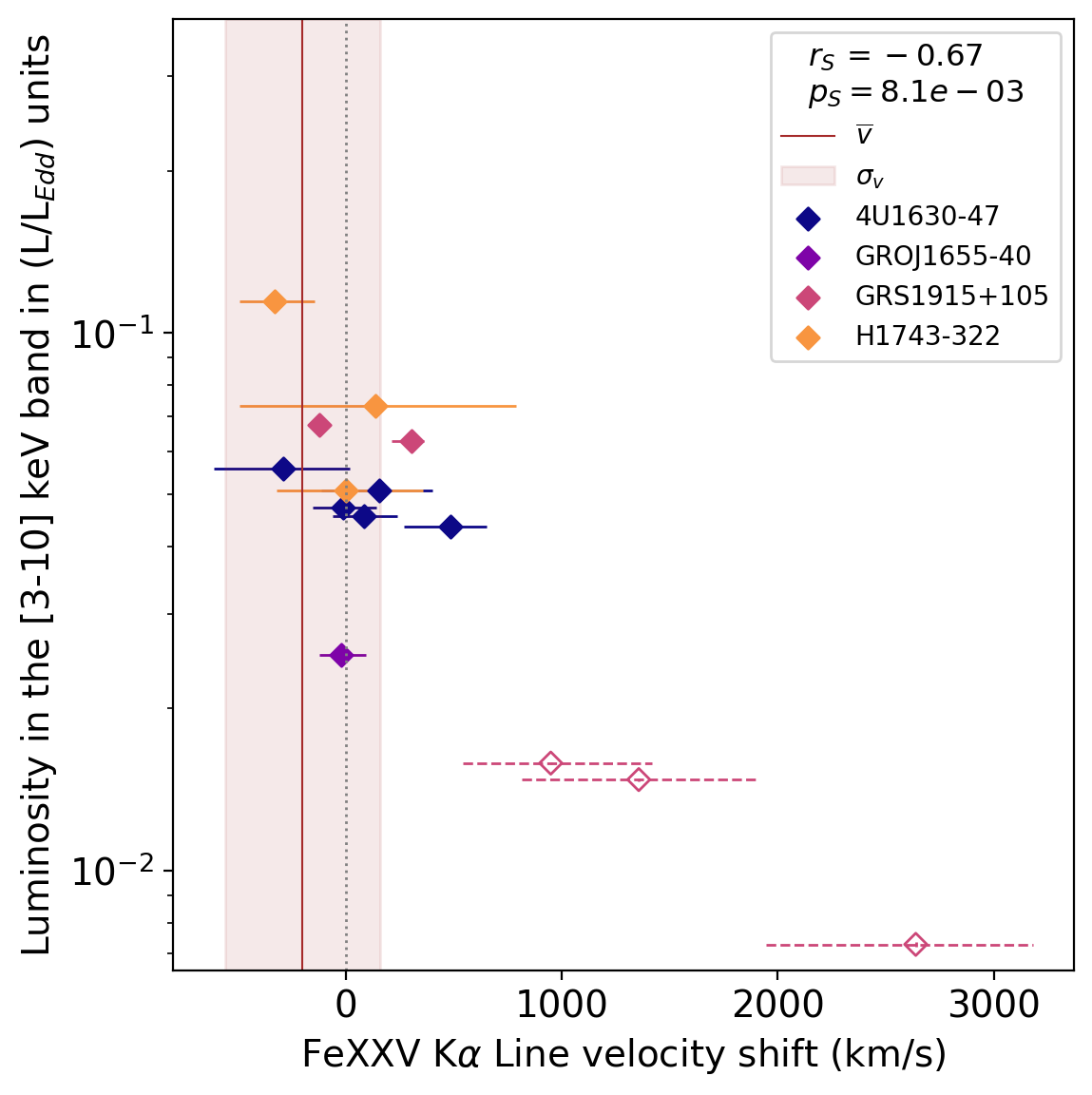}
\includegraphics[width=0.49\textwidth]{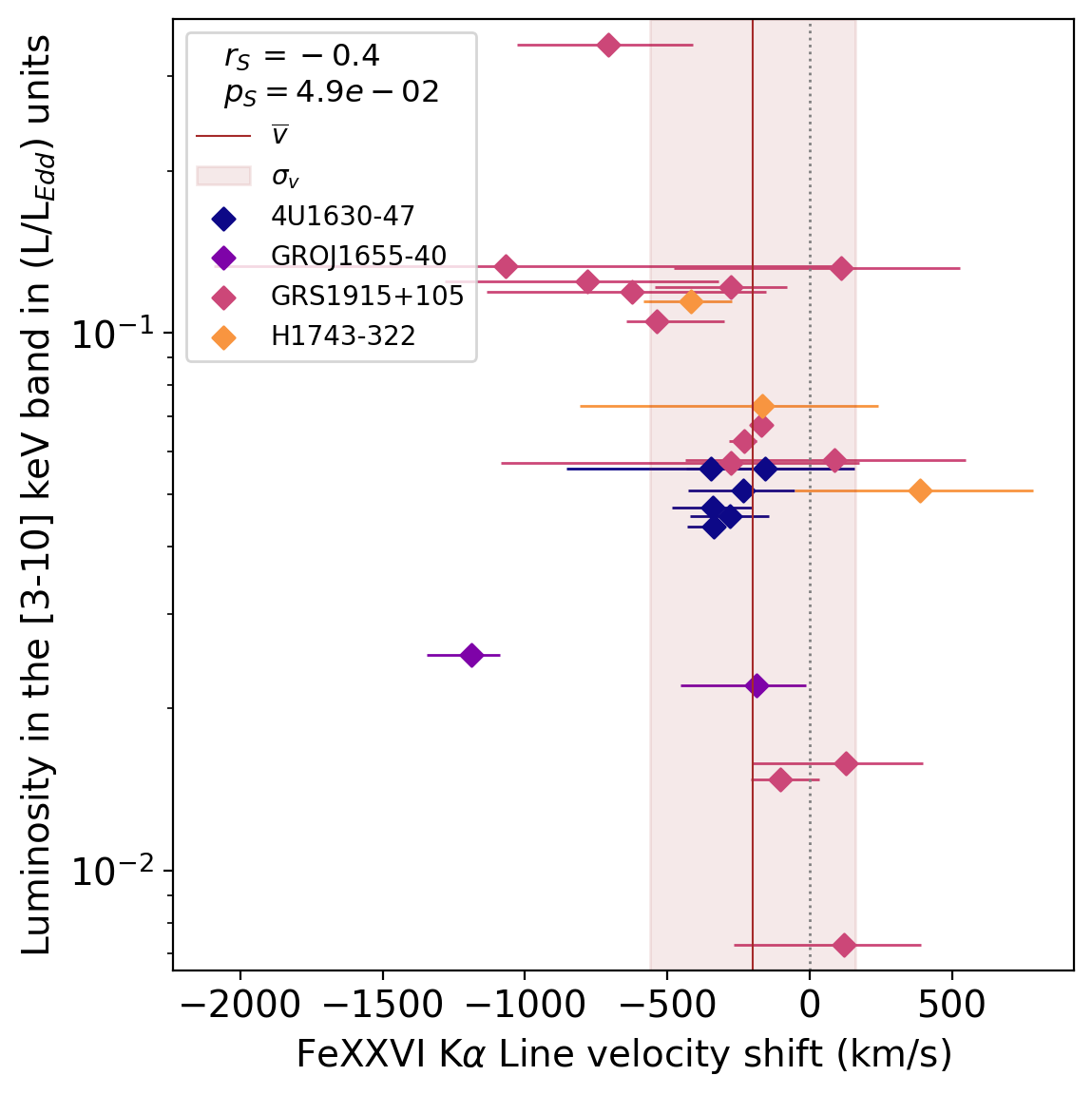}

\caption{Scatter plot of the \FeKav{} (left) and \FeKavi{} (right) velocity shifts against luminosity in \chandra{} observations, color coded according to the sources. The grey dotted line corresponds to 0 velocity, and the brown line to the mean of the curated K$\alpha$ blueshift distribution, whose standard deviation is visualized by the brown region. The biased \FeKav{} blueshifts measured in the obscured GRS 1915+105 observations, which are excluded from this distribution, are marked in dashes.}\label{fig:glob_correl_bshift_lum_Chandra}

\end{figure*}

\subsubsection{Parameter distribution}\label{sub:param_distrib}

We assess the main properties of the absorption features in our sample with the detection of each line, their EWs, and the velocity shifts for the better constrained K$\alpha$ complex. The distributions are presented in Fig.~\ref{fig:glob_distrib}. The left panels are split by source in order to show the properties of the absorption features in each object, but we stress that, except for a few outliers which will be discussed below, the number of detections is too limited for the differences between the distributions to be significant. The right panels, which are instead split by instrument, should exhibit mostly similar distributions as \xmm{} and \chandra{} observed similar portions of the HID. This is clearly the case for the distribution of line detections: both instruments show the largest number of detections for \FeKavi{}, followed by \FeKav{}, \FeKbv{} and \FeKbvi{}. Moreover, no K$\beta$ or K$\gamma$ lines are detected without the corresponding K$\alpha$. In addition, as can be seen in the list of detections in Tab. \ref{tab:obs_details}, the \FeKavi{} line is present in all observations where lines are detected except one where only \FeKav{} is detected. Meanwhile, the single significant detection of \FeKgvi{} is found in a \chandra{} spectrum. 

Although less apparent, the distribution of the EWs of both instruments are also broadly compatible, with a KS-test p-value of 0.46. The whole sample spans a range of $\sim 5-100$ eV, with XMM detections expectedly dropping below 15 eV due to more limited energy resolution.
The EW ratio between the \FeKavi{} and \FeKav{} line (hereafter called K$\alpha$ EW ratio) provides a proxy of the ionisation parameter $\xi$ in our sample \citep[e.g.][]{Bianchi2005_AGN_EW_ionisation}. As seen in the bottom right panel in Fig.~\ref{fig:glob_distrib}, in our sample, the majority of the K${\alpha}$ EW ratios is clustered between 1 and 2.5. This means that most exposures with line detections have sufficiently high ionization parameters for the \FeKavi{} line to be predominant. However, two objects (namely GRS 1915+105 and GRO J1655-40) show K${\alpha}$ EW ratios spread across also the entire observed range, with a number of detections significantly below 1 associated to lower $\xi$.

The velocity shift distributions for the strongest K$\alpha$ lines are clearly different between the two instruments, with a KS-test p-value of $1.7\times10^{-7}$ (see bottom right panel of Fig.~\ref{fig:glob_distrib}). \xmm{} shows a somewhat uniform distribution between ~-6500 and 2000 km s$^{-1}$, while the \chandra{} velocity shift distribution is much narrower and more symmetric around 0. The highest blueshift obtained with \chandra{} is around $\sim$1200 km s$^{-1}$, in accordance with the highest values previously reported in the literature for this observation \citep[][]{Miller2008_GROJ1655-40_wind_Chandra_2005_detailed}. 

This difference can be at least partly attributed to the limits of the EPIC-PN camera. Indeed, in timing mode, used for the vast majority of EPIC-PN observations in our sample, even after recent updates in energy-scale calibration\footnote{see bottom right panel of Fig.~3 in \href{https://xmmweb.esac.esa.int/docs/documents/CAL-SRN-0369-0-0.pdf}{https://xmmweb.esac.esa.int/docs/documents/CAL-SRN-0369-0-0.pdf}}, the energy accuracy remains limited, with a residual average shift of 18 eV ($\sim$ 800 km s$^{-1}$ for \FeKavi{}) and a standard deviation of 80 eV ($\sim$ 3500 km s$^{-1}$ for \FeKavi{}) at 12keV. The standard deviation of our measured distribution is of $\sim 2500 $km s$^{-1}$, and is thus compatible with the theoretical limits of the instrument's accuracy (which we can expect to somewhat better at 7kev). The mean value of (also) $\sim 2500 $km s$^{-1}$ is however significantly larger than the mean of post-calibration systematic energy accuracy, but this may be the consequence of our choice of restricting the allowed blueshift fitting range to [-10000,5000] km s$^{-1}$, which would introduce a bias in a distribution with such a significant spread. Moreover, this large average blueshift cannot be reconciled with the much smaller measurement of the more accurate \chandra{} HETG instrument, so we will only consider the \chandra{} blueshifts in the rest of the paper.

\begin{figure*}[]

\includegraphics[width=0.5\textwidth]{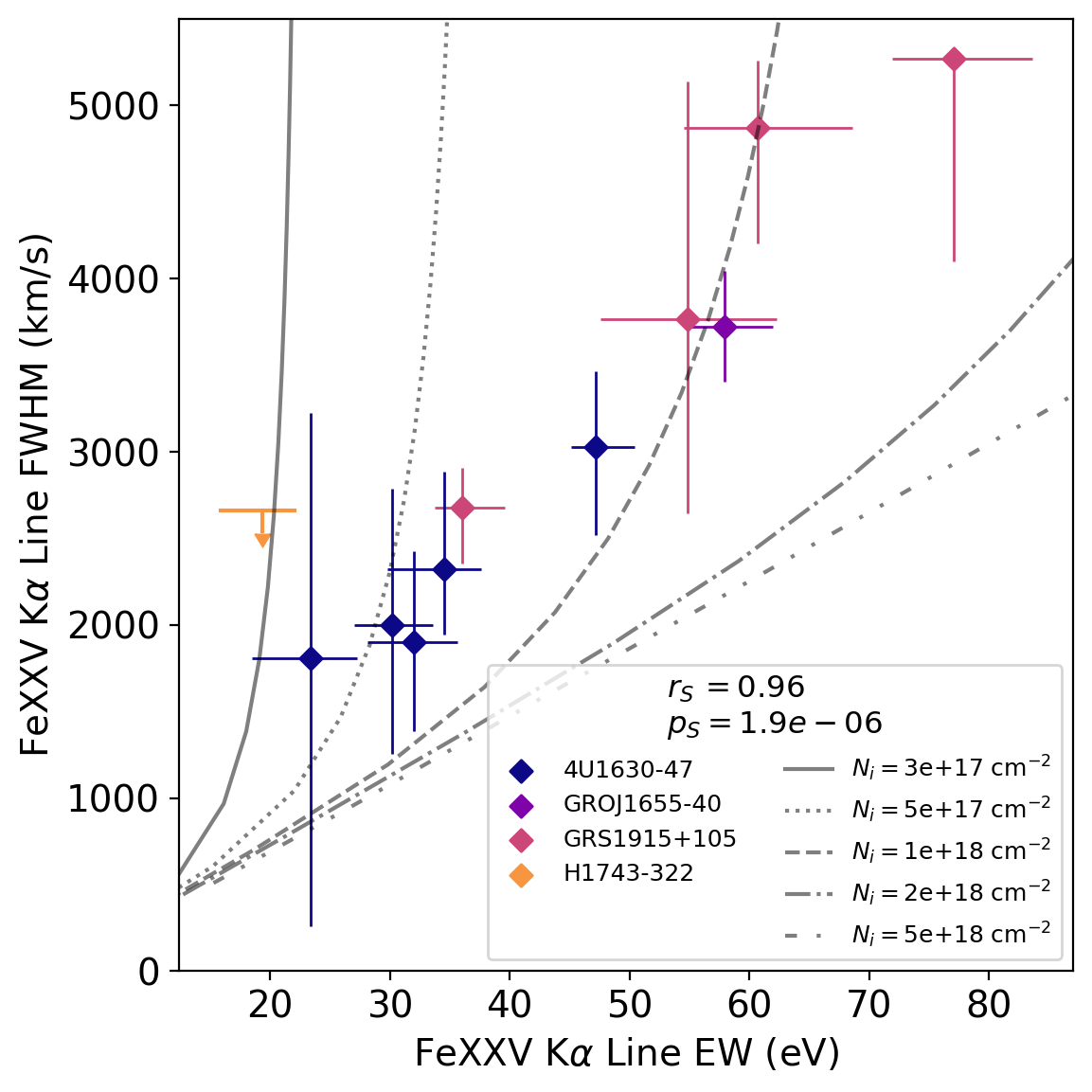}
\includegraphics[width=0.5\textwidth]{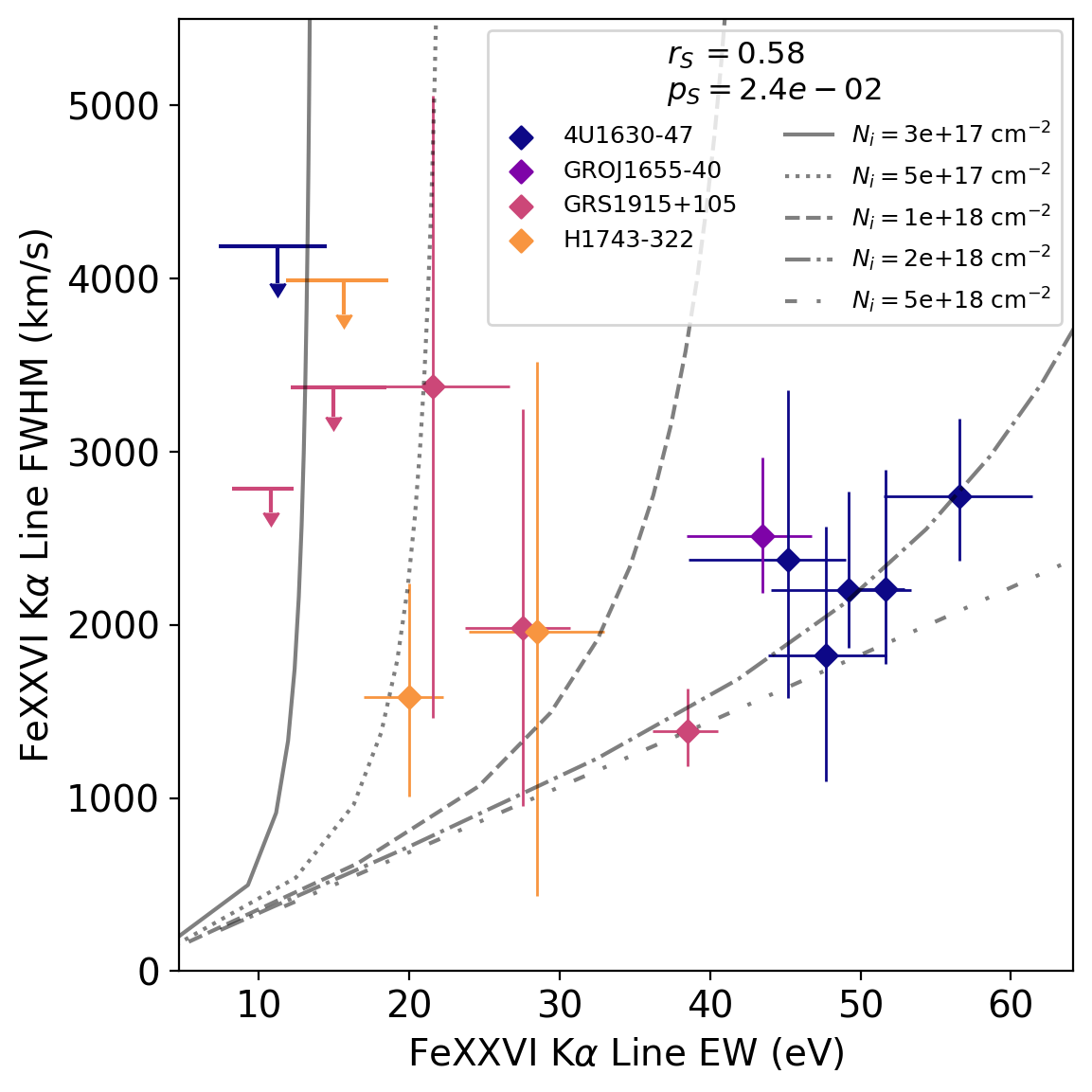}

\caption{Scatter plot of the EW and width for the \FeKav{} (left) and \FeKavi{} (right) lines in \chandra{} observations. The curves highlight the theoretical evolution of these parameters for a range of ionic column densities of the respective ions.}\label{fig:glob_width_EWa}
\end{figure*}

The observed \chandra{} velocity shift distribution is  within the expectations from a sample of intrinsically zero-velocity absorption lines, with an average value of $\mu\sim60\pm100$ km s$^{-1}$, and a standard deviation of $\sigma\sim630$ km s$^{-1}$. However, few observations have significant velocity beyond 2 $\sigma$ of the mean of this distribution. We report in Fig.~\ref{fig:glob_correl_bshift_lum_Chandra} the scatter plots of the \chandra{} velocity shifts of the \FeKav{} (left panel) and \FeKavi{} (right panel) lines against the 3-10kev luminosity in Eddington units, which highlights that the 3 faintest GRS 1915+105 exposures are the only one showing significant \FeKav{} positive shifts (i.e. redshifts). However, the \FeKav{} absorption line profiles observed in these three cases exhibit unusually asymmetric and broad absorption features (see the data panels of Fig.\ref{fig:fit_GRS} in appendix), while the \FeKavi{} lines energies are consistent with 0 velocity. 

According to \cite{Neilsen2020_GRS1915+105_obscured_NICER_winds_hard}, this apparent redshift might be caused by contributions from lines at lower energies, blended with \FeKav{}. We verified that with a simple fit with two photoionised slabs\footnote{We used the same \texttt{CLOUDY} absorption Tab. model described in \cite{Ratheesh2023_4U_pola}},  which we show in the lower panels of Fig.~\ref{fig:fit_GRS}. We find that the highest ionization component ($\log(\xi)\sim5-6$) models the \FeKavi{} and part of the \FeKav{} lines which shows a blueshift $\sim-250$ km s$^{-1}$, while a lower ionization phase ($\log(\xi)\sim2.5-3$) at 0 velocity produces some of the  \FeKav{} line, but is also heavily affected by absorption lines from \ion{Fe}{xxi} to \ion{Fe}{xxiv}, which reproduce the observed ``redshifted'' tail of the line profiles. 
We thus exclude these 3 observations from the velocity shift distribution, changing the distribution average to $\mu\sim-200\pm60$ km s$^{-1}$, and reducing the standard deviation to $\sigma\sim360$ km s$^{-1}$, as highlighted in Fig.~\ref{fig:glob_correl_bshift_lum_Chandra}. 

With this restriction, the only remaining outlier (more than $2\sigma$ away from the restricted mean) is found in the blueshifted \FeKavi{} line of the exceptional absorption signatures of GRO J1655-40's 2005 outburst \citep[][]{Miller2006_GROJ1655-40_winds}, and is in agreement with the extreme absorption features displayed in this observation (see \citealt{Miller2008_GROJ1655-40_wind_Chandra_2005_detailed} for a detailed study). 
We note that one exposure of 4U 1630-47 (obsid 13716) remains at the tail-end of the \FeKav{} velocity shift distribution, with a redshift of 500 km/s, distinct from 0  at more than $3\sigma$, as well as from the corresponding \FeKavi{} line (itself with a blueshift of $\sim 300$ km/s). This can once again be explained by a contamination from a lower ionization component, in line with more in depth analysis, such as the work of \cite{Trueba2019_4U1630-47_wind_2012-13Chandra}, who model the outflow with two photo-ionization components. In this observation, both components show a significant decrease in ionization parameter compared to rest of the coverage of the outburst, while maintaining low, negative velocity shifts, in accordance with our results for the other exposures.

The mean value of $-200\pm60$ km s$^{-1}$ is very low compared to the standard \chandra{} HETG absolute wavelength uncertainty of $\pm$ 0.006 \AA\footnote{see \href{https://cxc.harvard.edu/proposer/POG/html/chap8.html}{https://cxc.harvard.edu/proposer/POG/html/chap8.html}}, which translates to $\sim \pm$ 1000 km s$^{-1}$ at the \FeKavi{} energy ($\sim300$ km s$^{-1}$ at 2kev). However, empirical studies have shown that the "effective" absolute wavelength accuracy of HETG is significantly better, and reaches $\sim$ 25 km s$^{-1}$ at energies below $\sim$2kev \citep{Ishibashi2006_HETG_effective_accuracy,Bozzo2023_HETG_effective_accuracy_update}. This has been corroborated by other works making use of very precise spectral features \citep{Ponti2018_NS_mass_xray_spectro_HETG_accuracy}. The few existing BH wind studies that consider the effective HETG accuracy also estimate it to be up to 50-100 km s$^{-1}$ depending on the line considered (see \citealt{Miller2020_GRS1915+105_obscured_Chandra,Munoz-Darias2022_GS2023+338_wind_xray-optical_2015_details}). Thus, our sample is likely to exhibit a significant global blueshift, in agreement with the common association of these absorption lines to outflowing winds, although the average velocity is very low.

It is also possible to measure the widths of the of \FeKav{} and \FeKavi{} lines in the \chandra{} observations with the highest SNR. The distribution of the FHWM of the 21 lines with significant width measurements is reported in the lower left panel of Fig.~\ref{fig:glob_distrib}. While all significant line width measurements are in the 1500-5000 km s$^{-1}$ range, the highest values, found in the 3 GRS 1915+105 exposures with contamination from other line complexes discussed above, are probably overestimated. 

\subsubsection{Significant correlations}\label{sub:diff_line}

\begin{figure*}[]

\includegraphics[width=0.5\textwidth]{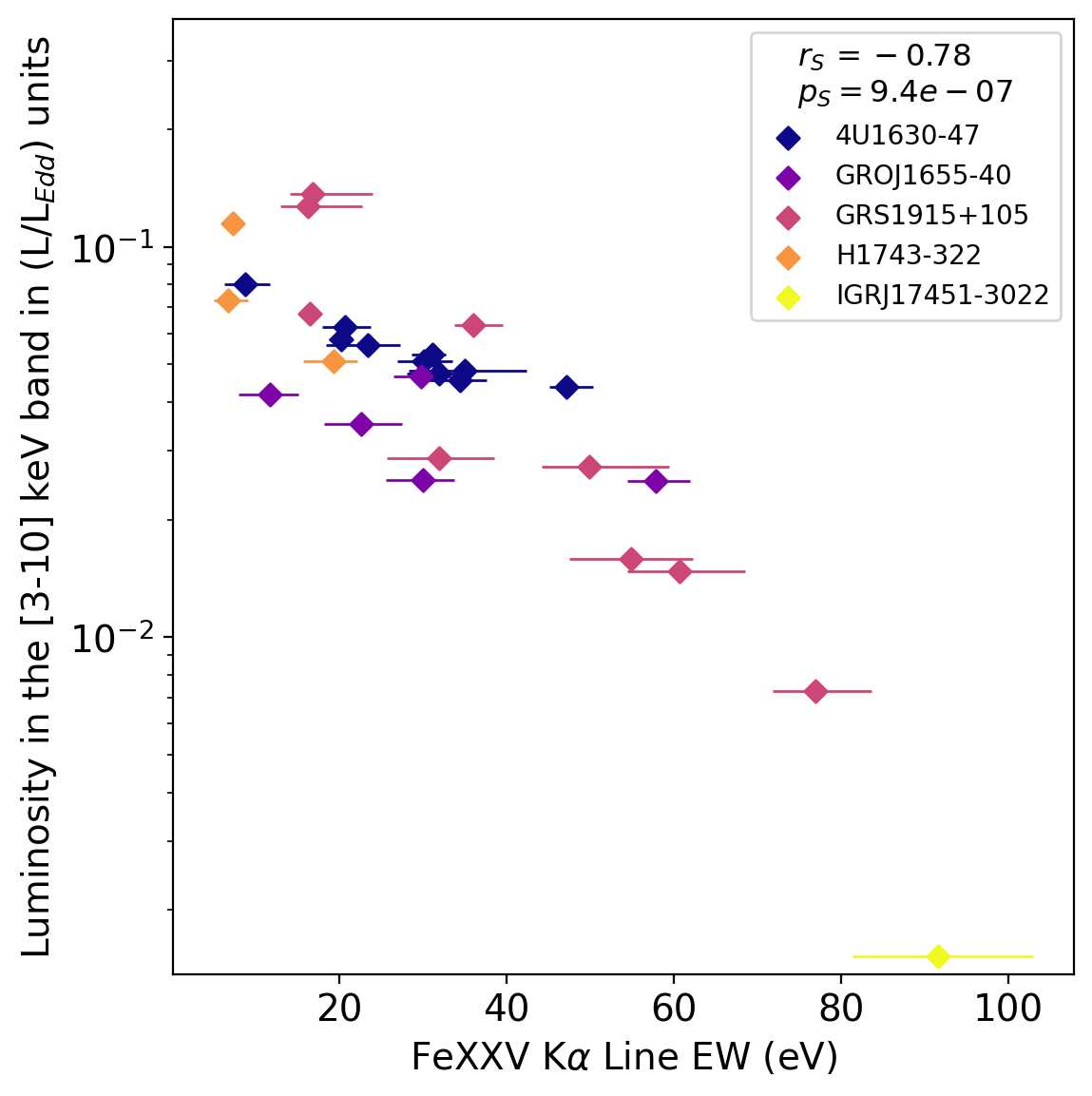}
\includegraphics[width=0.5\textwidth]{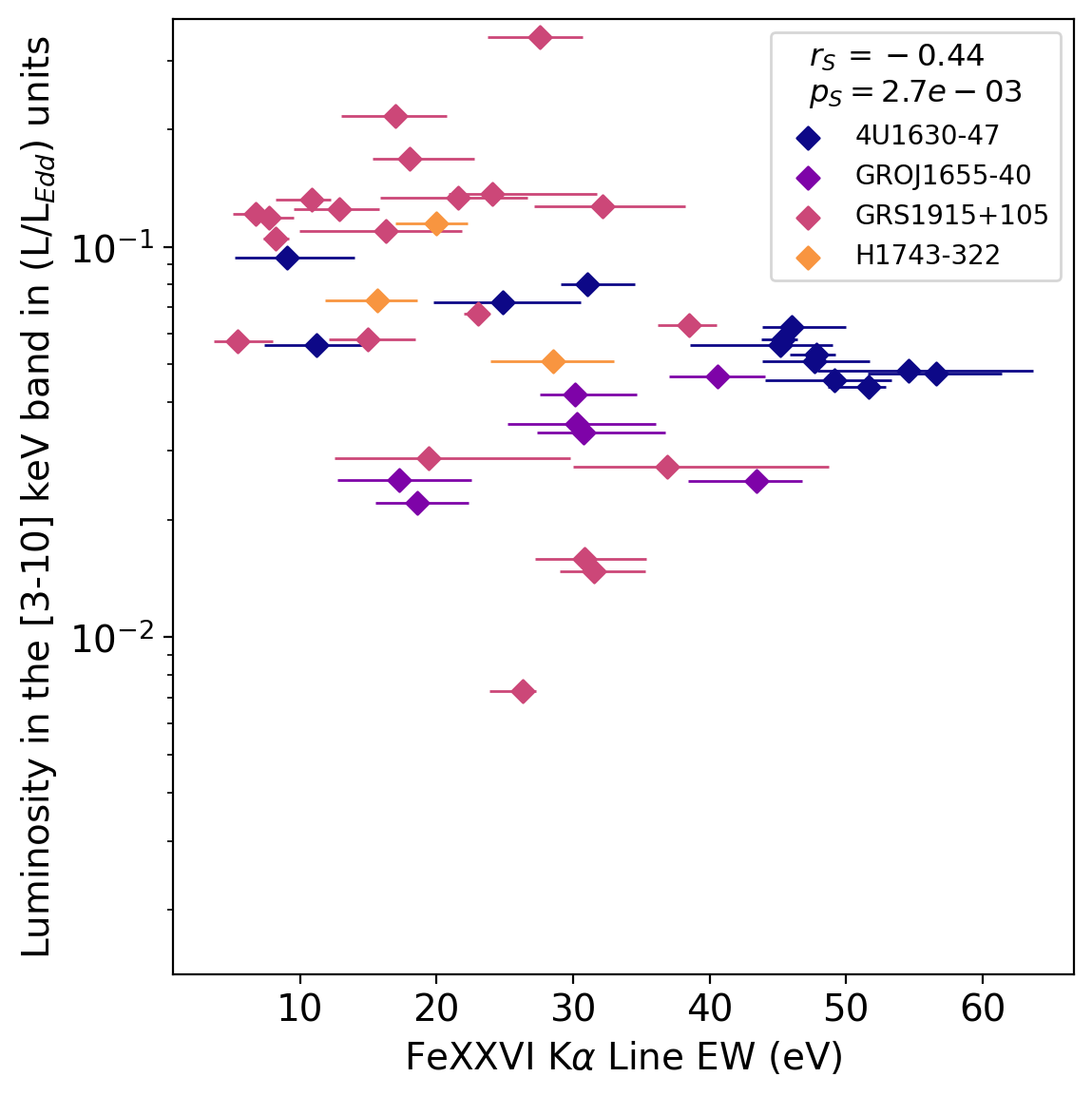}

\caption{Scatter plot of the \FeKav{} (left) and \FeKavi{} (right) EW against luminosity for the entire sample, color coded according to the sources.}\label{fig:glob_correl_EW_lumb}

\end{figure*}

\begin{figure}[]

\includegraphics[width=0.5\textwidth]{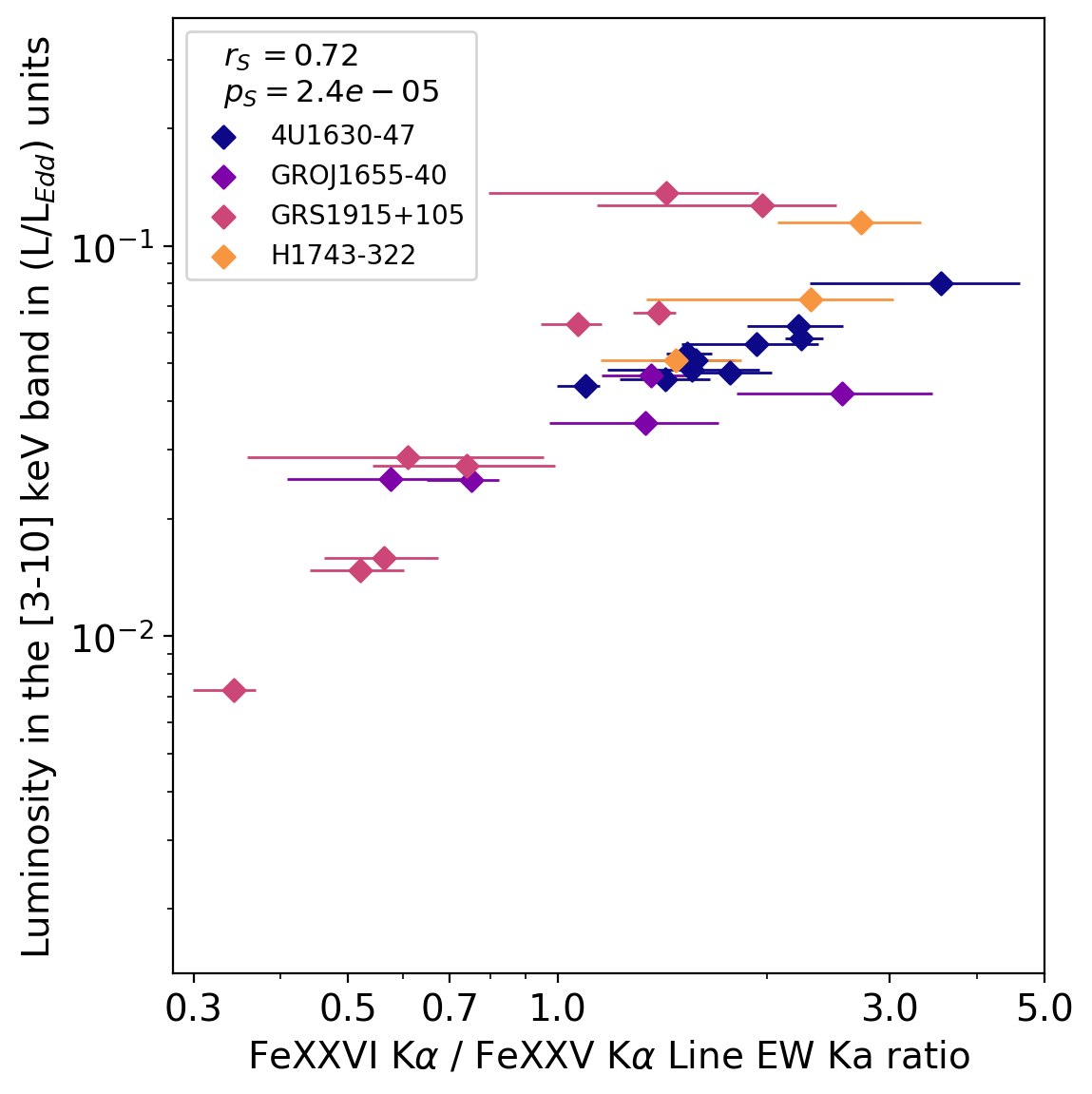}

\caption{Scatter plot of the \FeKavi{}/\FeKav{} EW ratio against luminosity for the entire sample.}\label{fig:glob_correl_eqw25_eqw26}

\end{figure}

\begin{figure*}[h]
\includegraphics[width=0.5\textwidth]{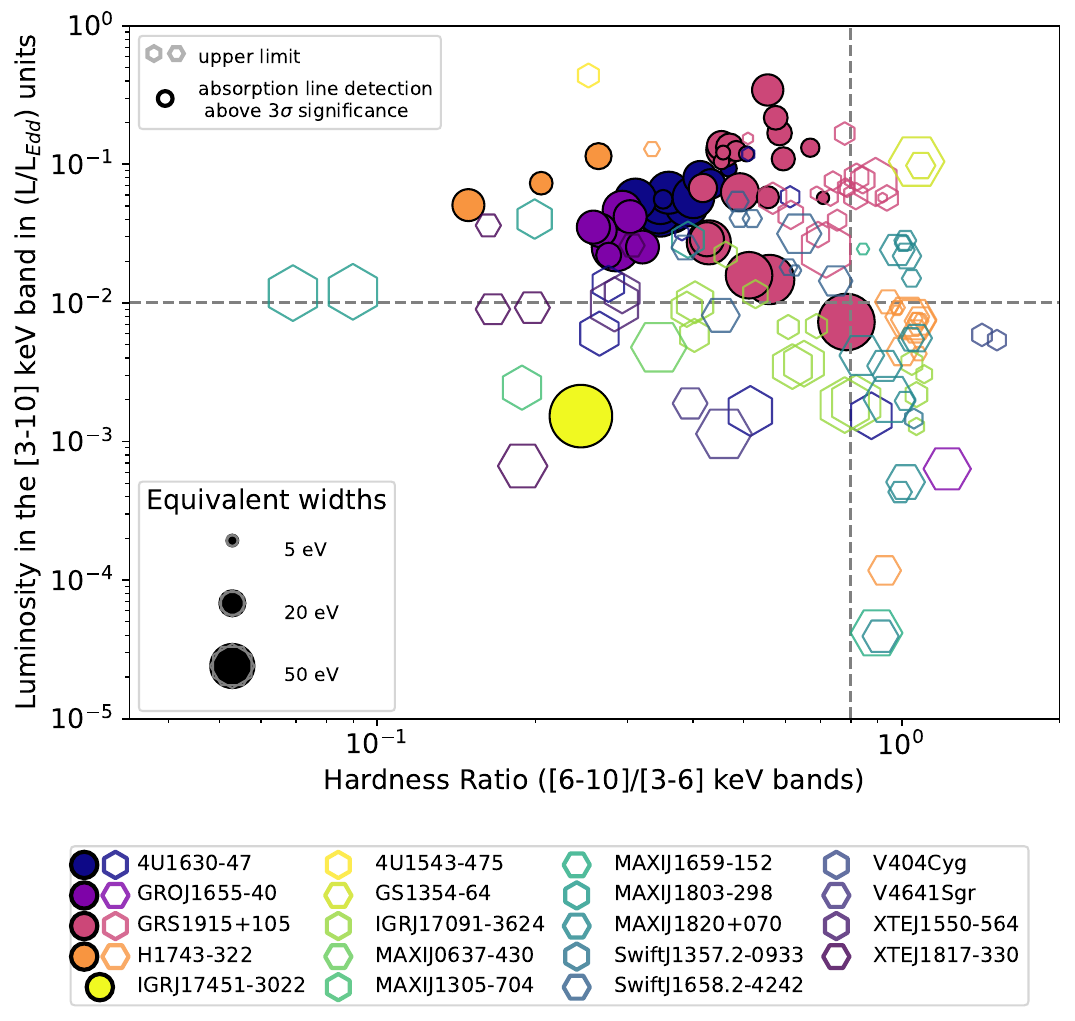}
\includegraphics[width=0.5\textwidth]{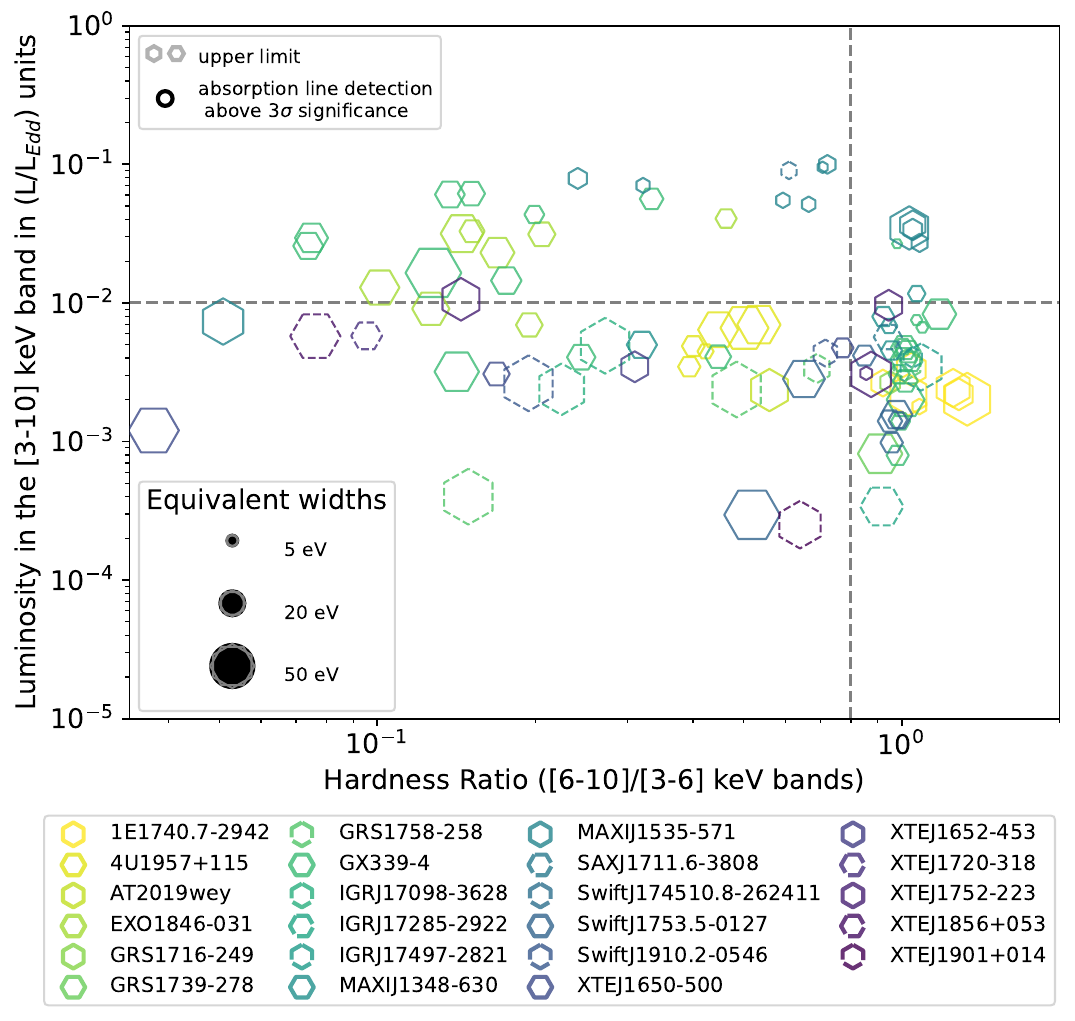}
\caption{HID diagram with the position of all detections in the sample, and \FeKavi{} upper limits when no line is detected, using the same inclination split as in Fig.\ref{fig:glob_hid_sources}. The vertical and horizontal line highlight the luminosity and HR thresholds proposed in   Sect.~\ref{sub:glob_results_dichotomy}. Sources with no inclination measurements in the right panel are dashed. }\label{fig:glob_hid_sources_withul}
\end{figure*}

The first significant correlation 
we find in our results is between the width and EW of the \FeKav{} line (p$\leq$0.0002), which we show in Fig.~\ref{fig:glob_width_EWa} and contrasts with the absence of correlation in the case of \FeKavi{}. Such a correlation may naturally arise because larger turbulence velocities delay the saturation at the line centre, allowing the EW to grow to larger values \citep[see e.g. the curve of growths presented in][]{Bianchi2005_AGN_EW_ionisation}. Moreover, the saturation itself at high column densities contribute to broaden the absorption lines. To test these effects, following the methodology detailed in \cite{Bianchi2005_AGN_EW_ionisation}, we computed the curve of growths for \FeKav{} and \FeKavi{} lines, as a function of the corresponding ionic column densities N$_i$ and different turbulence velocities. Moreover, we estimated the FWHM of each computed profile relative to the given N$_i$ (and therefore EW) and velocity. These computations allowed us to derive theoretical curves to superimpose on the data plotted in Fig.~\ref{fig:glob_width_EWa}. 

For both lines, all measurements are compatible with the expectations, appearing in the allowed portion of the parameter space. Indeed, the lower right corner of these plots is expected to be unpopulated, since the EW saturates at large N$_i$ and cannot grow further while the line width keeps increasing rapidly. On the other hand, we would also expect to populate the upper left corner, but there is likely a strong observational bias against broad lines with low EW. It is interesting to note that lower ionic column densities are needed for the majority of observed \FeKav{} lines with respect to \FeKavi{}, suggesting an average high ionization parameter, in accordance to the typical large \FeKavi{}/\FeKav{} EW ratio noted before in our sample. The few detections with the highest \FeKav{} EWs require higher \ion{Fe}{xxv} ionic column densities and thus a lower $\xi$, in accordance with their lower \FeKavi{}/\FeKav{} EW ratios.

We also observe a significant anti-correlation between the \FeKav{} EW versus the X-ray luminosity, as shown on the left panel of Fig.~\ref{fig:glob_correl_EW_lumb}. It is worth noting that the p-value remains below $10^{-5}$ even without including the uncertain luminosity measurement of IGR J17451-3022. This anti-correlation may naturally arise if we take the luminosity as a proxy for the ionization parameter (i.e. assuming a universal $nr^2$ factor for the whole sample): this is indeed what is expected if the average ionization parameter is just above the peak of the ionic fraction for \ion{Fe}{xxv} \citep[e.g.][]{Bianchi2005_AGN_EW_ionisation}.
In comparison, no such correlation is observed for the \FeKavi{} line (see right panel of Fig.~\ref{fig:glob_correl_EW_lumb}), as expected since its ionic fraction would instead be at its peak, for the same ionization parameter. An equivalent way to show these different behavior is via the significant correlation between the X-ray luminosity and the \FeKavi{}/\FeKav{} EW ratio for all the observations where both lines are detected (see Fig.~\ref{fig:glob_correl_eqw25_eqw26}). This ratio is expected to be a monotonic function of the ionization parameter \citep[e.g.][]{Bianchi2005_AGN_EW_ionisation}, and should thus correlate with luminosity.

\subsection{Favourable conditions for absorption line detections of \ion{Fe}{xxv} and \ion{Fe}{xxvi} in this sample}\label{sub:glob_results_dichotomy}

%UPPER LIMITS

Our HID diagrams in Fig.~\ref{fig:glob_hid_sources} show that absorption lines of He-like and H-like iron are mainly observed in luminous soft states of highly inclined sources. %(e.g., \citealt{Ponti2012_ubhw})
Indeed, we may further propose quantitative thresholds to define a ``favourable'' region for this type of wind detections, based on the hardness ratio, the inclination and the luminosity. 

Firstly, all absorption line detections in our sample occur below a Hardness Ratio (computed using unabsorbed flux) of HR$_{[6-10]/[3-10]}=0.8$. This cut remains nevertheless arbitrary because it depends on the blackbody temperature, which is affected by the mass and spin of the objects, and as such is expected to differ for each source. It also does not restrict to pure soft states, as this threshold also includes observations in soft and hard intermediate states. The two most notable exceptions are the two hardest detections in our sample, both exposures of the peculiar GRS 1915+105. One is in a bright, hard, jet-emitting state \cite{Klein-Wolt2002_GRS_hard_states_radio}, called $\chi$ state in \citealt[][]{Lee2002_GRS1915+105_winds_first}) in which winds signatures are normally undetected, although most $\chi$-state observations have a much higher HR (see \citealt{Neilsen2009_GRS1915+105_wind_jet_connection}). The other exposure occurred during the recent transition to a new obscured state, in which the source has spent the majority of the last few years \citep[][]{Miller2020_GRS1915+105_obscured_Chandra}. In this second observation, the observed HR is not an intrinsic property of the SED, but mostly an effect of absorption. A less conservative limit on the "soft" wind emitting states could be close to HR$_{[6-10]/[3-10]}=0.7$, when these two observations are excluded. We note that absorption line detections in
other sources than GRS 1915+105 are generally softer (HR$_{[6-10]/[3-10]}<0.5$), although this might simply be the result of a lack of
both softer GRS 1915+105 exposures and harder (but still below the previously defined threshold) observations for other sources, at least with \chandra{} and XMM.

Concerning the inclination, the 5 objects with detections of absorption lines, 4U 1630-47, GRO J1655-40, GRS 1915+105,
H 17432-322 and IGR J17451-3022, are all dippers (see Tab. \ref{table:sources}), and two (GRO J1655-40 and IGR J17451-3022) are eclipsing binaries \citep{Bailyn1995_GROJ1655-40_eclipses,Bozzo2016_IGRJ17451-3022_XMM_wind_2014}. Dipping behavior is traditionally associated to high inclination systems \citep[][]{Motta2015_inclination_dips}, and all independent inclination estimates for these 5 objects agree with values larger than 55 degrees. While estimates are too uncertain to propose this as a precise threshold, it suggests that the detection of X-ray wind signatures is restricted to the inclination range of dippers. 

On the other hand, none of the few non-dipping sources with inclination measurements below 55\degr{} show absorption lines (see right panel of Fig.~\ref{fig:glob_hid_sources}). However, the coverage of the soft state is very limited in these objects, with few sources with stringent upper limits. More importantly, not a single one of the remaining objects has a precise dynamical inclination measurement without conflicting reflection estimates. Thus, while dipping sources are definitely more prone to detection, a better coverage of low-inclined sources (and consensus on inclination estimates) would be preferable to conclude the opposite for low-inclined sources.

Finally, there are only two detections below $L_{X}\sim0.01L_{Edd}$. One is from IGR J17451, whose true Eddington ratio is highly uncertain as both its mass and distance are unknown, and the second is found in the faintest exposure of GRS 1915+105, whose luminosity is probably underestimated as it is in a semi-obscured state \citep[][]{Miller2020_GRS1915+105_obscured_Chandra}. This lack of detections below a certain luminosity threshold thus points to a certain Eddington ratio as a requirement to produce highly ionized iron absorption lines. However, our coverage of lower luminosity soft states is very limited, both in terms of number of sources and sampling. This, combined to intrinsically worse SNR (and thus a lack of constraining upper limits), prevents any definitive conclusion.

\subsection{Non detections in favourable conditions}\label{sub:results_nondet}

\begin{figure*}[h]
\includegraphics[width=0.5\textwidth]{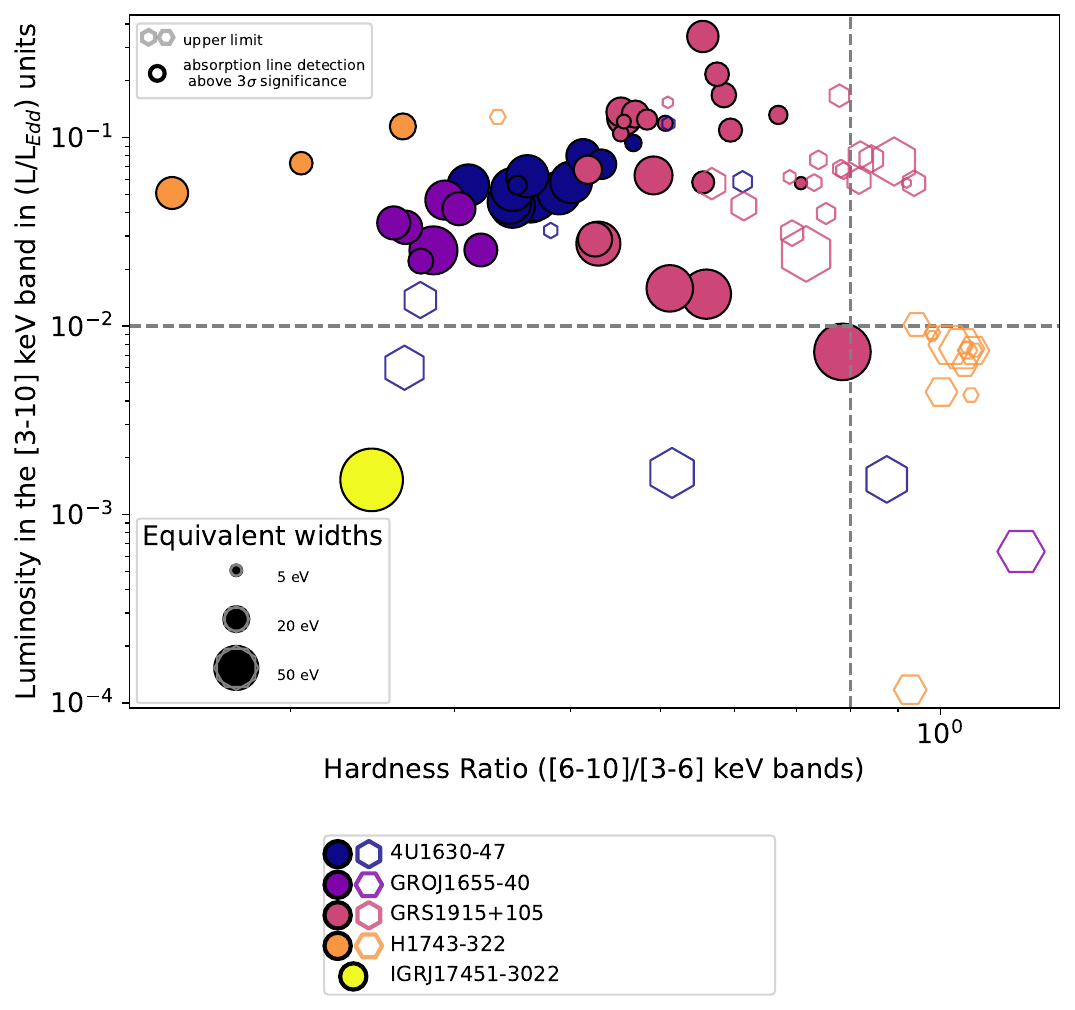}
\includegraphics[width=0.5\textwidth]{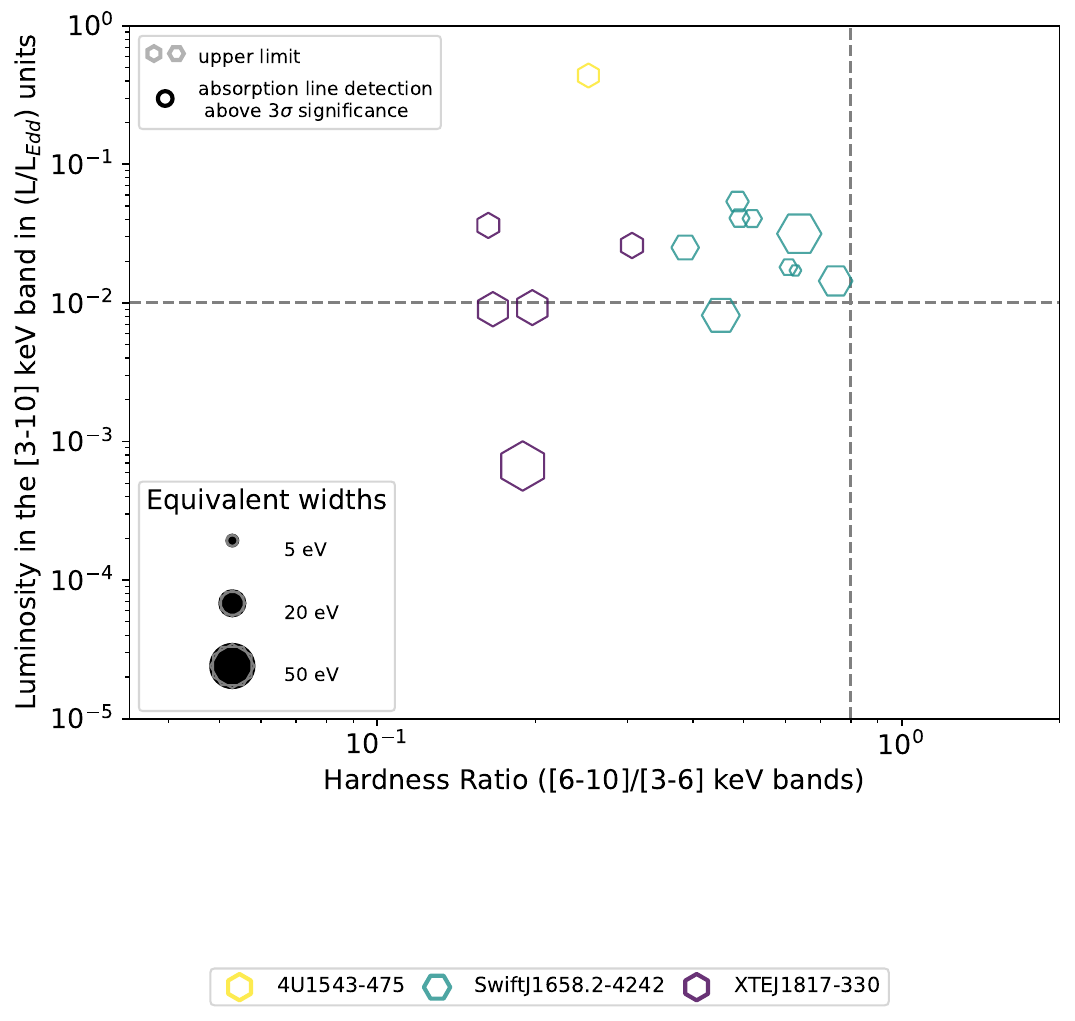}

\caption{HID diagrams of subsamples with relevant non-detections. (Left panel) Zoom on Sources with detection, and \FeKavi{} upper limits when no line is detected. (Right panel) Sources with constraining upper limits in the favourable zone, discussed in   Sect.~\ref{sub:results_nondet} }\label{fig:glob_hid_sources_details}
\end{figure*}

The presence of non-detections and stringent EW upper limits ($<$ 5 eV) in the wind-favourable region of the left panel of Fig.~\ref{fig:glob_hid_sources_withul} indicates that luminous soft states of high-inclined sources do not necessarily show absorption lines. Among the sources with detections, 4U 1630-47, GRS 1915+105, and H 17432-322 all have luminous soft state exposures without absorption lines, as can be seen in more details in the zoomed left panel of Fig.~\ref{fig:glob_hid_sources_details}. First looking at the case of GRS 1915+105, this source does not follow the standard state evolution, and instead evolves erratically in a limited part of the HID. Most of the lower EW upper limits obtained in this source concern observations with larger HR and luminosity than observations with  detection, but there is at least one observation, with HR $\sim$ 0.5, with a very stringent absorption line EW upper limit. This limit, being even lower than the absorption line EWs observed in all neighbouring detections, suggests different physical conditions for the wind between these observations, despite a similar SED. This behavior also reflects in the well known rapid variability of the lines themselves in this object (see e.g. \cite{Lee2002_GRS1915+105_winds_first,Neilsen2011_GRS1915+105_wind_heartbeat,Neilsen2020_GRS1915+105_obscured_NICER_winds_hard}). 

In the case of 4U 1630-47, there are at the least 3 exposures with stringent upper limits of 14 , 7 and 8 eV for ObsIDs 14441, 0670673201 and 15511 (see Table.  \ref{tab:obs_details} for details), with only observation 14441 being harder than the cluster of exposures with detections in this source. We note in the third observation the detection of a single, marginally significant (98.8\% significance at the Ftest), unidentified absorption feature at ~7.8kev. Finally, H 17432-322 shows a single very significant upper limit of 9 eV in ObsID 3804, which is spectrally relatively harder but remains both very soft and close (both in time and spectral distribution) to the 3 other detections in its 2004 outburst.

On the other hand, it is also important to assess whether non detection in other dipper/high-inclined sources in the favourable zone are constraining. To aid readability, we highlight the 3 sources with no detection despite stringent upper limits in this zone, 4U 1543-47, Swift J1658-4242 and XTE J1817-330, in the right panel of Fig.~\ref{fig:glob_hid_sources_details}. For 4U 1543-47, it is possible that the lack of lines is due to over-ionisation stemming from the extreme luminosity of this source, which is the brightest observed in our sample at $L_{X}/L_{Edd}\sim 0.45$ . We note that the bolometric luminosity of this source is expected to have surpassed the Eddington limit at the peak of its outburst, as seen by NICER and \nustar{} \citep{Prabhakar2023_4U1543_wind_x_soft_2021_nustar}. Another explanation would be that the peculiar dips detected in the source \citep[][]{Park2004_4U1543-47_dips} are not a consequence of high inclination. This would reconcile the geometry with the very low angle inferred from dynamical measurements \citep[][]{Orosz1998_4U1543-475_D,Orosz2003_4U1543-475_i_mass}, and the optical features reminiscing of low inclination recently detected in this source \citep[][]{Sanchez-Sierras2023_4U1543-47_optical_possible_outflow_lowi}. This would explain the lack of absorption lines. 

The same could be said for XTE J1817-330, which has a few stringent absorption line EW upper limits but no inclination constraints and lacks an actual mass estimate. It is worth noting that this source was even reported as low-inclined in previous works \citep[][]{Ponti2012_ubhw}, but lacks proper inclination measurements, and comparisons of its outburst evolution identify it to sources with mid to high inclination measurements \citep[][]{Munoz-Darias2013_HIDcolor_i}, in agreement with reports of erratic dips \citep[][]{Sriram2012_XTEJ1817-330_dips}. Finally, Swift J1658-4242, the only source with clear dipping behavior and no contradictory inclination measurement, shows a range of exposures with stringent upper limits at HID positions very close to detections in other sources. However, the lack of constraints on both its mass and distance prevents any definitive conclusion. Moreover, all constraining exposures are XMM observations with strong relativistic emission in the iron band, which are very complex to disentangle from possible absorption features and could completely hide a weak wind signature given XMM's limited spectral resolution.

\begin{table*}[h!]
\caption[blabla]{\raggedright Details of accretion states with reports of absorption line detection in both our work and the literature.}\label{table:sources_det_states}
\fontsize{10}{10}
\begin{center}
\begin{tabular}{c || c || c | c }

\hline
\hline
     Source
     & \multicolumn{3}{c}{accretion states with absorption lines reported}
     \T \B \\

\hline

     & this work
     & \multicolumn{2}{c}{other works}
     \T \B \\

\hline
     & iron band
     & iron band 
     & other energies 
     \T \B \\
\hline
\hline

\textbf{4U 1543-47} 
& X
& \textit{soft} \labelcref{ref_source_state:4U1543-47_winds_x_soft}
& X  \T \B \\

\textbf{4U 1630-47} 
& soft 
& soft \labelcref{ref_source_state:4U1630-47_winds} 
& soft$^{X}$ \labelcref{ref_source_state:4U1630-47_winds_softX}  \T \B \\

EXO 1846-031 
& X
& \textit{hard}\labelcref{ref_source_state:EXO1846-031_winds_hard_i} 
& X \T \B \\

\textbf{GRO J1655-40}
& soft 
& soft\labelcref{ref_source_state:GROJ1655-40_winds} 
& soft$^X$\labelcref{ref_source_state:GROJ1655-40_winds_softX} \T \B \\

GRS 1716-249 
& X
& X 
& hard$^V$\labelcref{ref_source_state:GRS1716-249_winds_hard_optical} \T \B \\

GRS 1758-258 
& X
& hard\labelcref{ref_source_state:GRS1758-258_winds_hard_COSPAR} 
& X \T \B \\

\textbf{GRS 1915+105}
& soft,hard
& soft:$\phi,\gamma,\rho,\beta$\labelcref{ref_source_state:GRS1915+105_wind_soft_phi_gamma_rho_beta},$\theta$\labelcref{ref_source_state:GRS1915+105_wind_soft_theta},$\kappa$\labelcref{ref_source_state:GRS1915+105_wind_soft_kappa},$\lambda$\labelcref{ref_source_state:GRS1915+105_wind_soft_lambda},hard:$\chi$\labelcref{ref_source_state:GRS1915+105_wind_hard_chi},obscured*\labelcref{ref_source_state:GRS1915+105_obscured_NICER_winds_hard}

& soft$^X$:$\phi$\labelcref{ref_source_state:GRS1915+105_wind_soft_phi_softX},obscured*:hard$^{IR}$\labelcref{ref_source_state:GRS1915+105_wind_obscured_IR} \T \B \\

GX 339-4 
& X
& X 
& soft$^V$\labelcref{ref_source_state:GX339-4_winds_hard_soft_visible},hard$^V$\labelcref{ref_source_state:GX339-4_winds_hard_soft_visible} \T \B \\

\textbf{H 1743-322}
& soft
& soft\labelcref{ref_source_state:H1743-322_winds_dips} 
& X \T \B \\

\textbf{IGR J17091-3624}
& X
& soft \labelcref{ref_source_state:IGRJ17091_winds},\textit{hard}\labelcref{ref_source_state:IGRJ17091_winds_hard_i_low}$\dagger$ 
& hard$^{X}$\labelcref{ref_source_state:IGRJ17091_absorber_static_hard_XMM_2016} \T \B \\

\textbf{IGR J17451-3022}
& soft 
& soft\labelcref{ref_source_state:IGRJ17451-3022_wind_dips_Suzaku_2014} 
& soft$^X$\labelcref{ref_source_state:IGRJ17451-3022_wind_dips_Suzaku_2014} \T \B \\

\textbf{MAXI J1305-704}
& X
& \textit{soft}\labelcref{ref_source_state:MAXIJ1305-704_wind_soft_soft+softX}\labelcref{ref_source_state:MAXIJ1305-704_wind_soft_hard_soft+softX_dip},\textit{hard} \labelcref{ref_source_state:MAXIJ1305-704_wind_soft_hard_soft+softX_dip} 
& soft$^X$\labelcref{ref_source_state:MAXIJ1305-704_wind_soft_soft+softX}\labelcref{ref_source_state:MAXIJ1305-704_wind_soft_hard_soft+softX_dip},hard$^X$\labelcref{ref_source_state:MAXIJ1305-704_wind_soft_hard_soft+softX_dip} \T \B \\

MAXI J1348-630 
& X
& \textit{soft}\labelcref{ref_source_state:MAXIJ1348-630_i_winds_soft_hard}, \textit{hard}\labelcref{ref_source_state:MAXIJ1348-630_i_winds_soft_hard}
& hard$^X$\labelcref{ref_source_state:MAXIJ1348-630_wind_hard_softX},soft$^{IR}$\labelcref{ref_source_state:MAXIJ1348-630_winds_optical},hard$^{V,IR}$\labelcref{ref_source_state:MAXIJ1348-630_winds_optical} \T \B \\

\textbf{MAXI J1803-298}
& X
& soft\labelcref{ref_source_state:MAXIJ1803-298_winds_soft}\labelcref{ref_source_state:MAXIJ1803-298_winds_xray_soft_Nustar}
& hard$^V$\labelcref{ref_source_state:MAXIJ1803-298_winds_hard_optical} \T \B \\

\textbf{MAXI J1820+070}
& X
& soft\labelcref{ref_source_state:MAXIJ1820+070_wind_xray_soft} 
& soft$^{IR}$\labelcref{ref_source_state:MAXIJ1820+070_winds_hard_visible},hard$^{V,IR}$\labelcref{ref_source_state:MAXIJ1820+070_winds_hard_visible} \T \B \\

\textbf{Swift J1357.2-0933}
& X
& X 
& hard$^V$\labelcref{ref_source_state:SwiftJ1357.2-0933_winds_visible_1}\labelcref{ref_source_state:SwiftJ1357.2-0933_winds_visible_2} \T \B \\

\textbf{Swift J1658.2-4242}
& X
& \textit{hard}\labelcref{ref_source_state:SwiftJ1658.2_i_winds} 
& X \T \B \\

V404 Cyg
& X
& hard \labelcref{ref_source_state:GS2023+338_wind_hard} 
& obscured*: hard$^X$ \labelcref{ref_source_state:GS2023+338_wind_hard_softX},hard$^V$\labelcref{ref_source_state:GS2023+338_wind_hard_visible} \T \B \\

V4641 Sgr
& X
& X 
& obscured*: hard$^V$\labelcref{ref_source_state:SAXJ1819.3-2525_winds_hard_optical}\T \B \\

XTE J1652-453 
& X
& \textit{hard}$\dagger$\labelcref{ref_source_state:XTEJ1652-453_i_winds_sigma}  
& X \T \B \\
\end{tabular}
\end{center}
\small
Notes: Bold source names indicate dippers. For "other" energies, $X$ superscripts indicate softer X-ray detections, $V$ visible and $IR$ infrared detections. Accretion states are reported in $italic$ for absorption lines embedded in reflection components. For all purposes, $\dagger$ indicates low significance detections. $\ast$ The observed HR value of the obscured state might not reflect the actual HR of the source. The list of reference papers is not exhaustive for objects with many wind detections. References:\\
\tiny

\begin{enumerate*}[label=\arabic{enumi}]

    %4U 1543-47
    \item \citep[][]{Prabhakar2023_4U1543_wind_x_soft_2021_nustar}\label{ref_source_state:4U1543-47_winds_x_soft}
    
    %4U 1630-47
    \item \citep[][]{Kubota2007_4U1630-47_winds}\label{ref_source_state:4U1630-47_winds}
    \item \citep[][]{Trueba2019_4U1630-47_wind_2012-13Chandra}\label{ref_source_state:4U1630-47_winds_softX}
    
    %EXO 1846-031
    \item \citep[][]{Wang2020_EXO1846-031_winds_hard_i}\label{ref_source_state:EXO1846-031_winds_hard_i}
    
    %GRO J1655-40
    \item \citep[][]{Miller2006_GROJ1655-40_winds}\label{ref_source_state:GROJ1655-40_winds}
    \item \citep[][]{Miller2008_GROJ1655-40_wind_Chandra_2005_detailed} \label{ref_source_state:GROJ1655-40_winds_softX}
    
    %GRS 1716-249
    \item \citep[][]{Cuneo2020_GRS1716-249_winds_hard_optical}\label{ref_source_state:GRS1716-249_winds_hard_optical}
    
    %GRS 1758-258
    \item \citep[][]{Reynolds2018_GRS1758-258_winds_hard_COSPAR}\label{ref_source_state:GRS1758-258_winds_hard_COSPAR}

    %GRS 1915+105
    %wind detections
    \item \citep[][]{Neilsen2009_GRS1915+105_wind_jet_connection}\label{ref_source_state:GRS1915+105_wind_soft_phi_gamma_rho_beta}
    \item \citep[][]{Ueda2010_GRS1915+105_wind_suzaku_class_change}\label{ref_source_state:GRS1915+105_wind_soft_theta}
    \item \citep[][]{Liu2022_GRS1915+105_HMXT_wind_soft_2017}\label{ref_source_state:GRS1915+105_wind_soft_kappa}
    \item \citep[][]{Neilsen2018_GRS1915+105_wind_NICER}\label{ref_source_state:GRS1915+105_wind_soft_lambda}
    \item \citep[][]{Lee2002_GRS1915+105_winds_first}\label{ref_source_state:GRS1915+105_wind_hard_chi}
    \item \citep[][]{Ueda2009_GRS1915+105_wind_2007}\label{ref_source_state:GRS1915+105_wind_soft_phi_softX}
    \item \citep[][]{Neilsen2020_GRS1915+105_obscured_NICER_winds_hard}\label{ref_source_state:GRS1915+105_obscured_NICER_winds_hard}

    \item \citep[][]{Sanchez-Sierras2023_GRS1915_winds_IR}
    \label{ref_source_state:GRS1915+105_wind_obscured_IR}
    %GX339-4
    \item \citep[][]{Rahoui2014_GX339-4_wind_visible_soft_hard}\label{ref_source_state:GX339-4_winds_hard_soft_visible}
    
    %H1743-322
    \item \citep[][]{Miller2006_H1743-322_winds}\label{ref_source_state:H1743-322_winds_dips}

    %IGR J17091-3624
    %soft winds
    \item \citep[][]{King2012_IGRJ17091_winds}\label{ref_source_state:IGRJ17091_winds}
    %hard wind
    \item \citep[][]{Wang2018_IGRJ17091_winds_hard_i_low}\label{ref_source_state:IGRJ17091_winds_hard_i_low}
    \item \citep[][]{Gatuzz2020_IGRJ17091-3624_absorber_static_XMM_2016}\label{ref_source_state:IGRJ17091_absorber_static_hard_XMM_2016}
    
    %IGR J17451-3022
    \item \citep[][]{Jaisawal2015_IGRJ17451-3022_wind_dips_Suzaku_2014}\label{ref_source_state:IGRJ17451-3022_wind_dips_Suzaku_2014}

    %MAXI J1305-704
    \item \citep[][]{Miller2014_MAXIJ1305-704_wind}\label{ref_source_state:MAXIJ1305-704_wind_soft_soft+softX}
    \item \citep[][]{Shidatsu2013_MAXIJ1305-704_wind_soft_hard_Swift_Suzaku}\label{ref_source_state:MAXIJ1305-704_wind_soft_hard_soft+softX_dip}
    
    %MAXIJ1348-630
    \item \citep[][]{Chakraborty2021_MAXIJ1348-630_i_winds_soft_hard}\label{ref_source_state:MAXIJ1348-630_i_winds_soft_hard}
    \item \citep[][]{Saha2021_MAXIJ1348-630_wind_hard_softX_2021}\label{ref_source_state:MAXIJ1348-630_wind_hard_softX}
    \item \citep[][]{Panizo-Espinar2022_MAXIJ1348-630-winds_optical}\label{ref_source_state:MAXIJ1348-630_winds_optical}
    
    %MAXI J1803-298
    \item \citep[][]{Miller2021_MAXIJ1803-298_wind_Swift_2021}\label{ref_source_state:MAXIJ1803-298_winds_soft}
    \item \citep[][]{Coughenour2023_MAXIJ1803-298_wind_xray_soft_Nustar}
    \label{ref_source_state:MAXIJ1803-298_winds_xray_soft_Nustar}
    \item \citep[][]{MataSanchez2022_MAXIJ1803-298_winds_hard_optical}\label{ref_source_state:MAXIJ1803-298_winds_hard_optical}
    
    %MAXIJ1820+070
    \item \citep[][]{Fabian2021_MAXIJ1820+070_wind_xray_soft}\label{ref_source_state:MAXIJ1820+070_wind_xray_soft}
    \item \citep[][]{Munoz-Darias2019_MAXIJ1820+070_winds_hard_visible}\label{ref_source_state:MAXIJ1820+070_winds_hard_visible}

    %Swift J1357.2-0933
    \item \citep[][]{Jimenez-Ibarra2019_SwiftJ1357.2-0933_winds_visible_1}\label{ref_source_state:SwiftJ1357.2-0933_winds_visible_1}
    \item \citep[][]{Charles2019_SwiftJ1357.2-0933_winds_visible_2}\label{ref_source_state:SwiftJ1357.2-0933_winds_visible_2}
    
    %Swift J1658.2-4242
    \item \citep[][]{Xu2018_SwiftJ1658.2_i_winds}\label{ref_source_state:SwiftJ1658.2_i_winds}

    %V404 Cyg (GS 2023+338)
    \item \citep[][]{Munoz-Darias2022_GS2023+338_wind_xray-optical_2015_details}\label{ref_source_state:GS2023+338_wind_hard}
    \item \citep[][]{King2015_GS2023+338_wind_x}\label{ref_source_state:GS2023+338_wind_hard_softX}
    \item \citep[][]{Munoz-Darias2016_GS2023+338_wind_hard_visible}\label{ref_source_state:GS2023+338_wind_hard_visible}

    %V 4641 Sgr (SAX J1819.3-2525)
    \item \citep[][]{Munoz-Darias2018_SAXJ1819.3-2525_winds_hard_optical}\label{ref_source_state:SAXJ1819.3-2525_winds_hard_optical}

    %XTE J1652-453
    \item \citep[][]{Chiang2012_XTEJ1652-453_i_winds_2sigma}\label{ref_source_state:XTEJ1652-453_i_winds_sigma}

\end{enumerate*}

\end{table*}

\section{Concluding remarks}\label{sec:discussion}

Our present study of \ion{Fe}{xxv} and \ion{Fe}{xxvi} absorption lines, in all publicly available XMM-pn and \chandra{}-HETG observations of BHLMXB candidates, gives results in good agreement with previous findings. All the wind signatures we found occur in luminous ($L_{X}>0.01L_{Edd}$) soft states (hardness ratio 
HR$_{[6-10]/[3-10]}<0.8$) of 5 dippers, 4U 1630-472, GRO J1655-40, GRS 1915+105, H 1743-322 and IGR J17451-3022. Existing inclination measurements are consistent with this behavior, with i>55°  in these 5 sources.

With the \chandra{} instrument, which proves to be the only one sufficiently precise to reliably measure the outflow velocity, the absorption signatures show a global trend of very small blueshifts. Indeed, the velocity shifts of our sample are of the order of minus a few hundreds of km s$^{-1}$, with a mean of $-200\pm{60}$ km s$^{-1}$. Moreover, only one detection (in GRO J1655-40) is significantly ($>$ 2 $\sigma$) below -1000 km s$^{-1}$. These values, although closer to the limits of HETG's absolute wavelength accuracy, remain consistent with past publications, and in particular with velocity shift measurements in lower energy lines (compared to \ion{Fe}{xxv} and \ion{Fe}{xxvi}) where HETG's accuracy is more well-studied (see e.g. \citealt{Ueda2009_GRS1915+105_wind_2007,Trueba2019_4U1630-47_wind_2012-13Chandra} and references therein). Other works claiming higher blueshift values make use of more complex fits using several photoionization models (see e.g. \citealt{Miller2015_HETG_analysis_multi}) and as such should not be directly compared to our results, although the main ionization zones generally remain in agreement with our findings.

We also get good constraints on a few line widths, with FWHMs of the order of a few thousands of km s$^{-1}$ for the broadest ones. The observed correlation between the line widths and \FeKav{} EW naturally arises in the presence of significant turbulence velocity in the wind, of the order of thousands km s$^{-1}$ when assuming a simple slab geometry (see  Sect.~\ref{sub:diff_line}). Reality is expected to be more complex, possibly with a radial distribution of density and velocity. A more precise modeling is certainly needed to better characterize the amount of turbulence.

We detect a very significant anti-correlation between the X-ray luminosity (in Eddington units) and the line EW in the case of \ion{Fe}{xxv} while no significant correlation is observed in the case of \ion{Fe}{xxvi}. This anti-correlation is present in single objects with multiple line detections but also in the entire set of sources showing absorption lines.  Although already found in the past in more restricted datasets \citep[][]{Miller2020_GRS1915+105_obscured_Chandra,Ponti2012_ubhw}, such a correlation observed in a sample of different sources would suggest a similar wind structure (i.e., a similar $nR^2$ factor) from source to source, at a given $L_{X}/L_{Edd}$. This anti-correlation would then be expected if the wind ionization is on average above the peak of the ionic fraction for \FeKav{}. While it predicts quite large \FeKav{} EWs ($\sim$ 100 eV) below our threshold of $0.01L_{Edd}$, the ionization at these luminosities could also go beyond the peak of \FeKav{} ion fraction, and shift to producing weaker lines from less-ionized ions. If this is not the case, the lack of detection at low flux may also be due to lower statistics or a more sparsely coverage, but could also be related to the physical processes producing the wind (e.g., thermal driven wind requiring high illuminating luminosity, \citealt{Done2018_thermal_winds_modeling_H1743_GROJ1655,Tomaru2019_H1743-322_radiathermalwind}).

The absence of \ion{Fe}{xxv} and \ion{Fe}{xxvi} absorption line detection in virtually all hard states in our sample agrees with recent theoretical studies suggesting that the ionisation range compatible with these ions could be thermally unstable when the gas is illuminated by a hard state SED  (e.g., \citealt{Chakravorty2013_thermal_stability,Chakravorty2016_JED-SAD_warm_wind_thermal_stability,Bianchi2017_stability_NS,Petrucci2021_outburst_wind_stability}). Thus, even if the wind itself were present, it could not be detectable through \ion{Fe}{xxv} and \ion{Fe}{xxvi} absorption lines. 

However, there have been recent reports in the literature of a few absorption line detections in hard states of different sources, as shown in Tab.~\ref{table:sources_det_states}, in which we list the reports of absorption lines in all wavebands and associated accretion states for sources in the sample. However, we must stress that the vast majority of these detections come from \nustar{} spectra blended with reflection. The limited spectral resolution of this instrument, combined with the model-dependent nature of the residuals of reflection components, means that special care should be put into computing the significance of these lines, especially when different reflection models disagree on their existence (see e.g. \citealt{Chakraborty2021_MAXIJ1348-630_i_winds_soft_hard} and \citealt{Jia2022_MAXIJ1348-630_refl_2021_nowind} for MAXIJ1348-630). In the meantime, other reports are either not well documented \citep[][]{Reynolds2018_GRS1758-258_winds_hard_COSPAR,Saha2021_MAXIJ1348-630_wind_hard_softX_2021}, or are associated to static or infalling material \citep[][]{Shidatsu2013_MAXIJ1305-704_wind_soft_hard_Swift_Suzaku}, and the only clear iron band hard state detections come from non-standard states of GRS 1915+105 and V404 Cyg \citep[][]{Lee2002_GRS1915+105_winds_first,Munoz-Darias2022_GS2023+338_wind_xray-optical_2015_details}.

The lack of standard X-ray detections in the hard state is still compatible with the increasing number of optical and infrared absorption lines detections in hard states seen in Table~\ref{table:sources_det_states}, which suggest that the outflow persists independently of the spectral states (see \citealt{Panizo-Espinar2022_MAXIJ1348-630-winds_optical} and references therein).  
They arise from the same category of high-inclined (mostly dipping) sources, except in the case of GX339-4, and provide different and complementary views of the outflow: visible lines are restricted to hard states while infrared detections have been obtained in the whole outburst. However, these detections generally have blueshifts in the range of few 1000 km s$^{-1}$, significantly higher than in X-rays. More critically, only two sources have clear reports of detections both in the X-rays and in the optical/infrared, V404 Cyg and MAXIJ1803-298. As of now, only the first one has been studied in detail, and shows properties consistent with being produced by the same outflowing material \citep[][]{Munoz-Darias2022_GS2023+338_wind_xray-optical_2015_details}, although in an obscured state with extremely strong emission lines and with short-term variability of absorption features in the iron band, which prevented the detection of absorption lines with our simple procedure.

It is difficult to assess whether the lack of common X-ray and optical/infrared absorption line detections is meaningful: in our study, the vast majority of sources with these features have very poor X-ray coverage in the favourable region. However, several objects have been extensively followed by other X-ray telescopes, such as MAXIJ1820+070 with NICER, with only a single tentative report of X-ray absorption detections up to now \citep[][]{Fabian2021_MAXIJ1820+070_wind_xray_soft}. On the other hand, the sources with X-ray detections in our sample lack either the optical counterpart or the high-quality optical data necessary to search for absorption lines. It is also possible that the physical conditions favoring X-ray and optical wind signatures do not perfectly match (see e.g., \citealt{Koljonen2023_MAXIJ1820+070_optical_modeling_wind_atmospheres_thermal}), but more optical/X-ray monitoring are required to conclude.\\

The results of this paper show that we can only put limited constraints on the evolution of the absorption lines with the current scarce sampling of each outburst. In this regard, the use of the new generation of telescopes with better monitoring capabilities, such as NICER, or of the next evolution of spectrometers, such as XRISM and Athena, will be paramount to separate the outflow evolution from the influence of the SED. We are currently performing a similar analysis on the NICER archive, which remains for the most part unpublished.

Through the analysis of the line parameters and HID positions, we also highlight some of the most critical exposures currently available, with well constrained and extreme or variable wind signatures that should be compared against existing and upcoming wind models.
In order to improve the current lack of coupling between disk and wind modeling, our next work will compare joint continuum and magnetic wind solutions arising from the JED-SAD framework \citep[][]{Jacquemin-Ide2019_wind_weak_magnetic_JEDSAD_modeling} to these datasets.
 
Finally, this work has not delved into the details of the behavior of each source. Although the results are directly available through the visualisation tool, we will address the most interesting sources individually in a follow-up paper. We will both compare their behavior with the global sample and highlight notable results in unpublished observations.

\begin{acknowledgements}
Part of this work has been done thanks to the financial supports from CNES and the French PNHE. SB and MP acknowledge support from PRIN MUR 2017 ``Black hole winds and the baryon life cycle of galaxies: the stone-guest at the galaxy evolution supper''. SB acknowledges support from the European Union Horizon 2020 Research and Innovation Framework Programme under grant agreement AHEAD2020 n. 871158. 
GP acknowledges financial support from the European Research Council (ERC) under the European Union’s Horizon 2020 research and innovation program “HotMilk” (grant agreement No. 865637) and support from Bando per il Finanziamento della Ricerca Fondamentale 2022 dell’Istituto Nazionale di Astrofisica (INAF): GO Large program.
This works uses data obtained from the \chandra{} Data Archive and software (CIAO and TGCat) provided by the\chandra{} X-ray Center (CXC), as well as data obtained through the HEASARC Online Service, provided by the NASA/GSFC, in support of NASA High Energy Astrophysics Programs. We especially thank the \chandra{}, TGCat and XMM helpdesks for their help and availability. 
\end{acknowledgements}
%\end{twocolumn}
% WARNING
%-------------------------------------------------------------------
% Please note that we have included the references to the file aa.dem in
% order to compile it, but we ask you to:
%
% - use BibTeX with the regular commands:
%   \bibliographystyle{aa} % style aa.bst
%   \bibliography{Yourfile} % your references Yourfile.bib
%
% - join the .bib files when you upload your source files
%-------------------------------------------------------------------

\bibliographystyle{aa}
\bibliography{biblio_windreview}

\begin{appendix}

\section{Visualisation tool}\label{sec:visualisation}

One of the secondary goal of this work is to complement current Black Hole candidates catalogs, which only list the physical parameters of the sources, with an inventory of the absorption features properties accessible with X-ray telescopes. In order to combine ease of access and visualisation of the data, we built an interactive webpage with the python library streamlit\footnote{\href{https://streamlit.io/}{https://streamlit.io/}}, accessible at \href{https://visual-line.streamlit.app/}{https://visual-line.streamlit.app/}. The dataset is loaded internally and the options chosen in the sidebar allow to navigate and display different information of any subsamples of the data and download results. All of the figures present in this paper except Fig.~\ref{fig:autofit_example} can be recreated using the online tool.\\

The main options of the tool are as follow:\\

\begin{enumerate*}
    \item \textbf{Sample selection:} The first options in the sidebar allow the user to restrict the data selection to any part of the sample. This can be achieved by manually selecting a subset of sources or via global constraints on inclination properties, using the values listed in Tab. \ref{table:sources}. Other options include the absorption lines to be considered, a time interval restriction, and the choice of significance threshold for features to be considered as detections (which uses the assessment of   Sect.~\ref{sub:sign_fakes}).\\
    
    \item\textbf{Hardness Intensity Diagram:} The main visualisation tool is the HID diagram in which both detections and non detections are displayed. Exposures can be colored according to several line parameters (in which case only extremal values are displayed for exposures with several lines), and several parameter specific to each observation or source. The fitting errors of both HID parameters can be displayed, and upper limits can be plotted for non detection, with different symbols in order to aid visibility for large subsamples, along with a range of other visualisation options.\\

    \item\textbf{Monitoring:} Whenever the sample selection is restricted to a single source, long-term lightcurves and HR evolution can be displayed using both RXTE-ASM and MAXI data with a 1-day binning. RXTE data is taken from a copy of the definitive products available at \href{http://xte.mit.edu/ASM_lc.html}{http://xte.mit.edu/ASM\_lc.html}. RXTE lightcurves use the sum of the intensity in all bands ($[1.5-12]$kev), corrected by a factor of 25 to match (visually) MAXI values, and HR values are built as the ratio of bands C and B+A, i.e. $[5.5-12]/[1.5-5]$kev. MAXI data is loaded on the fly from the official website at \href{http://maxi.riken.jp/top/slist.html}{http://maxi.riken.jp/top/slist.html}, in order to use the latest dataset available. MAXI lightcurves use the full $[2-20]$kev band, and the HR is built from the  $[4-10]/[2-4]$ kev bands.\\
    A transparency factor proportional to the quality of the data (estimated from the ratio of the HR values to their uncertainties) is applied to both HRs to aid visibility, and the dates of exposures with instruments used in the line detection sample are highlighted. The date restriction selected in the sample selection can be both highlighted and used to zoom the lightcurve display, while EW values and upper limits can be displayed on a secondary axis at the date of each exposure.\\

    \item\textbf{Parameter analysis:}
    The distribution and correlation of line parameters can be computed on the fly from the chosen data selection. Distributions are restricted to the main line parameters, and can be stacked/split according to sources and instruments. Scatter plots between various intrinsic parameters, as well as observation-level and source-level parameters can be displayed, with p-values computed according to the perturbation method discussed in   Sect.~\ref{sub:param_distrib}. Similarly to the HID, scatter plots can be color-coded according to various informations, and EW upper limits for currently selected sources can be included in the relevant plots, along with other secondary options.\\

    \item\textbf{Data display and download:} 
    the complete data of sources, observation and line parameters are displayed according to the current selection, and can be downloaded through separate csv file which can be loaded as multi-dimensional dataframes. 
    
\end{enumerate*}

\pagebreak
\begin{onecolumn}
\section{Results of the fitting procedure for the low luminosity observations of GRS 1915+105}\label{sec:appendix_GRS}

\begin{figure*}[h]
\centering
\includegraphics[width=1.\textwidth]{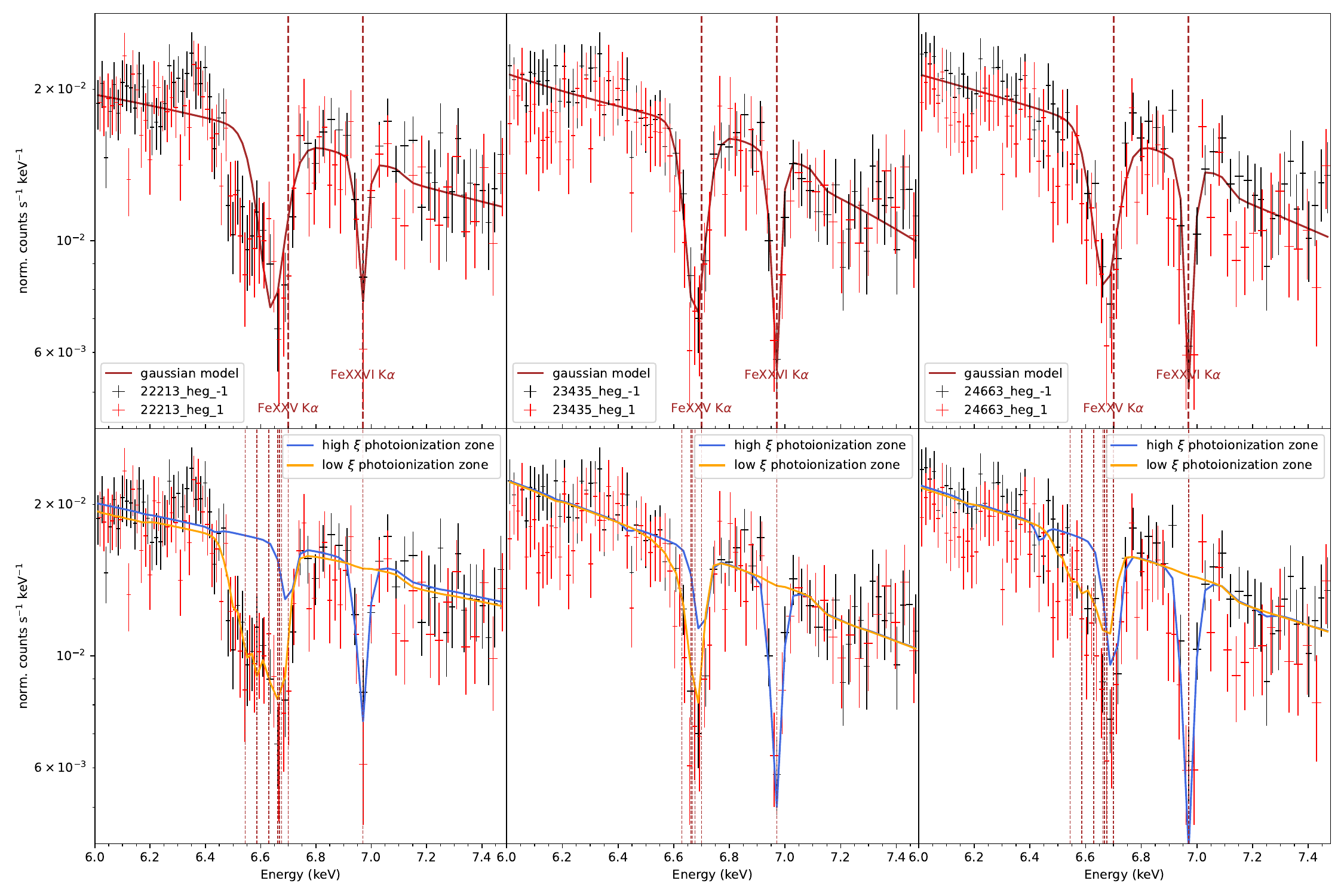}
\caption{Zoom of the fit around the K$\alpha$ lines for the 3 low-luminosity observations of GRS 195+105+105. The upper panels show the result of the autofit procedure with gaussians, and the lower panels a fit with two photoionization zones, the lower $\xi$ being fixed at zero velocity. The dashed lines show the energy of the main lines affecting the photoionization zones of each observation.}\label{fig:fit_GRS}
\end{figure*}

\section{Results of the line detection procedure for exposures analyzed in this work}\label{sec:table_exposures}

\subsection{Detections in all exposures}\label{tab:obs_details}

\medskip
Table C.1: List of exposures analyzed in this work ordered alphabetically and chronologically,  exposure time after data reduction, and either EWs results for detections or EWs upper limits for non detections of the main lines. Line EWs are only provided for detections above 3 $\sigma$ significance (see Sect.~\ref{sub:sign_fakes}), along with 90\% uncertainties. Upper limits above 100 eV are not reported. ObsIDs marked with a $\dagger$ have pile-up values between 5 and 7\% after the data reduction process.
\tiny

\centering
\begin{longtable}[h]{c c || c c c || c c c c c}

\hline 
\hline
\multirow{2}{*}{Source}
& \multirow{2}{*}{Date}
& \multirow{2}{*}{Instrument}
& \multirow{2}{*}{ObsID + identifier }
& \multirow{2}{*}{exp. time (ks)}
& \multicolumn{5}{c}{Fe line Equivalent Width / 3 $\sigma$ upper limit (eV)}
\T \B \\

& 
& 
& 
& 
& xxvK$\alpha$
& xxviK$\alpha$
& xxvK$\beta$
& xxviK$\beta$
& xxviK$\gamma$
\T \B \\
\hline
\hline
\endfirsthead
\caption{continued}\\
\hline
\hline
\multirow{2}{*}{Source}
& \multirow{2}{*}{Date}
& \multirow{2}{*}{Instrument}
& \multirow{2}{*}{ObsID + identifier }
& \multirow{2}{*}{exp. time (ks)}
& \multicolumn{5}{c}{Fe line Equivalent Width / 3 $\sigma$ upper limit (eV)}
\T \B \\

& 
& 
& 
& 
& xxvK$\alpha$
& xxviK$\alpha$
& xxvK$\beta$
& xxviK$\beta$
& xxviK$\gamma$
\T \B \\
\hline
\hline
\endhead

\multirow{11}{*}{1E1740.7-2942}&2000-09-15&XMM&0112970901\_S003&10.88&$\leq32$&$\leq69$&$\leq57$&$\leq57$&$\leq66$\T \B \\ 
&2000-09-21&XMM&0112970801\_S003&17.25&$\leq28$&$\leq28$&$\leq41$&$\leq46$&$\leq51$\T \B \\ 
&2001-04-01&XMM&0112971801\_S003&8.82&$\leq29$&$\leq39$&$\leq26$&$\leq28$&$\leq32$\T \B \\ 
&2001-09-14&Chandra&2491&61.16&$\leq37$&$\leq45$&/&/&/\T \B \\ 
&2003-09-11&XMM&0144630101\_S003&5.93&$\leq30$&$\leq30$&$\leq40$&$\leq41$&$\leq44$\T \B \\ 
&2005-10-02&XMM&0303210201\_S003&16.46&$\leq20$&$\leq21$&$\leq28$&$\leq29$&$\leq32$\T \B \\ 
&2012-04-03&XMM&0673550201\_S003&93.51&$\leq8$&$\leq5$&$\leq9$&$\leq9$&$\leq9$\T \B \\ 
\hline
\multirow{1}{*}{4U1543-475}&2021-06-21&Chandra&25079&4.59&$\leq15$&$\leq17$&$\leq57$&$\leq59$&$\leq90$\T \B \\ 
\hline
\multirow{30}{*}{4U1630-47}&2004-08-04&Chandra&4568&49.99&$\leq4$&\textbf{11}$_{-4}^{+3}$&$\leq19$&$\leq76$&/\T \B \\ 
&2012-01-17&Chandra&13714&28.92&$\textbf{32}\pm4$&$\textbf{57}\pm5$&/&/&/\T \B \\ 
&2012-01-20&Chandra&13715&29.28&\textbf{34}$_{-5}^{+3}$&\textbf{49}$_{-5}^{+4}$&\textbf{23}$_{-5}^{+7}$&/&/\T \B \\ 
&2012-01-26&Chandra&13716&29.28&\textbf{47}$_{-2}^{+3}$&\textbf{52}$_{-3}^{+1}$&\textbf{32}$_{-2}^{+11}$&\textbf{38}$_{-6}^{+9}$&/\T \B \\ 
&2012-01-30&Chandra&13717&29.44&$\textbf{30}\pm3$&$\textbf{48}\pm4$&\textbf{32}$_{-7}^{+14}$&\textbf{35}$_{-11}^{+10}$&\textbf{36}$_{-8}^{+13}$\T \B \\ 
&2012-03-04&XMM&0670671501\_S003&2.54&$\textbf{35}\pm7$&\textbf{55}$_{-7}^{+9}$&$\leq34$&/&/\T \B \\ 
&2012-03-04&XMM&0670671501\_U014&69.86&$\textbf{31}\pm2$&\textbf{48}$_{-2}^{+1}$&\textbf{21}$_{-1}^{+3}$&\textbf{22}$_{-0}^{+3}$&/\T \B \\ 
&2012-03-20&XMM&0670671301\_S003&22.26&$\textbf{21}\pm3$&\textbf{46}$_{-2}^{+4}$&$\leq16$&\textbf{18}$_{-5}^{+6}$&/\T \B \\ 
&2012-03-25&XMM&0670672901\_S003&62.81&$\textbf{20}\pm1$&\textbf{45}$_{-2}^{+1}$&$\textbf{9}\pm2$&$\textbf{19}\pm2$&/\T \B \\ 
&2012-06-03&Chandra&14441&19.0&$\leq12$&$\leq14$&$\leq34$&$\leq42$&$\leq62$\T \B \\ 
&2012-09-09&XMM&0670673001\_S003$^\dagger$&22.48&\textbf{9}$_{-2}^{+3}$&\textbf{31}$_{-2}^{+3}$&$\leq10$&$\leq10$&$\leq10$\T \B \\ 
&2012-09-10&XMM&0670673001\_U002&0.8&$\leq19$&\textbf{25}$_{-5}^{+6}$&$\leq17$&$\leq22$&$\leq23$\T \B \\ 
&2012-09-11&XMM&0670673101\_S003&0.93&$\leq14$&\textbf{9}$_{-4}^{+5}$&$\leq19$&$\leq16$&$\leq19$\T \B \\ 
&2012-09-28&XMM&0670673201\_S003&1.56&$\leq5$&$\leq7$&$\leq9$&$\leq11$&$\leq19$\T \B \\ 
&2013-04-25&Chandra&15511&49.39&$\leq8$&$\leq7$&$\leq26$&$\leq32$&$\leq35$\T \B \\ 
&2013-05-27&Chandra&15524&48.91&$\leq44$&$\leq62$&/&/&/\T \B \\ 
&2016-10-21&Chandra&19904&30.93&\textbf{23}$_{-5}^{+4}$&\textbf{45}$_{-7}^{+4}$&/&/&/\T \B \\ 
&2020-05-28&Chandra&22376&24.5&$\leq35$&$\leq35$&$\leq75$&/&/\T \B \\ 
&2020-06-06&Chandra&22377&24.5&$\leq41$&$\leq50$&/&/&/\T \B \\ 
&2020-06-13&Chandra&22378&23.54&$\leq39$&$\leq55$&/&/&/\T \B \\ 
\hline
\multirow{9}{*}{4U1957+115}&2004-09-07&Chandra&4552&65.6&$\leq14$&$\leq16$&/&/&/\T \B \\ 
&2004-10-16&XMM&0206320101\_S003&34.15&$\leq5$&$\leq9$&$\leq10$&$\leq10$&$\leq15$\T \B \\ 
&2008-12-07&Chandra&10659&9.87&$\leq41$&$\leq58$&/&/&/\T \B \\ 
&2008-12-07&Chandra&10660&13.44&$\leq44$&$\leq48$&/&/&/\T \B \\ 
&2008-12-08&Chandra&10661&9.82&$\leq26$&$\leq42$&/&/&/\T \B \\ 
&2013-11-17&XMM&0720940101\_S003&36.89&$\leq12$&$\leq15$&$\leq20$&$\leq27$&$\leq36$\T \B \\ 
\hline
\multirow{1}{*}{AT2019wey}&2020-09-20&Chandra&24651&24.51&$\leq26$&$\leq47$&$\leq85$&/&/\T \B \\ 
\hline
\multirow{11}{*}{EXO1846-031}&2019-08-13&Chandra&21235&27.99&$\leq12$&$\leq14$&$\leq29$&$\leq40$&$\leq66$\T \B \\ 
&2019-08-28&Chandra&21236&29.95&$\leq22$&$\leq32$&$\leq59$&$\leq83$&/\T \B \\ 
&2019-09-15&XMM&0851181101\_S009&0.3&$\leq34$&$\leq47$&$\leq66$&$\leq77$&/\T \B \\ 
&2019-09-15&XMM&0851181101\_S003&13.37&$\leq17$&$\leq19$&$\leq39$&$\leq40$&$\leq56$\T \B \\ 
&2019-09-19&Chandra&21237&29.4&$\leq15$&$\leq22$&$\leq48$&$\leq65$&$\leq86$\T \B \\ 
&2019-10-19&Chandra&21238&28.54&$\leq34$&$\leq41$&/&/&/\T \B \\ 
&2019-10-31&Chandra&20899&48.77&$\leq25$&$\leq37$&$\leq95$&/&/\T \B \\ 
&2019-11-09&Chandra&20900&45.85&$\leq17$&$\leq21$&$\leq54$&$\leq62$&$\leq99$\T \B \\ 
\hline
\multirow{11}{*}{GROJ1655-40}&2005-02-27&XMM&0112921301\_S003&1.23&$\leq56$&$\leq43$&$\leq67$&/&/\T \B \\ 
&2005-03-12&Chandra&5460&24.53&$\leq8$&\textbf{19}$_{-3}^{+4}$&$\leq65$&/&/\T \B \\ 
&2005-03-14&XMM&0112921401\_S003&0.44&$\leq16$&\textbf{31}$_{-3}^{+6}$&$\leq31$&$\leq29$&$\leq46$\T \B \\ 
&2005-03-15&XMM&0112921501\_S003&0.44&\textbf{23}$_{-4}^{+5}$&\textbf{30}$_{-5}^{+6}$&$\leq25$&$\leq30$&$\leq45$\T \B \\ 
&2005-03-16&XMM&0112921601\_S003&0.44&\textbf{12}$_{-4}^{+3}$&\textbf{30}$_{-3}^{+4}$&/&$\leq20$&$\leq38$\T \B \\ 
&2005-03-18&XMM&0155762501\_S001&0.69&\textbf{30}$_{-3}^{+4}$&\textbf{41}$_{-4}^{+3}$&$\textbf{23}\pm6$&$\leq32$&$\leq19$\T \B \\ 
&2005-03-27&XMM&0155762601\_S001&0.64&$\textbf{30}\pm4$&\textbf{17}$_{-4}^{+5}$&\textbf{26}$_{-6}^{+7}$&/&/\T \B \\ 
&2005-04-01&Chandra&5461&44.4&\textbf{58}$_{-3}^{+4}$&\textbf{43}$_{-5}^{+3}$&\textbf{58}$_{-4}^{+10}$&/&/\T \B \\ 
\hline
\multirow{1}{*}{GRS1716-249}&2017-02-06&Chandra&20008&29.95&$\leq12$&$\leq15$&$\leq28$&$\leq51$&$\leq50$\T \B \\ 
\hline
\multirow{1}{*}{GRS1739-278}&2016-09-24&Chandra&17791&29.39&$\leq47$&$\leq51$&/&/&/\T \B \\ 
\hline
\multirow{6}{*}{GRS1758-258}&2000-09-19&XMM&0112971301\_S003&8.97&$\leq24$&$\leq24$&$\leq32$&$\leq35$&$\leq39$\T \B \\ 
&2001-03-22&XMM&0136140201\_S001&18.43&$\leq73$&$\leq91$&/&/&/\T \B \\ 
&2002-03-18&Chandra&2750&26.47&$\leq51$&$\leq82$&/&/&/\T \B \\ 
&2002-09-28&XMM&0144630201\_S003&5.94&$\leq25$&$\leq26$&$\leq32$&$\leq37$&$\leq41$\T \B \\ 
\hline
\multirow{51}{*}{GRS1915+105}&2000-04-24&Chandra&660&29.76&$\leq5$&\textbf{5}$_{-2}^{+3}$&/&/&/\T \B \\ 
&2001-03-24&Chandra&1944&30.42&$\leq7$&$\leq5$&$\leq17$&$\leq17$&$\leq24$\T \B \\ 
&2001-05-23&Chandra&1945&30.04&$\leq3$&\textbf{8}$_{-1}^{+2}$&$\leq10$&$\leq12$&$\leq15$\T \B \\ 
&2001-08-05&Chandra&1946&28.44&$\leq8$&$\leq9$&$\leq22$&$\leq32$&$\leq33$\T \B \\ 
&2003-03-29&XMM&0112990101\_U002&0.23&$\leq13$&$\leq22$&$\leq17$&$\leq28$&$\leq34$\T \B \\ 
&2003-04-10&XMM&0112920701\_S007&0.18&$\leq14$&$\leq22$&$\leq26$&$\leq31$&$\leq36$\T \B \\ 
&2003-04-16&XMM&0112920801\_U002&0.04&$\leq55$&$\leq58$&$\leq88$&/&/\T \B \\ 
&2003-10-17&XMM&0112990501\_S008&0.48&$\leq10$&\textbf{18}$_{-3}^{+5}$&$\leq10$&$\leq13$&$\leq12$\T \B \\ 
&2003-10-22&XMM&0112920901\_S003&0.28&$\leq13$&$\textbf{16}\pm6$&$\leq15$&$\leq23$&$\leq20$\T \B \\ 
&2004-03-20&Chandra&4587&30.04&$\leq5$&$\leq11$&$\leq18$&$\leq18$&$\leq22$\T \B \\ 
&2004-03-30&Chandra&4588&27.17&$\leq8$&$\leq10$&$\leq18$&$\leq36$&$\leq39$\T \B \\ 
&2004-04-06&Chandra&4589&30.02&$\leq8$&$\leq9$&$\leq21$&$\leq28$&$\leq37$\T \B \\ 
&2004-04-17&XMM&0144090101\_U002&14.1&$\leq4$&$\leq4$&$\leq6$&$\leq6$&$\leq10$\T \B \\ 
&2004-04-21&XMM&0144090201\_S003&0.62&$\leq18$&$\leq19$&$\leq27$&$\leq26$&$\leq32$\T \B \\ 
&2004-05-03&XMM&0112921201\_U002&0.56&$\leq19$&$\leq21$&$\leq28$&$\leq30$&$\leq34$\T \B \\ 
&2005-12-01&Chandra&6579&12.3&$\leq10$&$\textbf{13}\pm3$&$\leq13$&$\leq90$&/\T \B \\ 
&2005-12-01&Chandra&6580&12.14&$\leq14$&\textbf{22}$_{-6}^{+5}$&$\leq22$&/&/\T \B \\ 
&2005-12-03&Chandra&6581&9.73&$\leq10$&\textbf{28}$_{-4}^{+3}$&/&/&/\T \B \\ 
&2007-08-14&Chandra&7485&47.38&\textbf{36}$_{-2}^{+4}$&$\textbf{38}\pm2$&\textbf{27}$_{-3}^{+8}$&\textbf{24}$_{-6}^{+5}$&/\T \B \\ 
&2007-09-24&XMM&0506160901\_U002&0.53&\textbf{17}$_{-3}^{+7}$&\textbf{24}$_{-3}^{+8}$&$\leq18$&$\leq16$&$\leq15$\T \B \\ 
&2007-09-26&XMM&0506161001\_U002&0.53&\textbf{16}$_{-3}^{+6}$&\textbf{32}$_{-5}^{+6}$&$\leq12$&$\leq17$&$\leq21$\T \B \\ 
&2007-09-28&XMM&0506161101\_S001&0.52&\textbf{50}$_{-6}^{+10}$&\textbf{37}$_{-7}^{+12}$&\textbf{45}$_{-11}^{+14}$&/&/\T \B \\ 
&2007-09-30&XMM&0506161201\_U002&0.59&\textbf{32}$_{-6}^{+7}$&\textbf{19}$_{-7}^{+10}$&$\textbf{31}\pm12$&/&/\T \B \\ 
&2011-06-21&Chandra&12462&116.4&$\leq2$&$\textbf{8}\pm1$&$\leq5$&$\leq6$&$\leq79$\T \B \\ 
&2015-02-23&Chandra&16709&39.91&$\leq2$&\textbf{7}$_{-2}^{+1}$&$\leq7$&$\leq10$&$\leq11$\T \B \\ 
&2015-03-19&Chandra&16710&38.04&$\leq4$&$\leq5$&$\leq7$&$\leq10$&$\leq19$\T \B \\ 
&2015-06-09&Chandra&16711&118.65&$\textbf{17}\pm1$&$\textbf{23}\pm1$&\textbf{17}$_{-3}^{+1}$&\textbf{16}$_{-1}^{+2}$&/\T \B \\ 
&2017-02-22&Chandra&19717&24.96&$\leq13$&$\leq14$&$\leq29$&$\leq36$&$\leq60$\T \B \\ 
&2017-03-27&Chandra&19718&25.01&$\leq13$&$\leq20$&$\leq31$&$\leq59$&$\leq68$\T \B \\ 
&2017-05-02&XMM&0804640201\_U002&0.06&/&$\leq100$&/&/&/\T \B \\ 
&2017-06-24&Chandra&19719&25.03&$\leq5$&\textbf{15}$_{-3}^{+4}$&$\leq12$&/&/\T \B \\ 
&2017-08-09&Chandra&19720&23.88&$\leq4$&\textbf{11}$_{-3}^{+1}$&$\leq10$&$\leq18$&$\leq99$\T \B \\ 
&2017-09-22&XMM&0804640501\_S003&0.4&$\leq14$&$\leq16$&$\leq20$&$\leq22$&$\leq30$\T \B \\ 
&2017-10-12&XMM&0804640601\_S003&0.43&$\leq10$&$\textbf{17}\pm4$&$\leq10$&$\leq12$&$\leq16$\T \B \\ 
&2018-04-10&XMM&0804640701\_S003&0.45&$\leq27$&$\leq28$&$\leq39$&$\leq43$&$\leq47$\T \B \\ 
&2018-04-19&XMM&0804640801\_S003&0.51&$\leq13$&$\leq23$&$\leq20$&$\leq30$&$\leq38$\T \B \\ 
&2019-04-30&Chandra&22213&29.08&\textbf{77}$_{-5}^{+7}$&\textbf{26}$_{-2}^{+1}$&/&/&/\T \B \\ 
&2021-07-14&Chandra&23435&24.5&$\textbf{55}\pm7$&\textbf{31}$_{-4}^{+5}$&\textbf{34}$_{-5}^{+9}$&/&/\T \B \\ 
&2021-07-15&Chandra&24663&23.5&\textbf{61}$_{-6}^{+8}$&\textbf{32}$_{-3}^{+4}$&\textbf{42}$_{-17}^{+23}$&/&/\T \B \\ 
\hline
\multirow{3}{*}{GS1354-64}&2015-08-06&XMM&0727961501\_S003&0.24&$\leq77$&$\leq84$&/&/&/\T \B \\ 
&2015-08-06&XMM&0727961501\_S004&10.99&$\leq21$&$\leq24$&$\leq30$&$\leq32$&$\leq36$\T \B \\ 
\hline
&2002-08-24&XMM&0093562701\_S005&1.28&$\leq24$&$\leq25$&$\leq36$&$\leq49$&$\leq59$\T \B \\ 
&2002-09-29&XMM&0156760101\_S001&2.25&$\leq10$&$\leq12$&$\leq15$&$\leq18$&$\leq22$\T \B \\ 
&2003-03-08&XMM&0148220201\_S001&12.75&$\leq16$&$\leq20$&$\leq34$&$\leq39$&$\leq49$\T \B \\ 
&2003-03-17&Chandra&4420&74.05&$\leq45$&$\leq50$&/&/&/\T \B \\ 
&2003-03-20&XMM&0148220301\_S001&3.98&$\leq23$&$\leq22$&$\leq34$&$\leq44$&$\leq47$\T \B \\ 
&2004-03-16&XMM&0204730201\_U002&101.25&$\leq5$&$\leq5$&$\leq8$&$\leq8$&$\leq8$\T \B \\ 
&2004-03-18&XMM&0204730301\_U002&88.92&$\leq5$&$\leq5$&$\leq7$&$\leq5$&$\leq8$\T \B \\ 
&2004-03-20&XMM&0204730301\_U003&5.07&$\leq31$&$\leq31$&$\leq38$&$\leq43$&$\leq33$\T \B \\ 
&2004-08-22&Chandra&4569&49.9&$\leq25$&$\leq26$&/&/&/\T \B \\ 
&2004-10-04&Chandra&4570&44.53&$\leq24$&$\leq25$&/&/&/\T \B \\ 
&2004-10-28&Chandra&4571&43.36&$\leq13$&$\leq30$&/&/&/\T \B \\ 
&2007-02-19&XMM&0410581201\_S001&0.45&$\leq18$&$\leq21$&$\leq48$&$\leq50$&$\leq62$\T \B \\ 
&2007-03-05&XMM&0410581301\_S001&0.48&$\leq17$&$\leq17$&$\leq28$&$\leq29$&$\leq45$\T \B \\ 
%no multicolumn here because the page change fucks it up
GX339\-4&2007-03-30&XMM&0410581701\_U002&0.26&/&/&/&/&/\T \B \\ 
&2009-03-26&XMM&0605610201\_S003&31.75&$\leq15$&$\leq14$&$\leq17$&$\leq22$&$\leq28$\T \B \\ 
&2010-03-28&XMM&0654130401\_S001&25.29&$\leq4$&$\leq4$&$\leq6$&$\leq5$&$\leq6$\T \B \\ 
&2013-09-29&XMM&0692341201\_S003&8.54&$\leq9$&$\leq10$&$\leq14$&$\leq14$&$\leq17$\T \B \\ 
&2013-09-30&XMM&0692341301\_S003&9.43&$\leq19$&$\leq21$&$\leq29$&$\leq30$&$\leq31$\T \B \\ 
&2013-10-01&XMM&0692341401\_S003&15.04&$\leq18$&$\leq18$&$\leq21$&$\leq22$&$\leq24$\T \B \\ 
&2015-08-28&XMM&0760646201\_S003&14.73&$\leq21$&$\leq18$&$\leq27$&$\leq33$&$\leq38$\T \B \\ 
&2015-09-02&XMM&0760646301\_S003&15.74&$\leq13$&$\leq14$&$\leq17$&$\leq19$&$\leq33$\T \B \\ 
&2015-09-07&XMM&0760646401\_S003&20.18&$\leq16$&$\leq18$&$\leq20$&$\leq22$&$\leq27$\T \B \\ 
&2015-09-12&XMM&0760646501\_S003&18.62&$\leq37$&$\leq38$&$\leq46$&$\leq55$&$\leq62$\T \B \\ 
&2015-09-17&XMM&0760646601\_S003&36.53&$\leq10$&$\leq12$&$\leq12$&$\leq13$&$\leq23$\T \B \\ 
&2015-09-30&XMM&0760646701\_S003&33.42&$\leq13$&$\leq15$&$\leq16$&$\leq17$&$\leq23$\T \B \\ 
\hline
\multirow{25}{*}{H1743-322}&2003-05-01&Chandra&3803&48.26&$\textbf{7}\pm1$&\textbf{20}$_{-3}^{+2}$&/&/&/\T \B \\ 
&2003-05-28&Chandra&3804&43.89&$\leq6$&$\leq9$&$\leq19$&$\leq24$&$\leq34$\T \B \\ 
&2003-06-23&Chandra&3805&49.87&$\textbf{7}\pm2$&\textbf{16}$_{-4}^{+3}$&/&/&/\T \B \\ 
&2003-07-30&Chandra&3806&50.0&\textbf{19}$_{-4}^{+3}$&\textbf{29}$_{-5}^{+4}$&/&/&/\T \B \\ 
&2008-09-29&XMM&0554110201\_S005&20.56&$\leq18$&$\leq19$&$\leq25$&$\leq24$&$\leq30$\T \B \\ 
&2010-08-08&Chandra&11048&60.29&$\leq16$&$\leq21$&$\leq46$&$\leq49$&$\leq69$\T \B \\ 
&2010-10-09&XMM&0553950201\_S003&59.96&$\leq24$&$\leq30$&$\leq51$&$\leq51$&$\leq49$\T \B \\ 
&2014-09-21&XMM&0724400501\_S001&135.08&$\leq5$&$\leq6$&$\leq6$&$\leq6$&$\leq10$\T \B \\ 
&2014-09-23&XMM&0724401901\_S001&77.74&$\leq10$&$\leq11$&$\leq15$&$\leq14$&$\leq18$\T \B \\ 
&2014-09-24&XMM&0740980201\_S003&48.61&$\leq8$&$\leq8$&$\leq8$&$\leq9$&$\leq15$\T \B \\ 
&2015-06-11&Chandra&16738&9.22&$\leq27$&$\leq28$&$\leq63$&$\leq91$&/\T \B \\ 
&2015-06-12&Chandra&17679&9.22&$\leq43$&$\leq52$&/&/&/\T \B \\ 
&2015-06-13&Chandra&17680&9.22&$\leq47$&$\leq50$&/&/&/\T \B \\ 
&2015-07-03&Chandra&16739&26.84&$\leq22$&$\leq28$&$\leq51$&$\leq79$&$\leq91$\T \B \\ 
&2016-03-13&XMM&0783540201\_S003&137.42&$\leq8$&$\leq8$&$\leq11$&$\leq11$&$\leq15$\T \B \\ 
&2016-03-15&XMM&0783540301\_U002&134.52&$\leq4$&$\leq5$&$\leq5$&$\leq6$&$\leq7$\T \B \\ 
&2018-09-26&XMM&0783540401\_S003&128.95&$\leq6$&$\leq8$&$\leq8$&$\leq9$&$\leq10$\T \B \\ 
\hline
\multirow{23}{*}{IGRJ17091-3624}&2011-03-27&XMM&0677980201\_S003&1.14&$\leq44$&$\leq46$&$\leq67$&$\leq76$&/\T \B \\ 
&2011-08-01&Chandra&12405&31.21&$\leq34$&$\leq32$&$\leq57$&$\leq93$&/\T \B \\ 
&2011-10-06&Chandra&12406&27.29&$\leq10$&$\leq23$&$\leq87$&/&/\T \B \\ 
&2012-09-29&XMM&0700381301\_S003&46.12&$\leq10$&$\leq10$&$\leq12$&$\leq13$&$\leq15$\T \B \\ 
&2016-03-07&XMM&0743960201\_S003&57.98&$\leq10$&$\leq11$&$\leq17$&$\leq17$&$\leq21$\T \B \\ 
&2016-03-09&XMM&0744361501\_S003&38.16&$\leq15$&$\leq19$&$\leq20$&$\leq21$&$\leq40$\T \B \\ 
&2016-03-11&XMM&0744361801\_S003&28.59&$\leq15$&$\leq18$&$\leq17$&$\leq18$&$\leq26$\T \B \\ 
&2016-03-23&XMM&0744361701\_S003&61.24&$\leq13$&$\leq18$&$\leq22$&$\leq23$&$\leq29$\T \B \\ 
&2016-03-30&Chandra&17787&39.48&$\leq16$&$\leq18$&$\leq40$&$\leq58$&$\leq95$\T \B \\ 
&2016-04-30&Chandra&17788&38.75&$\leq19$&$\leq30$&$\leq69$&/&/\T \B \\ 
&2016-05-26&Chandra&17789&20.05&$\leq48$&$\leq53$&/&/&/\T \B \\ 
&2016-05-27&Chandra&18855&19.97&$\leq31$&$\leq51$&/&/&/\T \B \\ 
&2016-06-24&Chandra&17790&19.97&$\leq66$&$\leq77$&/&/&/\T \B \\ 
&2016-06-25&Chandra&18874&19.86&$\leq60$&$\leq91$&/&/&/\T \B \\ 
&2022-06-16&Chandra&26435&29.09&$\leq16$&$\leq20$&$\leq51$&$\leq55$&$\leq76$\T \B \\ 
\hline
\multirow{3}{*}{IGRJ17098-3628}&2006-08-25&XMM&0406140101\_U002&3.74&$\leq83$&/&/&/&/\T \B \\ 
&2007-02-19&XMM&0406140401\_S003&7.02&$\leq47$&$\leq66$&/&/&/\T \B \\ 
\hline
\multirow{1}{*}{IGRJ17285-2922}&2010-09-09&XMM&0405182701\_S003&18.5&$\leq49$&$\leq48$&$\leq60$&$\leq69$&$\leq72$\T \B \\ 
\hline
\multirow{1}{*}{IGRJ17451-3022}&2015-03-06&XMM&0748391201\_S001$^\dagger$&36.45&\textbf{92}$_{-10}^{+11}$&$\leq77$&/&/&/\T \B \\ 
\hline
\multirow{3}{*}{IGRJ17497-2821}&2006-09-22&XMM&0410580401\_S001&31.18&$\leq13$&$\leq14$&$\leq14$&$\leq17$&$\leq23$\T \B \\ 
&2006-10-01&Chandra&6613&19.7&$\leq46$&$\leq56$&/&/&/\T \B \\ 
\hline
\multirow{1}{*}{MAXIJ0637-430}&2019-11-17&XMM&0853980801\_S001&0.6&/&/&/&/&/\T \B \\ 
\hline
\multirow{1}{*}{MAXIJ1305-704}&2012-04-29&Chandra&14425&29.38&$\leq50$&$\leq30$&/&/&/\T \B \\ 
\hline
\multirow{8}{*}{MAXIJ1348-630}&2019-02-01&XMM&0831000101\_S001&7.85&$\leq10$&$\leq10$&$\leq13$&$\leq14$&$\leq21$\T \B \\ 
&2019-02-26&XMM&0831000301\_S001&3.58&$\leq24$&$\leq26$&$\leq37$&$\leq42$&$\leq54$\T \B \\ 
&2019-06-21&Chandra&21239&19.04&$\leq12$&$\leq15$&$\leq43$&$\leq36$&$\leq55$\T \B \\ 
&2019-06-26&Chandra&21240&20.04&$\leq8$&$\leq9$&$\leq22$&$\leq30$&$\leq31$\T \B \\ 
&2019-07-07&Chandra&21241&20.05&$\leq11$&$\leq14$&$\leq29$&$\leq33$&$\leq37$\T \B \\ 
\hline
\multirow{16}{*}{MAXIJ1535-571}&2017-09-07&XMM&0795711801\_S014&4.6&$\leq6$&$\leq11$&$\leq9$&$\leq11$&$\leq14$\T \B \\ 
&2017-09-07&XMM&0795711801\_S003&0.57&$\leq14$&$\leq14$&$\leq20$&$\leq24$&$\leq29$\T \B \\ 
&2017-09-08&XMM&0795711801\_U014&0.05&$\leq48$&$\leq49$&$\leq68$&$\leq75$&$\leq82$\T \B \\ 
&2017-09-08&XMM&0795711801\_U015&0.25&$\leq23$&$\leq24$&$\leq31$&$\leq34$&$\leq39$\T \B \\ 
&2017-09-13&Chandra&20203&22.97&$\leq8$&$\leq8$&$\leq23$&$\leq44$&$\leq45$\T \B \\ 
&2017-09-14&XMM&0795712001\_S003&0.82&$\leq4$&$\leq5$&$\leq6$&$\leq7$&$\leq8$\T \B \\ 
&2017-09-15&XMM&0795712101\_S003&0.46&$\leq11$&$\leq11$&$\leq16$&$\leq20$&$\leq21$\T \B \\ 
&2017-09-27&Chandra&20204&18.85&$\leq12$&$\leq14$&$\leq37$&$\leq63$&/\T \B \\ 
&2017-10-08&Chandra&20205&20.7&$\leq7$&$\leq8$&$\leq22$&$\leq37$&$\leq60$\T \B \\ 
&2017-10-24&Chandra&20206&27.22&$\leq7$&$\leq8$&$\leq20$&$\leq26$&$\leq41$\T \B \\ 
&2017-12-31&Chandra&20169&21.19&$\leq46$&$\leq55$&/&/&/\T \B \\ 
\hline
\multirow{3}{*}{MAXIJ1659-152}&2010-09-27&XMM&0656780601\_S003&22.88&$\leq5$&$\leq6$&$\leq8$&$\leq6$&$\leq8$\T \B \\ 
&2011-03-22&XMM&0677980101\_U002&20.51&$\leq66$&$\leq66$&/&/&/\T \B \\ 
\hline
\multirow{6}{*}{MAXIJ1803-298}&2021-05-17&Chandra&25039&10.02&$\leq30$&$\leq41$&/&/&/\T \B \\ 
&2021-05-23&Chandra&25040&10.24&$\leq28$&$\leq36$&/&/&/\T \B \\ 
&2021-06-17&Chandra&25041&6.31&$\leq85$&/&/&/&/\T \B \\ 
&2021-06-18&Chandra&25063&7.91&$\leq92$&/&/&/&/\T \B \\ 
\hline
\multirow{21}{*}{MAXIJ1820+070}&2018-03-17&XMM&0830190201\_S001&5.37&$\leq24$&$\leq26$&$\leq33$&$\leq36$&$\leq42$\T \B \\ 
&2018-03-17&XMM&0830190201\_S002&2.04&$\leq12$&$\leq10$&$\leq15$&$\leq14$&$\leq17$\T \B \\ 
&2018-03-19&XMM&0820880201\_S003&0.3&$\leq21$&$\leq23$&$\leq29$&$\leq29$&$\leq35$\T \B \\ 
&2018-03-19&XMM&0820880201\_S011&3.85&$\leq17$&$\leq20$&$\leq22$&$\leq25$&$\leq50$\T \B \\ 
&2018-03-22&XMM&0820880301\_S003&0.6&$\leq13$&$\leq15$&$\leq17$&$\leq20$&$\leq23$\T \B \\ 
&2018-03-27&XMM&0820880401\_S003&0.85&$\leq5$&$\leq10$&$\leq8$&$\leq9$&$\leq10$\T \B \\ 
&2018-04-12&XMM&0820880501\_S003&0.11&$\leq33$&$\leq31$&$\leq41$&$\leq45$&$\leq56$\T \B \\ 
&2018-09-28&XMM&0820880601\_S003&0.3&$\leq46$&$\leq51$&$\leq64$&$\leq71$&$\leq80$\T \B \\ 
&2018-09-30&XMM&0820881101\_S003&0.24&$\leq33$&$\leq33$&$\leq44$&$\leq49$&$\leq71$\T \B \\ 
&2018-10-05&XMM&0830191901\_S001&0.15&$\leq62$&$\leq66$&$\leq94$&$\leq91$&$\leq79$\T \B \\ 
&2018-10-05&XMM&0830191901\_S002&5.24&$\leq13$&$\leq14$&$\leq21$&$\leq20$&$\leq31$\T \B \\ 
&2019-03-22&XMM&0844230201\_S003&8.49&$\leq40$&$\leq40$&$\leq56$&$\leq61$&$\leq73$\T \B \\ 
&2019-03-26&XMM&0844230301\_S003&11.34&$\leq17$&$\leq17$&$\leq21$&$\leq23$&$\leq34$\T \B \\ 
&2019-09-20&XMM&0851181301\_S003&56.28&$\leq35$&$\leq33$&$\leq51$&$\leq56$&$\leq60$\T \B \\ 
\hline
\multirow{1}{*}{SAXJ1711.6-3808}&2001-03-02&XMM&0135520401\_S001&6.03&$\leq17$&$\leq23$&$\leq23$&$\leq26$&$\leq26$\T \B \\ 
\hline
\multirow{1}{*}{SwiftJ1357.2-0933}&2011-02-05&XMM&0674580101\_U014&33.48&$\leq13$&$\leq11$&$\leq22$&$\leq19$&$\leq21$\T \B \\ 
\hline
&2018-02-25&XMM&0802300201\_S003&41.06&$\leq13$&$\leq10$&$\leq17$&$\leq18$&$\leq26$\T \B \\ 
&2018-02-27&XMM&0811213401\_S003&28.58&$\leq15$&$\leq15$&$\leq25$&$\leq26$&$\leq30$\T \B \\ 
&2018-03-04&XMM&0805200201\_S007$^\dagger$&0.66&$\leq45$&$\leq50$&$\leq52$&$\leq77$&$\leq82$\T \B \\ 
&2018-03-04&XMM&0805200201\_S003$^\dagger$&30.96&$\leq11$&$\leq12$&$\leq21$&$\leq21$&$\leq27$\T \B \\
%no multi-column to avoid issue with page change
SwiftJ1658.2\-4242&2018-03-11&XMM&0805200301\_S003&29.45&$\leq9$&$\leq10$&$\leq13$&$\leq15$&$\leq29$\T \B \\ 
&2018-03-11&XMM&0805200301\_S014&0.44&$\leq27$&$\leq31$&$\leq65$&$\leq65$&$\leq53$\T \B \\ 
&2018-03-15&XMM&0805200401\_S003&32.95&$\leq4$&$\leq5$&$\leq9$&$\leq9$&$\leq12$\T \B \\ 
&2018-03-28&XMM&0805201301\_S003&33.72&$\leq20$&$\leq22$&$\leq34$&$\leq37$&$\leq44$\T \B \\ 
&2018-04-28&Chandra&21083&29.08&$\leq26$&$\leq38$&$\leq63$&/&/\T \B \\ 
\hline
\multirow{1}{*}{SwiftJ174510.8-262411}&2012-09-28&XMM&0693020301\_S003&1.11&$\leq9$&$\leq10$&$\leq10$&$\leq12$&$\leq13$\T \B \\ 
\hline
\multirow{10}{*}{SwiftJ1753.5-0127}&2006-03-24&XMM&0311590901\_S001&40.11&$\leq13$&$\leq15$&$\leq20$&$\leq23$&$\leq26$\T \B \\ 
&2009-09-29&XMM&0605610301\_U002&25.19&$\leq15$&$\leq16$&$\leq18$&$\leq19$&$\leq24$\T \B \\ 
&2012-05-03&Chandra&14428&19.63&$\leq46$&$\leq45$&/&/&/\T \B \\ 
&2012-09-10&XMM&0691740201\_S001&37.42&$\leq9$&$\leq11$&$\leq13$&$\leq14$&$\leq25$\T \B \\ 
&2012-10-08&XMM&0694930501\_S001&28.38&$\leq28$&$\leq28$&$\leq39$&$\leq44$&$\leq48$\T \B \\ 
&2014-09-13&XMM&0744320201\_S001&46.18&$\leq15$&$\leq18$&$\leq26$&$\leq27$&$\leq33$\T \B \\ 
&2015-03-19&XMM&0770580201\_S003&31.37&$\leq74$&$\leq79$&/&/&/\T \B \\ 
\hline
\multirow{3}{*}{SwiftJ1910.2-0546}&2012-09-22&Chandra&14634&29.96&$\leq77$&$\leq89$&/&/&/\T \B \\ 
&2012-10-17&XMM&0691271401\_S001&40.49&$\leq19$&$\leq21$&$\leq27$&$\leq32$&$\leq37$\T \B \\ 
\hline
\multirow{3}{*}{V404Cyg}&2015-06-22&Chandra&17696&20.76&$\leq10$&$\leq16$&$\leq25$&$\leq31$&$\leq37$\T \B \\ 
&2015-06-23&Chandra&17697&25.25&$\leq12$&$\leq13$&$\leq33$&$\leq48$&$\leq59$\T \B \\ 
\hline
\multirow{3}{*}{V4641Sgr}&2020-02-14&Chandra&22389&44.0&$\leq20$&$\leq33$&/&/&/\T \B \\ 
&2020-02-15&Chandra&23158&29.35&$\leq48$&$\leq77$&/&/&/\T \B \\ 
\hline
\multirow{3}{*}{XTEJ1550-564}&2000-05-03&Chandra&680&2.14&$\leq33$&$\leq43$&$\leq99$&/&/\T \B \\ 
&2000-05-06&Chandra&681&2.13&$\leq63$&$\leq72$&/&/&/\T \B \\ 
\hline
\multirow{5}{*}{XTEJ1650-500}&2001-09-13&XMM&0136140301\_S001&0.69&$\leq10$&$\leq14$&$\leq16$&$\leq17$&$\leq18$\T \B \\ 
&2001-10-05&Chandra&2699&22.51&$\leq16$&$\leq20$&/&/&/\T \B \\ 
&2001-10-29&Chandra&2700&26.36&$\leq47$&$\leq63$&/&/&/\T \B \\ 
\hline
\multirow{1}{*}{XTEJ1652-453}&2009-08-22&XMM&0610000701\_U002&38.22&$\leq22$&$\leq27$&$\leq40$&$\leq39$&$\leq43$\T \B \\ 
\hline
\multirow{1}{*}{XTEJ1720-318}&2003-02-20&XMM&0154750501\_S001&7.73&$\leq24$&$\leq27$&$\leq42$&$\leq50$&$\leq67$\T \B \\ 
\hline
\multirow{6}{*}{XTEJ1752-223}&2009-11-01&Chandra&10069&30.55&$\leq15$&$\leq27$&$\leq42$&$\leq51$&$\leq73$\T \B \\ 
&2010-02-08&Chandra&10070&21.31&$\leq40$&$\leq47$&/&/&/\T \B \\ 
&2010-04-06&XMM&0653110101\_S003&18.17&$\leq6$&$\leq6$&$\leq11$&$\leq11$&$\leq12$\T \B \\ 
&2010-04-07&XMM&0653110101\_S008&0.57&$\leq36$&$\leq53$&$\leq58$&$\leq59$&$\leq73$\T \B \\ 
\hline
\multirow{8}{*}{XTEJ1817-330}&2006-02-13&Chandra&6615&29.07&$\leq15$&$\leq19$&$\leq38$&$\leq66$&$\leq97$\T \B \\ 
&2006-02-24&Chandra&6616&38.96&$\leq19$&$\leq19$&$\leq48$&$\leq75$&$\leq91$\T \B \\ 
&2006-03-13&XMM&0311590501\_S003&0.6&$\leq30$&$\leq32$&$\leq59$&$\leq56$&$\leq68$\T \B \\ 
&2006-03-15&Chandra&6617&46.53&$\leq27$&$\leq34$&$\leq57$&$\leq92$&/\T \B \\ 
&2006-05-22&Chandra&6618&50.77&$\leq60$&$\leq61$&/&/&/\T \B \\ 
\hline
\multirow{1}{*}{XTEJ1856+053}&2007-03-14&XMM&0510010101\_U002&1.5&$\leq53$&$\leq62$&/&/&/\T \B \\ 
\hline
\multirow{1}{*}{XTEJ1901+014}&2006-10-14&XMM&0402470401\_S003&8.73&$\leq56$&$\leq57$&$\leq88$&$\leq81$&/\T \B \\ 
\hline 
\end{longtable}

%%%%%%%%%%%%%%%%%%%%%%%%%%%%%%%%%
% RESULTS TABLE
%%%%%%%%%%%%%%%%%%%%%%%%%%%%%%%%%

\pagebreak

\subsection{Parameters of K$\alpha$ detections}

\begin{table*}[h]
\centering
\scriptsize
\caption[*]{Main characteristics of significant K$\alpha$ line detections from the sample. Uncertainties on the luminosity are not quoted due to being negligible}\label{tab:bshift_details}
\begin{tabular}{c c c || c c || c  c  c | c  c  c }
\hline
\hline
\multirow{2}{*}{Source}
& \multirow{2}{*}{Date}
& \multirow{2}{*}{ObsID}
& \multirow{2}{*}{HR$_{[6-10]/[3-10]}$}
& \multirow{2}{*}{$L_{[3-10]}/L_{Edd}$}
& \multicolumn{3}{c}{\FeKav{}}
& \multicolumn{3}{c}{\FeKavi{}}
\T \B \\

&
&
&
& $\times 10^{-2}$
& EW
& blueshift
& width
& EW
& blueshift
& width
\T \B \\
\hline
\hline
\multirow{21}{*}{4U1630-47}&2004-08-04&4568&$0.351_{-0.003}^{+0.003}$&$5.6$&/&/&/&\textbf{11}$_{-4}^{+3}$&\textbf{-300}$_{-500}^{+500}$&\textbf{0}$^{+4200}$\T \B \\ 
&2012-01-17&13714&$0.362_{-0.003}^{+0.003}$&$4.7$&$\textbf{32}\pm4$&\textbf{0}$_{-100}^{+200}$&\textbf{1900}$_{-500}^{+500}$&$\textbf{57}\pm5$&\textbf{-300}$_{-100}^{+100}$&\textbf{2700}$_{-400}^{+400}$\T \B \\ 
&2012-01-20&13715&$0.344_{-0.002}^{+0.002}$&$4.6$&\textbf{34}$_{-5}^{+3}$&\textbf{100}$_{-100}^{+200}$&\textbf{2300}$_{-400}^{+600}$&\textbf{49}$_{-5}^{+4}$&\textbf{-300}$_{-100}^{+100}$&\textbf{2200}$_{-300}^{+600}$\T \B \\ 
&2012-01-26&13716&$0.347_{-0.003}^{+0.002}$&$4.4$&\textbf{47}$_{-2}^{+3}$&\textbf{500}$_{-200}^{+200}$&\textbf{3000}$_{-500}^{+400}$&\textbf{52}$_{-3}^{+1}$&\textbf{-300}$_{-100}^{+0}$&\textbf{2200}$_{-400}^{+700}$\T \B \\ 
&2012-01-30&13717&$0.389_{-0.003}^{+0.003}$&$5.1$&$\textbf{30}\pm3$&\textbf{200}$_{-300}^{+200}$&\textbf{2000}$_{-700}^{+800}$&$\textbf{48}\pm4$&\textbf{-200}$_{-200}^{+200}$&\textbf{1800}$_{-700}^{+700}$\T \B \\ 
&2012-03-04&0670671501\_S003&$0.366_{-0.002}^{+0.002}$&$4.8$&$\textbf{35}\pm7$&\textbf{-5000}$_{-1900}^{+2200}$&/&\textbf{55}$_{-7}^{+9}$&\textbf{-5800}$_{-1200}^{+1600}$&/\T \B \\ 
&2012-03-04&0670671501\_U014&$0.347_{-0.0}^{+0.0}$&$5.3$&$\textbf{31}\pm2$&\textbf{-5200}$_{-300}^{+200}$&/&\textbf{48}$_{-2}^{+1}$&\textbf{-5200}$_{-200}^{+100}$&/\T \B \\ 
&2012-03-20&0670671301\_S003&$0.36_{-0.001}^{+0.001}$&$6.2$&$\textbf{21}\pm3$&\textbf{-3900}$_{-800}^{+900}$&/&\textbf{46}$_{-2}^{+4}$&\textbf{-4300}$_{-400}^{+400}$&/\T \B \\ 
&2012-03-25&0670672901\_S003&$0.401_{-0.0}^{+0.0}$&$5.8$&$\textbf{20}\pm1$&\textbf{-6000}$_{-500}^{+500}$&/&\textbf{45}$_{-2}^{+1}$&\textbf{-5900}$_{-300}^{+200}$&/\T \B \\ 
&2012-09-09&0670673001\_S003&$0.413_{-0.001}^{+0.001}$&$8.0$&\textbf{9}$_{-2}^{+3}$&\textbf{-4600}$_{-3000}^{+2600}$&/&\textbf{31}$_{-2}^{+3}$&\textbf{-4300}$_{-800}^{+1000}$&/\T \B \\ 
&2012-09-10&0670673001\_U002&$0.432_{-0.002}^{+0.002}$&$7.2$&/&/&/&\textbf{25}$_{-5}^{+6}$&\textbf{-3500}$_{-3000}^{+2700}$&/\T \B \\ 
&2012-09-11&0670673101\_S003&$0.467_{-0.002}^{+0.002}$&$9.4$&/&/&/&\textbf{9}$_{-4}^{+5}$&\textbf{-1200}$_{-6400}^{+5800}$&/\T \B \\ 
&2016-10-21&19904&$0.311_{-0.002}^{+0.002}$&$5.6$&\textbf{23}$_{-5}^{+4}$&\textbf{-300}$_{-300}^{+300}$&\textbf{1800}$_{-1500}^{+1400}$&\textbf{45}$_{-7}^{+4}$&\textbf{-200}$_{-300}^{+300}$&\textbf{2400}$_{-800}^{+1000}$\T \B \\ 
\hline
&2005-03-12&5460&$0.276_{-0.004}^{+0.003}$&$2.2$&/&/&/&\textbf{19}$_{-3}^{+4}$&\textbf{-200}$_{-300}^{+200}$&/\T \B \\ 
&2005-03-14&0112921401\_S003&$0.266_{-0.001}^{+0.002}$&$3.3$&/&/&/&\textbf{31}$_{-3}^{+6}$&\textbf{900}$_{-1200}^{+700}$&/\T \B \\ 
&2005-03-15&0112921501\_S003&$0.258_{-0.001}^{+0.002}$&$3.5$&\textbf{23}$_{-4}^{+5}$&\textbf{-2400}$_{-1500}^{+1500}$&/&\textbf{30}$_{-5}^{+6}$&\textbf{-500}$_{-1100}^{+1200}$&/\T \B \\ 
GROJ1655-40&2005-03-16&0112921601\_S003&$0.304_{-0.002}^{+0.002}$&$4.2$&\textbf{12}$_{-4}^{+3}$&\textbf{2300}$_{-2300}^{+900}$&/&\textbf{30}$_{-3}^{+4}$&\textbf{1100}$_{-500}^{+600}$&/\T \B \\ 
&2005-03-18&0155762501\_S001&$0.293_{-0.001}^{+0.001}$&$4.7$&\textbf{30}$_{-3}^{+4}$&\textbf{-100}$_{-600}^{+800}$&/&\textbf{41}$_{-4}^{+3}$&\textbf{300}$_{-500}^{+600}$&/\T \B \\ 
&2005-03-27&0155762601\_S001&$0.32_{-0.002}^{+0.002}$&$2.5$&$\textbf{30}\pm4$&\textbf{-200}$_{-600}^{+600}$&/&\textbf{17}$_{-4}^{+5}$&\textbf{-3500}$_{-1600}^{+1400}$&/\T \B \\ 
&2005-04-01&5461&$0.285_{-0.001}^{+0.001}$&$2.5$&\textbf{58}$_{-3}^{+4}$&\textbf{0}$_{-100}^{+100}$&\textbf{3700}$_{-300}^{+300}$&\textbf{43}$_{-5}^{+3}$&\textbf{-1200}$_{-200}^{+100}$&\textbf{2500}$_{-300}^{+500}$\T \B \\ 
\hline
\multirow{31}{*}{GRS1915+105}&2000-04-24&660&$0.708_{-0.005}^{+0.008}$&$5.7$&/&/&/&\textbf{5}$_{-2}^{+3}$&\textbf{-300}$_{-800}^{+500}$&/\T \B \\ 
&2001-05-23&1945&$0.506_{-0.003}^{+0.003}$&$11.9$&/&/&/&\textbf{8}$_{-1}^{+2}$&\textbf{-600}$_{-500}^{+500}$&/\T \B \\ 
&2003-10-17&0112990501\_S008&$0.585_{-0.002}^{+0.002}$&$16.8$&/&/&/&\textbf{18}$_{-3}^{+5}$&\textbf{-2900}$_{-1100}^{+1100}$&/\T \B \\ 
&2003-10-22&0112920901\_S003&$0.595_{-0.004}^{+0.004}$&$11.0$&/&/&/&$\textbf{16}\pm6$&\textbf{500}$_{-2400}^{+1700}$&/\T \B \\ 
&2005-12-01&6579&$0.484_{-0.007}^{+0.005}$&$12.5_{-0.1}$&/&/&/&$\textbf{13}\pm3$&\textbf{-800}$_{-500}^{+500}$&/\T \B \\ 
&2005-12-01&6580&$0.47_{-0.006}^{+0.005}$&$13.3_{-0.1}^{+0.1}$&/&/&/&\textbf{22}$_{-6}^{+5}$&\textbf{-1100}$_{-1000}^{+1100}$&\textbf{3400}$_{-1900}^{+1700}$\T \B \\ 
&2005-12-03&6581&$0.555_{-0.003}^{+0.003}$&$34.4_{-0.1}^{+0.1}$&/&/&/&\textbf{28}$_{-4}^{+3}$&\textbf{-700}$_{-300}^{+300}$&\textbf{2000}$_{-1000}^{+1300}$\T \B \\ 
&2007-08-14&7485&$0.491_{-0.003}^{+0.003}$&$6.3$&\textbf{36}$_{-2}^{+4}$&\textbf{300}$_{-100}^{+100}$&\textbf{2700}$_{-300}^{+200}$&$\textbf{38}\pm2$&\textbf{-200}$_{-100}^{+0}$&\textbf{1400}$_{-200}^{+200}$\T \B \\ 
&2007-09-24&0506160901\_U002&$0.453_{-0.002}^{+0.002}$&$13.7$&\textbf{17}$_{-3}^{+7}$&\textbf{-600}$_{-1300}^{+1700}$&/&\textbf{24}$_{-3}^{+8}$&\textbf{-1400}$_{-700}^{+1500}$&/\T \B \\ 
&2007-09-26&0506161001\_U002&$0.457_{-0.002}^{+0.002}$&$12.7$&\textbf{16}$_{-3}^{+6}$&\textbf{-4100}$_{-1700}^{+1900}$&/&\textbf{32}$_{-5}^{+6}$&\textbf{-2900}$_{-500}^{+1100}$&/\T \B \\ 
&2007-09-28&0506161101\_S001&$0.429_{-0.004}^{+0.004}$&$2.7$&\textbf{50}$_{-6}^{+10}$&\textbf{300}$_{-900}^{+900}$&/&\textbf{37}$_{-7}^{+12}$&\textbf{-2000}$_{-1400}^{+1700}$&/\T \B \\ 
&2007-09-30&0506161201\_U002&$0.425_{-0.004}^{+0.004}$&$2.9$&\textbf{32}$_{-6}^{+7}$&\textbf{700}$_{-1600}^{+1100}$&/&\textbf{19}$_{-7}^{+10}$&\textbf{-6600}$_{-2300}^{+3700}$&/\T \B \\ 
&2011-06-21&12462&$0.453_{-0.002}^{+0.001}$&$10.5$&/&/&/&$\textbf{8}\pm1$&\textbf{-500}$_{-100}^{+200}$&/\T \B \\ 
&2015-02-23&16709&$0.457_{-0.002}^{+0.002}$&$12.2$&/&/&/&\textbf{7}$_{-2}^{+1}$&\textbf{-300}$_{-300}^{+200}$&/\T \B \\ 
&2015-06-09&16711&$0.418_{-0.002}^{+0.002}$&$6.7$&$\textbf{17}\pm1$&\textbf{-100}$_{-0}^{+0}$&/&$\textbf{23}\pm1$&\textbf{-200}$_{-0}^{+0}$&/\T \B \\ 
&2017-06-24&19719&$0.556_{-0.006}^{+0.006}$&$5.8$&/&/&/&\textbf{15}$_{-3}^{+4}$&\textbf{100}$_{-500}^{+500}$&\textbf{0}$^{+3400}$\T \B \\ 
&2017-08-09&19720&$0.669_{-0.004}^{+0.005}$&$13.2$&/&/&/&\textbf{11}$_{-3}^{+1}$&\textbf{100}$_{-600}^{+400}$&\textbf{0}$^{+2800}$\T \B \\ 
&2017-10-12&0804640601\_S003&$0.575_{-0.002}^{+0.002}$&$21.7$&/&/&/&$\textbf{17}\pm4$&\textbf{-5000}$_{-1400}^{+1400}$&/\T \B \\ 
&2019-04-30&22213&$0.784_{-0.013}^{+0.013}$&$0.7$&\textbf{77}$_{-5}^{+7}$&\textbf{2600}$_{-700}^{+500}$&\textbf{5300}$_{-1200}$&\textbf{26}$_{-2}^{+1}$&\textbf{100}$_{-400}^{+300}$&/\T \B \\ 
&2021-07-14&23435&$0.512_{-0.006}^{+0.007}$&$1.6$&$\textbf{55}\pm7$&\textbf{900}$_{-400}^{+500}$&\textbf{3800}$_{-1100}^{+1400}$&\textbf{31}$_{-4}^{+5}$&\textbf{100}$_{-300}^{+300}$&/\T \B \\ 
&2021-07-15&24663&$0.56_{-0.007}^{+0.007}$&$1.5$&\textbf{61}$_{-6}^{+8}$&\textbf{1400}$_{-500}^{+600}$&\textbf{4900}$_{-700}^{+400}$&\textbf{32}$_{-3}^{+4}$&\textbf{-100}$_{-100}^{+100}$&/\T \B \\ 
\hline
&2003-05-01&3803&$0.264_{-0.002}^{+0.002}$&$11.5$&$\textbf{7}\pm1$&\textbf{-300}$_{-200}^{+200}$&/&\textbf{20}$_{-3}^{+2}$&\textbf{-400}$_{-200}^{+100}$&\textbf{1600}$_{-600}^{+700}$\T \B \\ 
H1743-322&2003-06-23&3805&$0.205_{-0.002}^{+0.002}$&$7.3$&$\textbf{7}\pm2$&\textbf{100}$_{-600}^{+600}$&/&\textbf{16}$_{-4}^{+3}$&\textbf{-200}$_{-600}^{+400}$&\textbf{0}$^{+4000}$\T \B \\ 
&2003-07-30&3806&$0.149_{-0.002}^{+0.002}$&$5.1$&\textbf{19}$_{-4}^{+3}$&\textbf{0}$_{-300}^{+400}$&\textbf{0}$^{+2700}$&\textbf{29}$_{-5}^{+4}$&\textbf{400}$_{-400}^{+400}$&\textbf{2000}$_{-1500}^{+1600}$\T \B \\ 
\hline
IGRJ17451-3022&2015-03-06&0748391201\_S001&$0.244_{-0.003}^{+0.003}$&$0.2$&\textbf{92}$_{-10}^{+11}$&\textbf{-1200}$_{-1100}^{+800}$&/&/&/&/\T \B \\ 
\hline

\end{tabular}
\end{table*}

\end{onecolumn}
\end{appendix}
\end{document}